\documentclass{aa}

\usepackage{verbatim}
\usepackage{graphicx}
\usepackage{epstopdf}
\usepackage{subfigure}
\usepackage{mathtools}
\usepackage{color,soul}

\newcommand{\mdot}{M$_\odot$ yr$^{-1}$}

\newcommand{\kms}{km~s$^{-1}$}

\title{Stellar Winds on the Main-Sequence II: the Evolution of Rotation and Winds}

\titlerunning{Stellar Winds on the Main-Sequence II}

\author{C. P. Johnstone\inst{\ref{vienna}} \and M. G\"{u}del\inst{\ref{vienna}} \and I. Brott\inst{\ref{vienna}} \and T. L\"{u}ftinger\inst{\ref{vienna}}}
\institute{
University of Vienna, Department of Astrophysics, T\"{u}rkenschanzstrasse 17, 1180 Vienna, Austria \label{vienna}
}

\abstract{}{
We study the evolution of stellar rotation and wind properties for low-mass main-sequence stars.
Our aim is to use rotational evolution models to constrain the mass loss rates in stellar winds and to predict how their properties evolve with time on the main-sequence.
}{
We construct a rotational evolution model that is driven by observed rotational distributions of young stellar clusters.
Fitting the free parameters in our model allows us to predict how wind mass loss rate depends on stellar mass, radius, and rotation.
We couple the results to the wind model developed in Paper~I of this series to predict how wind properties evolve on the main-sequence. 
}{
We estimate that wind mass loss rate scales with stellar parameters as $\dot{M}_\star \propto R_\star^2 \Omega_\star^{1.33} M_\star^{-3.36}$. 
We estimate that at young ages, the solar wind likely had a mass loss rate that is an order of magnitude higher than that of the current solar wind.
This leads to the wind having a higher density at younger ages; however, the magnitude of this change depends strongly on how we scale wind temperature. 
Due to the spread in rotation rates, young stars show a large range of wind properties at a given age.
This spread in wind properties disappears as the stars age. 
}{
There is a large uncertainty in our knowledge of the evolution of stellar winds on the main-sequence, due both to our lack of knowledge of stellar winds and the large spread in rotation rates at young ages.
Given the sensitivity of planetary atmospheres to stellar wind and radiation conditions, these uncertainties can be significant for our understanding of the evolution of planetary environments. 
}

\begin{document}

\maketitle


\section{Introduction}


Low-mass main-sequence stars slowly lose mass and angular momentum through magnetised stellar winds, though the properties of such winds are currently not well understood. 
These winds influence stellar rotation by removing angular momentum, causing their rotation to slow down as they age (\citealt{1967ApJ...148..217W}; \citealt{1967ApJ...150..551K}) and leading to older stars having weaker magnetic fields and lower levels of high energy radiation (\citealt{1997ApJ...483..947G}; \citealt{2005ApJ...622..680R}; \citealt{2012LRSP....9....1R}; \citealt{2014MNRAS.441.2361V}). 
The evolution of planetary atmospheres is highly sensitive to the surrounding stellar environments.
For example, high levels of \mbox{X-ray} and EUV radiation incident on a planetary atmosphere lead to inflation of the atmosphere and in some cases, hydrodynamic escape (\citealt{2003ApJ...598L.121L}; \citealt{2008JGRE..113.5008T}).
At the same time, strong winds compress the magnetosphere and can cause non-thermal erosion of the atmosphere a significant amount of atmospheric gas lies above the magnetospheric obstacle (\citealt{2010Icar..210....1L}; \citealt{2014A&A...562A.116K}).   
Therefore, the spin down of stars is highly significant in the study of the evolution of planetary atmospheres and the development of habitable planetary environments (\citealt{2014arXiv1407.8174G}).
Stellar winds can therefore severely influence the atmospheres of planets, both directly and indirectly through their influence on the rotation rates of their host stars.

Since the initial confirmation of the spin down of main-sequence stars as they age (\citealt{1967ApJ...150..551K}), a detailed observational understanding of the rotational evolution of main-sequence stars has been achieved, though many questions still remain (for summaries, see \citealt{1997ApJ...480..303K}, \citealt{2007prpl.conf..297H} and \citealt{2013arXiv1309.7851B}). 
Rotation periods are now known for tens of thousands of stars, with particularly interesting studies being surveys of rotation periods in stellar clusters of different ages. 
Stars start out their lives on the main-sequence with rotation rates that span two orders of magnitude, as can be seen in young clusters such as NGC~2547 ($\sim$40~Myr; \citealt{2008MNRAS.383.1588I}) and Pleiades ($\sim$120~Myr; \citealt{2010MNRAS.408..475H}). 
This spread in rotation rates can be traced back to the early pre-main sequence and has the appearance that most stars lie on one of two district branches (\citealt{1993ApJ...409..624S}; \citealt{2003ApJ...586..464B}). 
As stars age, their rotation rates quickly converge, as can be seen in older clusters such as M37 ($\sim$550~Myr; \citealt{2009ApJ...691..342H}) and NGC~6811 ($\sim$1~Gyr; \citealt{2011ApJ...733L...9M}).
This convergence happens quicker for higher mass stars (\citealt{1997ApJ...480..303K}).
At 100~Myr, most solar mass stars already lie on the slowly rotating track, and by 500~Myr, almost all solar mass stars have converged.
On the other hand, for 0.5~M$_\odot$ stars, rotation rates are much more distributed at 100~Myr, and much less convergence has taken place by 500~Myr.
Going to masses below 0.35~M$_\odot$, there is evidence that the distribution of rotation rates in fact broadens between the ZAMS and 10~Gyrs (\citealt{2011ApJ...727...56I}), though this is difficult to study due to the difficulty in measuring stellar ages for non-cluster stars.
We do not consider such low stellar masses in this paper.
After the convergence of stellar rotation rates there exists a one-to-one relation between stellar age and rotation rate, at a given mass, given approximately by $\Omega_\star \propto t^{-0.5}$ (\citealt{1972ApJ...171..565S}; \citealt{2007ApJ...669.1167B}; \citealt{2008ApJ...687.1264M}).

A likely consequence of the spin down of stars and the decay of their magnetic dynamos is that their winds evolve in time. 
The link between magnetic activity and winds is poorly understood, so constraining the exact dependence of winds on stellar mass and rotation is difficult. 
Intuitively, we would expect that the solar wind mass loss rate was higher in the past as a result of stronger magnetic activity.  
Observationally, there is some evidence for this (\citealt{2002ApJ...574..412W}; \citealt{2005ApJ...628L.143W}), though the current observational picture is unclear. 
There have been some theoretical studies of the past solar wind properties. 
\citet{1976ApJ...210..498B} assumed that the processes responsible for the acceleration of the current solar wind, which they called simply `thermal', have remained constant with age, and estimated that the early solar wind could have had velocities of $\sim$4000~\kms\hspace{0mm} simply due to magneto-rotational acceleration.
However, they found little influence on the mass loss rate. 
In reality, the `thermal' acceleration is unlikely to be constant, but will depend on rotationally driven changes in the star's magnetic field. 
More recent models for the evolution of the solar wind have concentrated on the influences of these mechanisms (\citealt{2004A&A...425..753G}; \citealt{2007A&A...463...11H}; \citealt{2011JGRA..116.1217S}; \citealt{2013PASJ...65...98S}).
A general property of these models is that the mass flux in the solar wind was higher at previous times than it is currently, though it is unclear by how much this was the case. 
What is also unclear is how the wind speed has changed in time.
Observations of the coronae of young solar analogues show that coronal temperatures decay with magnetic activity as stars spin down (\citealt{1997ApJ...483..947G}; \citealt{2005ApJ...622..653T}), which suggests that the solar wind temperatures, and therefore speeds, were higher at young ages.
However, the link between coronal and wind properties is currently unclear, and so we should be cautious when making such conclusions (see Section~4.1 of Paper~I of this series for a detailed discussion on this issue).

Another open question about stellar winds is how they depend on stellar mass.
Naively, we would expect that lower mass stars have lower mass loss rates due simply to the smaller surface areas. 
However, the situation is unlikely to be so simple; for example, at a given rotation rate, low-mass stars have higher \mbox{X-ray} surface fluxes, $F_\text{X}$, than high-mass stars in the unsaturated regime. 
\citet{2011MNRAS.412..351V} used 3D magnetohydrodynamic (MHD) simulations of the wind from the $\sim$0.3~M$_\odot$ rapid-rotator V374~Peg and argued that the star is likely to have a more powerful wind than the current Sun. 
\citet{2014ApJ...790...57C} modelled the wind of the $\sim$0.3~M$_\odot$ moderate-rotator EV~Lac and derived a mass loss rate per unit surface area that is an order of magnitude above the current solar wind. 
A significant open question when it comes to the mass dependence of stellar winds is the possible saturation of mass loss rates at fast rotation. 
In \mbox{X-rays}, the saturation threshold for low-mass stars is lower than for high-mass stars, meaning that the most rapidly rotating low-mass stars are not able to become as active as the most rapidly rotating high-mass stars.
If something similar is happening for winds, then the mass loss rate could have significantly different mass dependences in the saturated and unsaturated regimes.

In this paper, we study the rotational evolution of low-mass main-sequence stars and the influence that this has on the properties of stellar winds. 
We concentrate on rotational evolution for two reasons: firstly, by fitting our rotational evolution model to the observational constraints, we attempt to derive a scaling law for wind mass loss rates as a function of stellar parameters, and secondly, we use rotational evolution to predict the time evolution of wind properties on the main-sequence.
This study is made possible by the recent derivation of the dependence of wind torque on mass loss rate (\citealt{2012ApJ...754L..26M}), and a recent series of excellent observational campaigns measuring rotation in young stellar clusters (e.g. \citealt{2007MNRAS.377..741I}; \citealt{2009ApJ...691..342H}; \citealt{2009MNRAS.392.1456I}; \citealt{2009ApJ...695..679M}; \citealt{2010MNRAS.408..475H}).
In Section~\ref{sect:rotevo}, we develop our rotational evolution model, where the mass loss rate is incorporated into the model as a free parameter.
In Section~\ref{sect:obsconstraints}, we collect observed rotation periods for over 2000 stars in seven young clusters to derive observational constraints on the rotational evolution of stars on the main-sequence. 
In Section~\ref{sect:rotevofitting}, we fit the free parameters in our rotational evolution model to the observational constraints, allowing us to derive wind mass loss rates as a function of stellar mass, radius, and rotation rate.
In Section~\ref{sect:rotevoresults}, we summarise the results of our rotational evolution model. 

The results of our rotational evolution models are also that we can predict both the rotational evolution of individual stars and the distributions of rotation rates at the ages and stellar masses that we consider. 
In the second part of this paper, we couple the results of the rotational evolution model to our wind model derived in Paper~I. 
In Section~\ref{sect:windevo}, we show how stellar wind properties, such as mass fluxes and wind speeds, evolve on the main-sequence.
Finally, in Section~\ref{sect:summary}, we summarise our results and discuss their implications for the evolution of planetary habitability.

\section{Rotational Evolution: the Physical Model} \label{sect:rotevo}

In order to study the evolution of stellar winds from stars of different masses and ZAMS rotation rates, we first need a good understanding of the rotational evolution of main-sequence stars. 
For this purpose, we develop a rotational evolution model for stars of masses between 0.4~M$_\odot$ and 1.1~M$_\odot$ and between ages of 100~Myr and 5~Gyrs. 
In order to constrain the free parameters in our rotational evolution model, we collect measured rotation periods for several young clusters from the literature. 


As summarised by  \citet{1997ApJ...480..303K}, the basic ingredients of any rotational evolution model are the initial rotation rate, the internal structure of the star, and the rate at which angular momentum is removed from the star by the wind. 
Since we do not consider rotational evolution at ages younger than 100~Myr, we constrain our initial conditions using observations of rotation in young clusters with ages close to 100~Myr.
To determine the internal structure of the star, and most importantly, the moments of inertia as a function of stellar mass and age, $I_\star(M_\star, t)$, we run stellar evolution models for each stellar mass of interest.
For the stellar evolution calculations, we use the 1D hydrodynamic stellar evolution code described in \citet{2000ApJ...528..368H}, \citet{2011A&A...530A.115B}, and \citet{2011A&A...530A.116B}.
For the simulations, we assume solar metallicity and neglect the effects of stellar rotation and mass loss (which would anyway only have negligible effects on the models). 
In addition to the stellar moments of inertia, we use these simulations to derive stellar radii and bolometric luminosities as a function of stellar mass and age, which we use as important input in the rest of the calculations in this paper. 

An important ingredient in several rotational evolution models is core-envelope decoupling. 
Core-envelope decoupling happens when the timescale for angular momentum transport within the star is not negligible compared to the timescale over which the moment of inertia of the star changes and the timescale over which angular momentum is removed from the stellar surface by the wind. 
The result would be that stars arrive at the ZAMS with cores that rotate more rapidly than the surfaces. 
Several previous studies have included a core-envelope coupling timescale as a free parameter in their models and found that short coupling timescales of a few 10~Myr are required to reproduce the observational constraints (e.g. \citealt{1991ApJ...376..204M}; \citealt{2011MNRAS.416..447S}; \citealt{2013A&A...556A..36G}), which leads to moderate differential rotation between the core and the envelope.
In order to properly take into account core-envelope decoupling, we would ideally have to model rotational evolution before the ZAMS in order to determine the initial core rotation rates for ZAMS stars.
Given the short coupling timescales found in these previous studies and the fact that we are able to fit our models to the observational constraints without assuming core-envelope decoupling, for simplicity, we assume solid-body rotation\footnotemark.

\footnotetext{
Recently, \citet{2015arXiv150205801G} argued that low mass stars ($\approx$0.5~M$_\odot$) have much longer core-envelope coupling timescales of >100~Myr.
}

The fact that a rotating magnetised star that loses mass through an ionised winds will spin down was recognised immediately after the first predictions of the existence of the solar wind (\citealt{1958ApJ...128..664P}; \citealt{1962AnAp...25...18S}).
The reason for this spin down is that the wind has a larger specific angular momentum than the material in the star, mostly due to the angular momentum contained in the stresses of the magnetic field (\citealt{1967ApJ...148..217W}).
As the wind propagates outwards, the angular momentum held in the magnetic field is then transferred to the gas (see Fig.~3 of \citealt{2014MNRAS.438.1162V}).
Another way to think about it is that in the absence of magnetic fields, the angular momentum per unit mass of the wind in the equatorial plane at the surface of the star is \mbox{$\Omega_\star R_\star^2$}.
However, in the presence of a magnetic field, the wind torque is equivalent to what it would be if the material was held in strict corotation with the star out to the Alfv\'{e}n radius, $R_\text{A}$, and then released, and therefore the angular momentum per unit mass lost in the wind in the equatorial plane is \mbox{$\Omega_\star R_\text{A}^2$}.

The rate at which stars lose angular momentum depends on the stellar magnetic field, the wind mass loss rate, the stellar mass and radius, and the angular velocity of the star.
Relating the wind torque to these parameters is difficult.
Most rotational evolution models have used a formula for calculating wind torques derived by \citet{1988ApJ...333..236K} based on simple 1D approximations for the wind and the magnetic field. 
In most formulations of the \citet{1988ApJ...333..236K} formula that have been used, the wind torque does not depend on the mass loss rate, which is not reasonable.
A more realistic formula for wind torques was derived by \citet{2012ApJ...754L..26M} using a grid of 2D MHD wind models.
The main advantage of this approach is that simplifying assumptions about the Alfv\'{e}n surface do not need to be made since this is solved in the MHD simulations self-consistently.

Formulating rotational evolution models using these torque formulae is a tricky business that requires detailed knowledge of the field strengths, mass loss rates (in models where the wind torque has a mass loss rate dependence), and the internal structure of the star.
In order to explain the fast spin down of young rapidly rotating stars, it is necessary that their winds carry away angular momentum faster than the current solar wind (\citealt{1976ApJ...210..498B}), which can easily be explained by higher magnetic field strengths and mass loss rates.
In general, Skumanich style spin down requires that the wind torque depend on $\Omega_\star^3$.
However, it is difficult to explain the slow spin down of the most rapidly rotating stars with such a strong dependence.
Instead, a much weaker dependence between wind torque and rotation above a certain threshold is required (e.g. \citealt{1991ApJ...376..204M}).
\citet{1996ApJ...462..746B} and \citet{1997A&A...326.1023B} showed that this threshold must be lower for low-mass stars in order to explain the slower spin down of rapidly rotating low-mass stars.
The reason for this threshold is likely the saturations of the wind mass loss rates and magnetic field strengths at high rotation rates.

In the absence of core-envelope decoupling, the rate at which a star spins down is given by

\begin{equation} \label{eqn:angularmomentumall}
\frac{d\Omega_\star}{dt} = \frac{1}{I_\star} \left( \tau_{\text{w}} - \frac{dI_\star}{dt} \Omega_\star \right),
\end{equation}

\noindent where $I_\star$ is the star's moment of inertia and $\tau_{\text{w}} =dJ/dt$ is the torque on the star by the wind, where $J$ is the star's angular momentum.
On the pre-main-sequence, the term involving $dI_\star/dt$ is very important; on the main-sequence, this term is negligible. 
Therefore, in order to predict how a star's rotation evolves in time, we require a model for calculating the wind torque.

To calculate the wind torque, we use the formula derived by \citet{2012ApJ...754L..26M}, where the torque is related to the stellar mass, $M_\star$, stellar radius, $R_\star$, stellar angular velocity, $\Omega_\star$, magnetic field strength, $B_{\text{dip}}$,  and mass loss rate, $\dot{M}_\star$, as

\begin{equation} \label{eqn:matttorque}
\tau' = K_1^2 B_{\text{dip}}^{4m} \dot{M}_\star^{1-2m} R_\star^{4m+2} \frac{\Omega_\star}{(K_2^2 v_{\text{esc}}^2 + \Omega_\star^2 R_\star^2)^m},
\end{equation}

\noindent where $K_1 = 1.3$, $K_2 = 0.0506$, $m = 0.2177$, $v_{\text{esc}}$ is the surface escape velocity, and all quantities should be in cgs units.
This torque formula is calculated from the results of 50 MHD wind simulations and implies that when all other parameters are held constant, the wind torque depends on $\dot{M}_\star^{0.56}$.
The fact that the $\dot{M}_\star$ dependence should be weaker than linear can be seen from simpler considerations.
The torque associated with the wind in the equatorial plane is given by $\tau_\text{w} = \dot{M}_\star \Omega_\star R_\text{A}^2$.
The mass flux in a wind can be increased either by increasing the wind speed or by increasing the wind density.
Increasing the mass loss rate while keeping the magnetic field strength constant leads to a lower value of the Alfv\'{e}n radius, either because the wind accelerates faster or because the higher wind density leads to lower Alfv\'{e}n speeds.
Therefore, increasing the mass loss rate leads to a slower than linear increase in the wind torque. 

This formulation represents a significant advance in our ability to predict stellar wind torques and is likely more realistic than the torque formula from \citet{1988ApJ...333..236K}. 
However, as we discuss in more detail in Section~\ref{sect:uncertainty}, the simulations of  \citet{2012ApJ...754L..26M} did not take into account all of the detailed physical mechanisms operating in stellar winds.
Furthermore, our constraints on the values of $B_\text{dip}$ and $\dot{M}_\star$ are likely to contain uncertainties.
To take into account these uncertainties, we add an extra variable into the torque formula, such that the actual torque that we use in our model, $\tau_{\text{w}}$, is related to the torques calculated using Eqn.~\ref{eqn:matttorque} by

\begin{equation} \label{eqn:modifyMatttorque}
\tau_{\text{w}} = K_\tau \tau',
\end{equation}

\noindent where $K_\tau$ is a free parameter in our model.
A similar factor was introduced into Eqn.~\ref{eqn:matttorque} of \citet{2013A&A...556A..36G} (though they did so by changing the value of $K_1$), who found $K_\tau \approx 2$.

In order to calculate the torque on a star from the wind, we therefore need to be able to calculate $\dot{M}_\star$ and $B_{\text{dip}}$. 
We derive the mass loss rates by fitting our rotational evolution model to the observational constraints.
We make the assumption that in the unsaturated regime, the mass loss rate per unit surface area of a star has power-law dependences on its mass and rotation rate, such that

\begin{equation} \label{eqn:Mdotassumption}
\dot{M}_\star = \dot{M}_\odot \left( \frac{R_\star}{R_\odot} \right)^2  \left( \frac{\Omega_\star}{\Omega_\odot} \right)^a  \left( \frac{M_\star}{M_\odot} \right)^b,
\end{equation}

\noindent where $a$ and $b$ are free parameters.
This assumption is likely a simplification; in real stellar winds it might not be possible to genuinely separate the stellar mass and rotation dependences in this way. 
We discuss our fits to these parameters in detail in Section~\ref{sect:rotevofitting}.
We choose to include a surface area dependence in the above assumption for mass loss rate since this is likely to be more realistic. 
However, this is unlikely to influence our results significantly given that we determine the strength of the mass dependence of $\dot{M}_\star$ by fitting the rotational evolution models.
Removing the surface area dependence would simply lead to us deriving a different value of $b$.
As discussed below, we assume that the dependence of $\dot{M}_\star$ on $\Omega_\star$ saturates at fast rotation.

How the magnetic field strength should be calculated is a difficult issue.
It has been known since \citet{1967ApJ...150..551K} and \mbox{\citet{1972ApJ...171..565S}} that rapidly rotating stars have higher levels of magnetic activity than slowly rotating stars. 
At a given stellar mass, the magnetic field strength has approximately a power law dependence on rotation rate, such that $B_\star \propto \Omega_\star^d$ for slow rotators, and saturates at fast rotation (e.g. \citealt{2012LRSP....9....1R}).
However, exact quantification of this relation is difficult and requires knowledge of the index in the power law, the constant of proportionality, and the (mass-dependent) location of the saturation threshold. 


Most studies of magnetic field strengths in the past have focused on the surface average field strength, often denoted $fB$, where $f$ is the surface magnetic filling factor and $B$ is the magnetic field strength.
\citet{1996IAUS..176..237S} and \citet{2001ASPC..223..292S} found that $fB$ is proportional to $P_\text{rot}^{-1.7}$ and $Ro^{-1.2}$ respectively.
The Rossby number, $Ro$, is given by $Ro \equiv P_\text{rot} / \tau_\star$, where $\tau_\star$ is the convective turnover time. 
In the latter case especially, this relation is likely to be made artificially shallower due to the inclusion of stars in the saturated regime and the lack of real measurements for slow rotators (the measurements for two of the three stars slower than 12~days were upper limits in the sample of \citealt{1996IAUS..176..237S}). 
At the same time, measurements of $fB$ based on Zeeman broadening are very difficult for weak magnetic fields and can be difficult to distinguish from rotational broadening (\citealt{2014IAUS..302..156R}). 
A measure of magnetic field strength that is likely to be more reliable is X-ray emission.
Combining the relation between magnetic flux, $\Phi_\text{B}$, and \mbox{X-ray} luminosity, $L_\text{X}$, of $L_\text{X} \propto \Phi_\text{B}^{1.15}$ from \citet{2003ApJ...598.1387P} with the relation between $R_\text{X}$ (where \mbox{$R_\text{X} = L_\text{X} / L_\text{bol}$}) and Rossby number of $R_\text{X} \propto P_{\text{rot}}^{-2.18}$ from \citet{2011ApJ...743...48W} would imply for a given stellar mass that $fB \propto P_{\text{rot}}^{-1.9}$, though using the steeper relation of $R_\text{X} \propto P_{\text{rot}}^{-2.7}$ from \citet{2011ApJ...743...48W} would imply that $fB \propto P_{\text{rot}}^{-2.3}$. 

We stress, however, that the magnetic field term in the torque formula (Eqn.~\ref{eqn:matttorque}) is the \emph{equatorial} strength of the dipole component of the field and does not correspond to the surface average field strength.
Naively, we would expect that the strengths of the different components of the field scale with rotation in the same way.
However, using the results of recent Zeeman-Doppler Imaging studies of over 70 stars, \citet{2014MNRAS.441.2361V} showed that the large scale field strength averaged over the stellar surface\footnotemark, $\bar{B}_V$, scales with rotation as $\bar{B}_V \propto P_{\text{rot}}^{-1.32}$, which is weaker than the relations found for $fB$.
Although this is not the dipole component of the field, there is an approximately linear scaling between $\bar{B}_V$ and $B_\text{dip}$ in the sample of stars (Aline Vidotto, private communication).
One thing that is unclear is whether the magnetic field strength scales with rotation rate directly or with some other parameter, such as Rossby number. 
We consider the correlation between $\bar{B}_V$ and Rossby number shown in \citet{2014MNRAS.441.2361V} to be better than the correlation between $\bar{B}_V$ and rotation period (compare their Fig.~3 and Fig.~4), and so we assume that $B_{\text{dip}}$ scales with Rossby number.
Therefore, when $\Omega_\star \le \Omega_{\text{sat}}$, we get

\footnotetext{
The subscript $V$ here represents magnetic field strengths that have been derived from Stokes~V measurements, as opposed to Stokes~I measurements which are used to derive $fB$. 
}

\begin{equation} \label{eqn:dipoleRossby}
B_{\text{dip}} = B_{\text{dip},\odot} \left( \frac{\Omega_\star \tau_\star}{\Omega_\odot \tau_\odot} \right)^{1.32},
\end{equation}

\noindent and when $\Omega_\star \ge \Omega_{\text{sat}}$, the dipole field strength remains at the saturation value. 
The convective turnover times, $\tau_\star$ and $\tau_\odot$, should not be confused with the wind torque, $\tau_\text{w}$.

A very important part of rotational evolution models is the saturation of the magnetic field strength and the mass loss rate.
In the saturated regime, the dependence of the wind torque on $\Omega_\star$ must be much weaker than in the unsaturated regime.
Without saturation, the most rapidly rotating stars would spin down much faster than is observed. 
Based on the slow rotational evolution of young rapidly rotating stars, there can be no serious doubt that both mass loss rate and the dipole component of the field saturate at fast rotation. 
However, it is not clear that saturation takes place at the exact same rotation rates for the mass loss rate and the magnetic field.
Similarly, it is not clear that saturation takes place at the same rotation rates for different components of the magnetic field (e.g. $B_\text{dip}$, $fB$, etc.), and for indicators of magnetic activity such as \mbox{X-ray} emission. 
For simplicity, we assume that both $\dot{M}_\star$ and $B_\text{dip}$ saturate at the same rotation rate for a given stellar mass. 
The exact rotation rate at which saturation takes place is not well constrained, although good hints on the saturation level can be derived from \mbox{X-ray} emission. 
For example, \citet{2003A&A...397..147P} estimated that the rotation period for saturation depends on stellar luminosity as $P_{\text{sat}} \approx 1.2 \left( L_{\text{bol}} / L_\odot \right)^{-1/2}$, where $P_\text{sat}$ is in days.
A similar result of $P_{\text{sat}} \approx 1.6 \left( L_{\text{bol}} / L_\odot \right)^{-1/2}$ was estimated using a larger sample of stars by \citet{2014arXiv1408.6175R}.
Similarly, \citet{2011ApJ...743...48W} determined that saturation in \mbox{X-ray} emission takes place for all stars at approximately a Rossby number of 0.13.
How the saturation rotation rate is then determined from the saturation Rossby number depends on the exact form of the convective turnover time.
The convective turnover times derived by \citet{2011ApJ...743...48W} imply that for 1~M$_\odot$ and 0.5~M$_\odot$ stars, saturation occurs at rotation periods of 1.9~days and 4.7~days respectively. 
In our model, we make the assumption that the saturation rotation rate for the magnetic field and the mass loss rate has a power law dependence on stellar mass,

\begin{equation} \label{eqn:saturation}
\Omega_{\text{sat}} \left( M_\star \right) = \Omega_{\text{sat},\odot} \left( \frac{M_\star}{M_\odot} \right)^c,
\end{equation}

\noindent where $\Omega_{\text{sat},\odot}$ is the saturation rotation rate for solar mass stars.
From the results of \citet{2003A&A...397..147P} and \citet{2014arXiv1408.6175R}, we would expect that $\Omega_{\text{sat},\odot}$ is 21~$\Omega_\odot$ and 17~$\Omega_\odot$ respectively.
On the other hand, the results of \citet{2011ApJ...743...48W} suggest that $\Omega_{\text{sat},\odot}$ is 14.5~$\Omega_\odot$.
In our models, we make the assumption that $\Omega_{\text{sat},\odot}$ is 15~$\Omega_\odot$ and treat $c$ as a free parameter.
The results of \citet{2003A&A...397..147P} and \citet{2014arXiv1408.6175R} suggest that $c \sim 2$. 
On the other hand, the results of \citet{2011ApJ...743...48W} suggest that $c \sim 1.3$.
In Section~\ref{sect:rotevoresults}, we derive the values of the other free parameters in our model by fitting them to observational constraints discussed in Section~\ref{sect:obsconstraints}.
We find that in order to properly reproduce the rotational evolution of lower mass stars, values of $c$ larger than 2 are required, and we therefore set $c=2.3$ in our rotational evolution model. 
We use this value since it is the closest value to the above \mbox{X-ray} constraints that allows us to get visually acceptable fits to the observational rotational evolution.

\begin{figure}
\centering
\includegraphics[trim=15mm 10mm 10mm 10mm,width=0.45\textwidth]{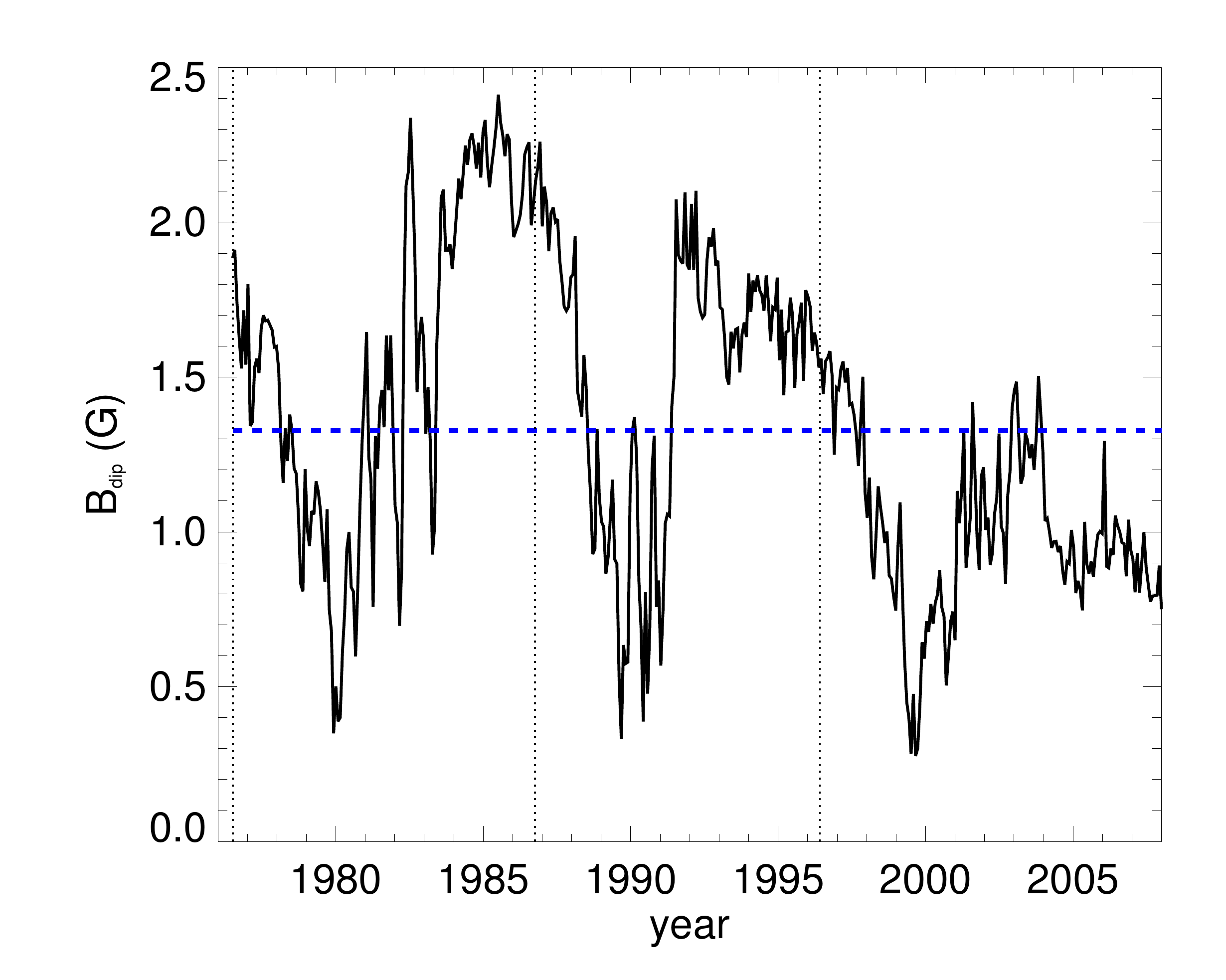}
\caption{
Plot showing the time variations in the equatorial strength of the dipole component of the solar magnetic field as measured by the Wilcox Solar Observatory. 
The dashed horizontal line shows \mbox{$B_{\text{dip}} = 1.35$~G}.
The dotted vertical lines show the beginnings of the previous three activity cycles.
}
 \label{fig:solarBdip}
\end{figure}

Clearly, to use Eqn.~\ref{eqn:dipoleRossby}, we need to know the equatorial strength of the dipole component of the Sun's magnetic field.
This is difficult because the dipole component on the Sun varies over the solar cycle, with a maximum strength of a few G around cycle minimum.
Therefore, given that the mass loss rate of the solar wind is approximately constant over the solar cycle, we expect that the wind torque on the Sun has a time dependence due to changes in the dipole field strength. 
To see how the strength of the dipole component of the field varies over the solar cycle, we use dipole field strengths derived from magnetic maps of the solar surface by the Wilcox Solar Observatory (WSO)\footnotemark.
WSO has been measuring the solar magnetic field daily for almost forty years, and has derived synoptic magnetic maps for Cycle~21, Cycle~22, and Cycle~23.
In Fig.~\ref{fig:solarBdip}, we show their measurements of the dipole field strength over these three cycles. 
Since we are interested in the evolution of stellar rotation rates over hundreds of millions of years, we are only interested in the average torque exerted on the star, and not variations over short time scales.
Since the torque is proportional to $B_{\text{dip}}^{0.87}$, we calculate the mean value of $B_{\text{dip},\odot}^{0.87}$ over these cycles, which corresponds to $B_{\text{dip},\odot} \sim 1.35$~G, which we then take as $B_{\text{dip},\odot}$ in Eqn.~\ref{eqn:dipoleRossby}.
Since the $B_{\text{dip},\odot}^{0.87}$ term acts simply as a multiplicative constant in Eqn.~\ref{eqn:matttorque}, the exact value of $B_{\text{dip},\odot}$ does not matter and changing it would just lead to a different value of the free parameter $K_\tau$. 

\footnotetext{
The data from the Wilcox Solar Observatory can be obtained from \url{wso.stanford.edu}.
}

We now have a rotational evolution model with four free parameters. 
These are $K_\tau$ in Eqn.~\ref{eqn:modifyMatttorque}, $a$ and $b$ in Eqn.~\ref{eqn:Mdotassumption}, and $c$ in Eqn.~\ref{eqn:saturation}.
We constrain these parameters in Section~\ref{sect:rotevofitting}.

\begin{figure*}
\includegraphics[trim=10mm 5mm 10mm 10mm,clip=true,width=0.49\textwidth]{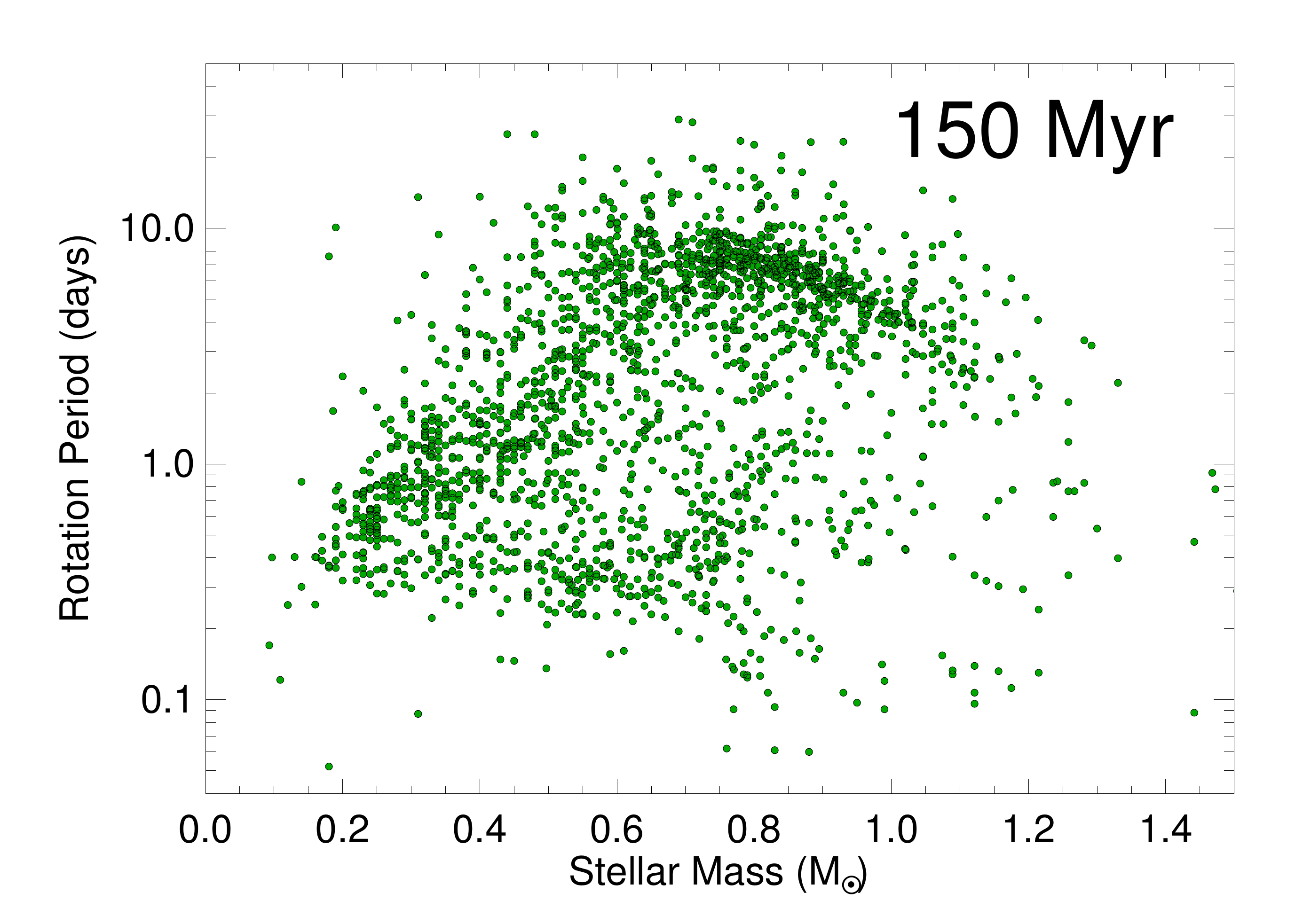}
\includegraphics[trim=10mm 5mm 10mm 10mm,clip=true,width=0.49\textwidth]{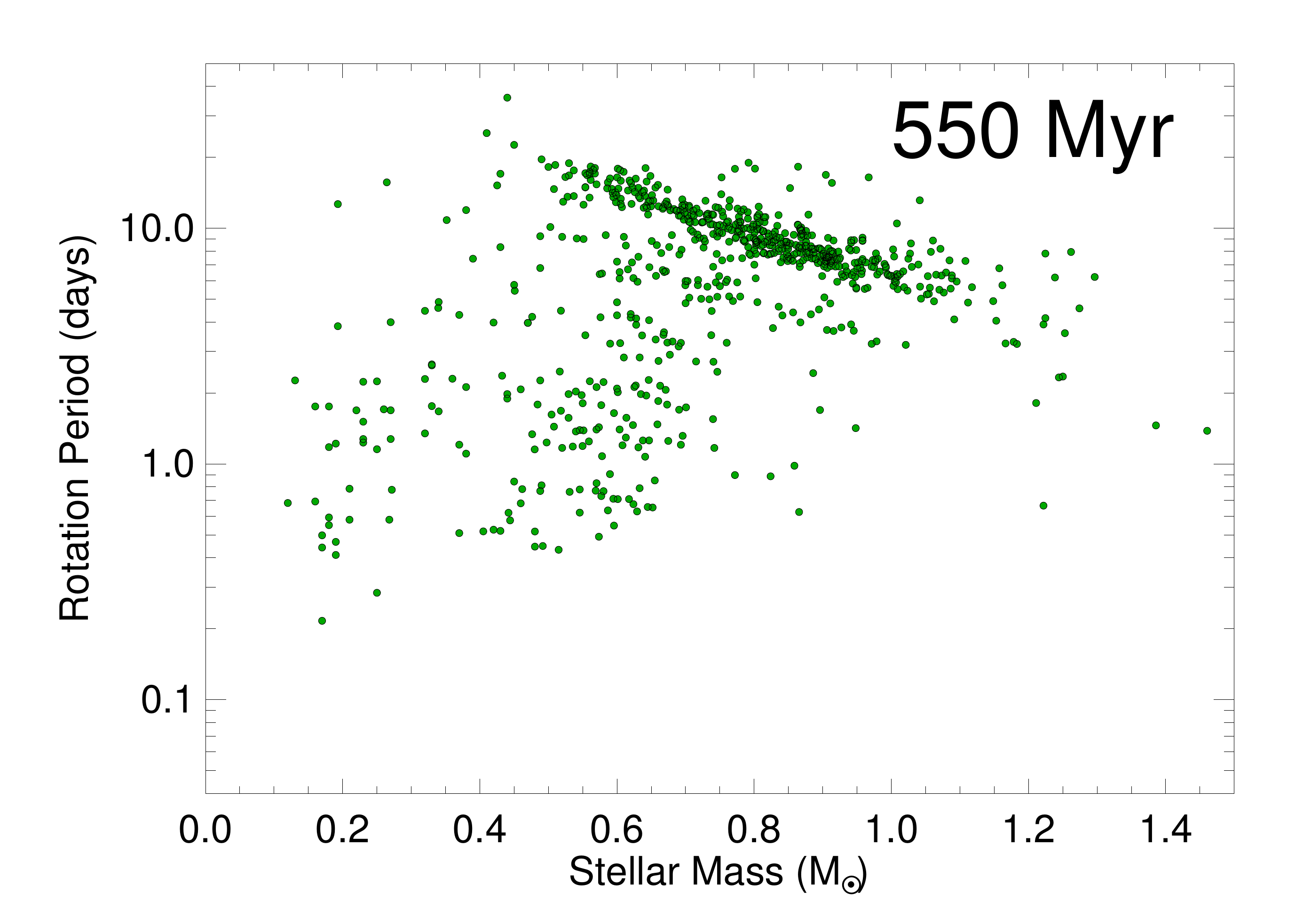}
\caption{
Plots showing the distributions of stellar rotation rates at $\sim$150~Myr (\emph{left panel}) and at $\sim$550~Myr (\emph{right panel}).
The distribution at $\sim$150~Myr is a combination of measurements in the Pleiades, M50, M35, and NGC~2516 and the distribution at $\sim$550~Myr is a combination of measurements from M37 and Praesepe, as discussed in the text.
Histograms showing the distributions of rotation periods in different mass bins for these two sets of stars are shown in Fig.~\ref{fig:supercluster100hist}.
}
 \label{fig:supercluster100}
\end{figure*}

\begin{figure*}
\includegraphics[trim=15mm 5mm 15mm 15mm,clip=true,width=0.49\textwidth]{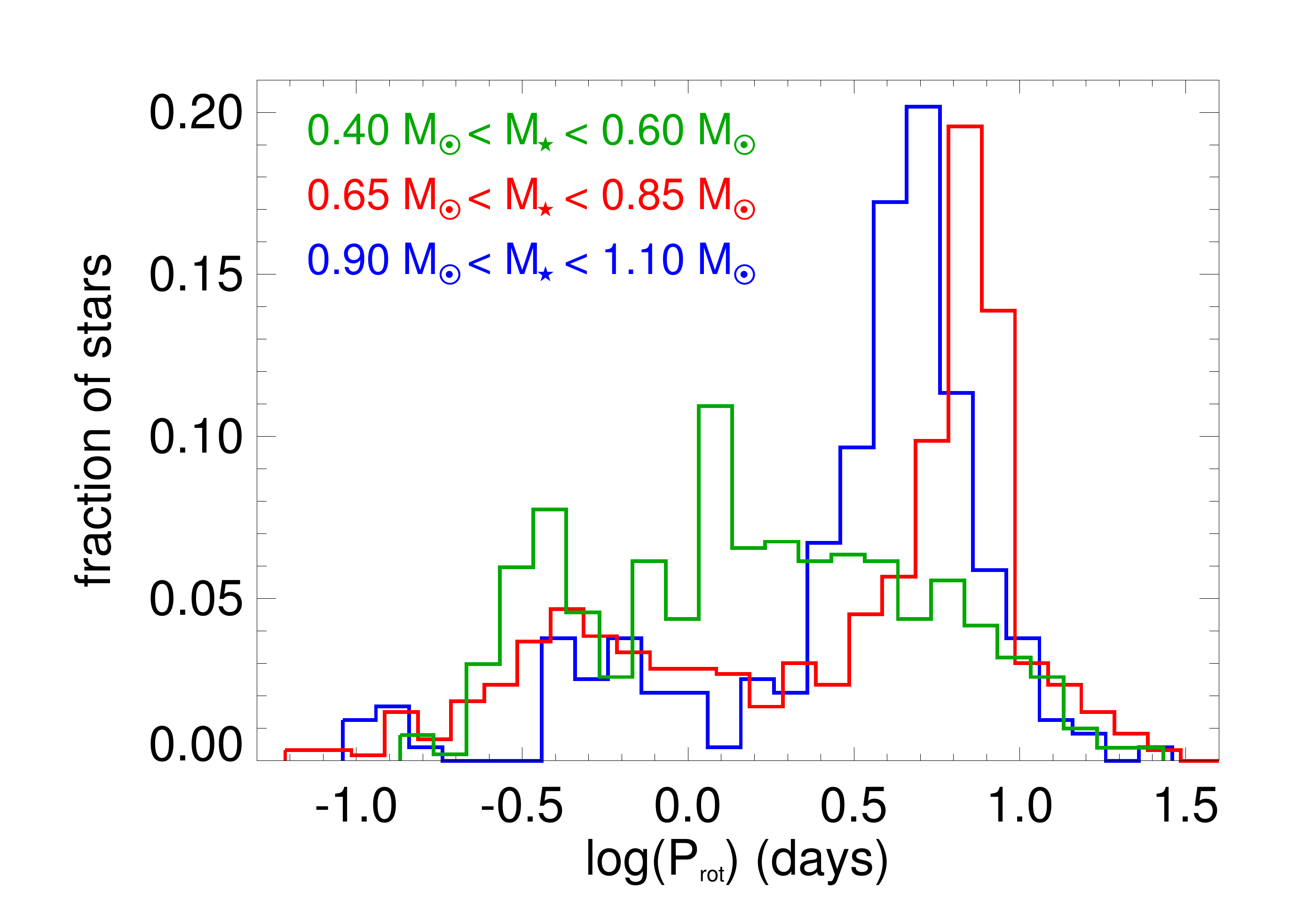}
\includegraphics[trim=15mm 5mm 15mm 15mm,clip=true,width=0.49\textwidth]{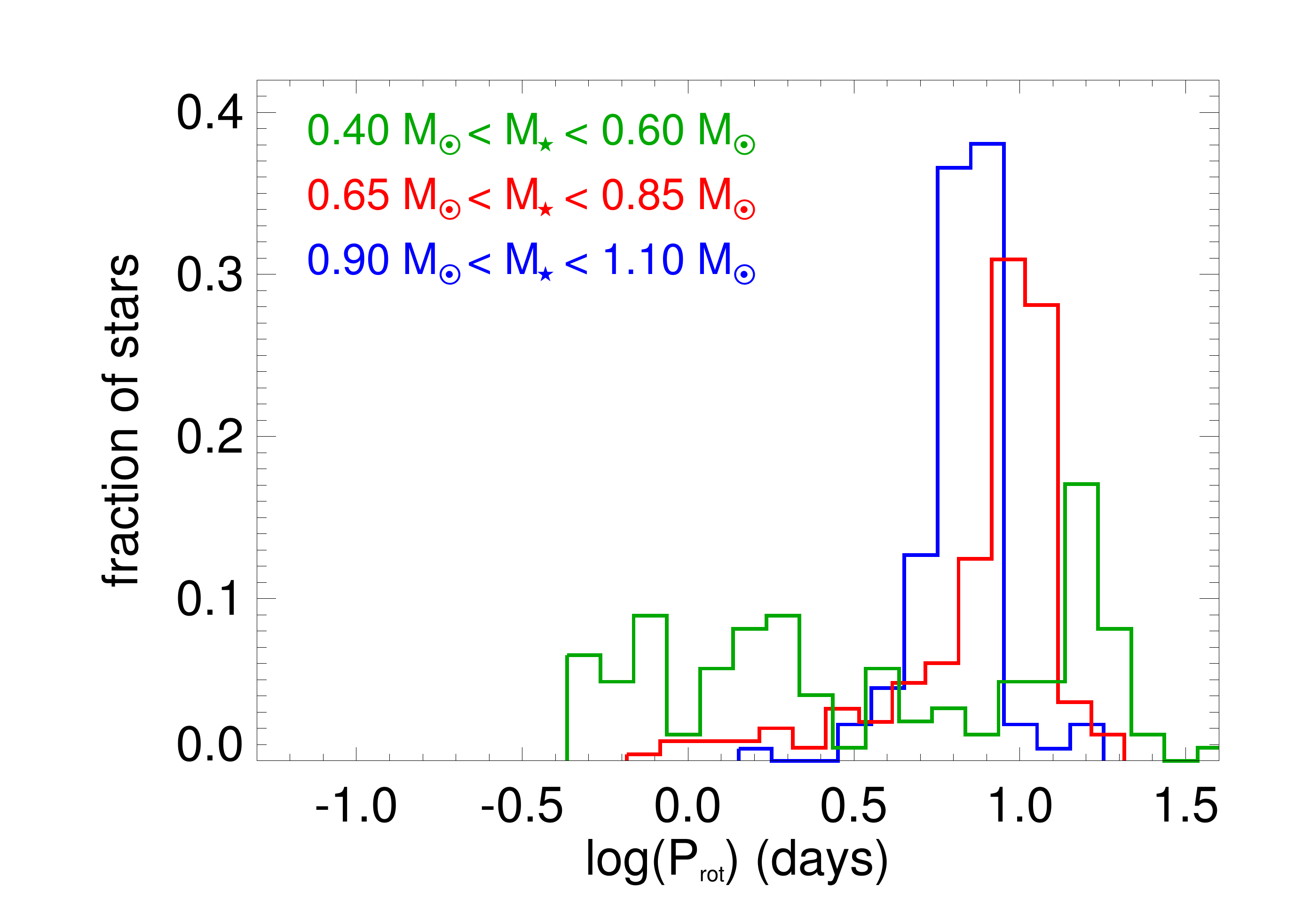}
\caption{
Histograms showing distributions of stellar rotation rates at $\sim$150~Myr (\emph{left panel}) and at $\sim$550~Myr (\emph{right panel}) at three different mass bins.
The stellar samples from which these distributions are derived are shown in Fig.~\ref{fig:supercluster100}.
}
 \label{fig:supercluster100hist}
\end{figure*}

\section{Rotational Evolution: Observational Constraints} \label{sect:obsconstraints}

\begin{table*}
\centering 
\begin{tabular}{ccccccc}
Mass Bin ($M_\odot$) & $n_\star$ & $\bar{\Omega}_\star$ ($\Omega_\odot$)& $\sigma_\Omega$ ($\Omega_\odot$) & $\Omega_{10}$ ($\Omega_\odot$) & $\Omega_{50}$  ($\Omega_\odot$) & $\Omega_{90}$  ($\Omega_\odot$)\\
\hline
\hline
Pleiades + M50 + M35 + NGC~2516 \\
($\sim$150~Myr) \\
\hline
All & 1556 & $26.00 \pm 1.08$ & $42.73 \pm 2.79$ & $3.13 \pm 0.06$ & $7.66 \pm 0.38$ & $74.01 \pm 2.54$ \\
0.50 & 463 & $30.62 \pm 1.54$ & $33.14 \pm 1.82$  & $3.61\pm 0.27$ & $17.35 \pm 1.42$ & $80.58 \pm 3.37$ \\
0.75 & 573 & $26.75 \pm 2.05$ & $48.92 \pm 5.09$ & $2.97 \pm 0.07$ & $5.30 \pm 0.33$ & $75.38 \pm 5.32$ \\
1.00 & 230 & $20.27 \pm 2.87$ & $43.65 \pm 6.97$ & $3.72 \pm 0.36$ & $6.66 \pm 0.26$ & $48.64 \pm 6.83$ \\
\hline
M37 + Praesepe\\
($\sim$550~Myr)\\
\hline
All & 639 & $7.17 \pm 0.39$ & $10.02 \pm 0.67$ & $1.97 \pm 0.07$ & $3.49 \pm 0.07$ & $19.02 \pm 1.86$ \\
0.50 & 119 & $15.67 \pm 1.52$ & $16.87 \pm 1.17$ & $1.57 \pm 0.04$ & $8.41 \pm 3.00$ & $43.88 \pm 5.05$ \\
0.75 & 249 & $4.59 \pm 0.33$ & $5.17 \pm 0.75$ & $2.22 \pm 0.03$ & $3.00 \pm 0.07$ & $8.33 \pm 1.41$ \\
1.00 & 134 & $4.30 \pm 0.15$ & $1.74 \pm 0.47$ & $3.14 \pm 0.12$ & $3.96 \pm 0.09$ & $5.42 \pm 0.69$ \\
\hline
NGC~6811 \\
($\sim$1000~Myr)\\
\hline
All & 51 & $2.69 \pm 0.12$ & $1.17 \pm 0.56$ & $2.41 \pm 0.06$ & $2.53 \pm 0.02$ & $2.75 \pm 0.10$ \\
1.00 & 36 & $2.75 \pm 0.16$ & $1.39 \pm 0.68$ & $2.31 \pm 0.08$ & $2.53 \pm 0.03$ & $2.83 \pm 0.96$ \\
\hline
\end{tabular}
\caption{
Table showing statistical properties of the distributions of rotation rates that we use to constrain our rotational evolution model.
All of the angular velocities are given in units of the solar angular velocity, $\Omega_\odot$, which we define as the Carrington rotation rate of $2.67 \times 10^{-6}$~rad~s$^{-1}$. 
As described in the text, the first two distributions are constructed by combining measured rotation periods from different clusters with approximately identical ages.
Both of these aggregate clusters are shown in Fig.~\ref{fig:supercluster100}.
For each cluster, we bin the stars by mass assuming a bin width of 0.2~M$\odot$ and calculate for each mass bin, the number of stars in each bin, $n_\star$, the mean angular velocity, $\bar{\Omega}_\star$, the standard deviation in the angular velocity, $\sigma_\Omega$, and the 10th, 50th, and 90th percentiles of the distributions, given by $\Omega_{10}$, $\Omega_{50}$, and $\Omega_{90}$ respectively.
The values marked as `All' are calculated from all stars with masses between 0.4~M$_\odot$ and 1.1~M$_\odot$.
The errors on the percentiles are calculated using the bootstrap method. 
We do not report statistics for NGC~6811 below the highest mass bin because very few stars lie below 0.9~M$_\odot$.
}
\label{tbl:percentiles}
\end{table*}

To constrain the free parameters in the rotational evolution model, and to study the distribution of stellar rotation rates (and therefore wind properties) at any given age, we collect measured rotation periods from the literature. 
The data that we collect are for seven young clusters: these are the Pleiades ($\sim$125~Myr), M50 ($\sim$130~Myr), M35 ($\sim$150~Myr), NGC~2516 ($\sim$150~Myr), M37 ($\sim$550~Myr), Praesepe ($\sim$580~Myr), and NGC~6811 ($\sim$1000~Myr). 
Although rotation data is available for other young clusters, we choose to use only these clusters due to the large number of measured rotation periods available for these clusters.
The only exception is NGC~6811, for which many fewer rotation periods are known, and all of which are for stars with masses near 1~M$_\odot$.
However, given the advanced age of this cluster, including it gives us essential information about the rotational distribution at 1~Gyr.
As can clearly be seen in Fig.~1 of \citet{2013arXiv1309.7851B}, the rotational distributions in the other young clusters that have been studied are consistent with the rotational distributions that we consider. 
A significant uncertainty in our analysis comes from uncertainties in the ages of the clusters that we use. 
Determining the ages of young clusters can be challenging (\citealt{2013arXiv1311.7024S}).
However, we do not consider uncertainties in the ages in the remainder of this paper, and just assume the given ages as correct.

We follow the method of several previous studies (\citealt{2007MNRAS.377..741I}; \citealt{2008A&A...489L..53B}; \citealt{2009ApJ...691..342H}; \citealt{2013A&A...556A..36G}) by using percentiles of the rotational distributions to fit our rotational evolution models.
As described below, we combine the measured rotation periods from stars in several clusters to get rotation period distributions at $\sim$150~Myr and $\sim$550~Myr, and we use rotation data from a cluster at 1000~Myr to constrain the rotation of solar mass stars at this age. 
We bin the stars in these distributions by mass and then calculate the 10th, 50th, and 90th percentiles of the distribution of angular velocities, $\Omega_\star$, for each mass at each age. 
We produce bins for stellar masses of 0.50, 0.75 and 1.00~M$_\odot$, assuming a bin width of 0.2~M$_\odot$ (such that the 0.5~M$_\odot$ bin contains all stars with masses between 0.40~M$_\odot$ and 0.60~M$_\odot$). 
The result is that for each stellar mass, we have time sequences of the rotation rates for stars at the 10th, 50th, and 90th percentiles of the rotational distribution.
We make the assumption that  a star starting at a certain position in the rotational distribution remains at this position as the distribution evolves, which means the time series of a given percentile can be used as observational constraints on the real spin down tracks of stars at that percentile.
This is likely to be reasonable given that a star's magnetic activity, and therefore the rate at which it spins down, is primarily determined by its rotation rate, which means that a star at a given rotation rate is unlikely to overtake a slower rotating star as it spins down.

The percentiles for each mass bin and each cluster are given in Table.~\ref{tbl:percentiles}.
The uncertainties in each value are calculated using the bootstrap method and are mostly very small given the large number of stars in each mass bin. 
However, these uncertainties should be taken with caution; they represent the uncertainties in the percentiles due to the fact that they are calculated from a finite number of stars, and do not take into account errors from other sources such as uncertainties in the rotation period measurements, biases in the distributions of period measurements, uncertainties in the stellar mass determinations, and uncertainties in to what extent the time sequences of percentiles can truly be used as a proxy for the rotation histories of stars. 
We do not consider the uncertainties in the percentiles when calculating the best fit values for the free parameters in the rotational evolution model.

The two youngest clusters in our sample are the Pleiades and M50 with approximate ages of 125~Myr and 130~Myr respectively (\citealt{1998ApJ...499L.199S}; \citealt{2003AJ....126.1402K}). 
\citet{2010MNRAS.408..475H} derived rotation periods for 383 Pleiades stars, of which 332 are in the mass range of interest to us. 
Rotation in M50 was studied by \citet{2009MNRAS.392.1456I} who measured rotation periods for 812 stars, of which 601 stars are in the mass range of interest to us.
At approximately the same age are M35 and NGC~2516, both with ages of 150~Myr (\citealt{2002AJ....124.1555V}; \citealt{2001A&A...375..863J}). 
\citet{2009ApJ...695..679M} determined rotation periods for 441 stars in M35, of which 324 stars are in the mass range of interest to us and have reported $(B-V)_0$ values that we use to calculate the stellar mass. 
The stars in the sample of \citet{2009ApJ...695..679M} all have masses above 0.6~M$_\odot$.
Luckily, the measurements of M35 are nicely complemented by the rotation periods derived for stars in NGC~2516 by \citet{2007MNRAS.377..741I}. 
They derived rotation periods for 362 stars, of which 140 are in the mass range of interest to us, all of which are below 0.7~M$_\odot$.
It is quite remarkable that the distributions of rotation rates at every stellar mass derived for these four clusters are completely consistent.
\citet{2010MNRAS.408..475H} provided a comparison of these four clusters and showed that they have almost identical rotation period distributions (see Fig.~14 and Fig.~15 of their paper). 
This provides good support for the assumption that a time series of percentiles in the rotational distributions can be used as a proxy for the evolution of individual stars. 
By combining these clusters, we have a distribution of $\sim$1500 stars which we assume represents rotation at 150~Myr. 

In Fig.~\ref{fig:supercluster100}, we show the distribution of rotation rates against stellar mass for stars at 150~Myr based on combining these four clusters.
The structure of this distribution is well known in the literature (e.g., \citealt{2007prpl.conf..297H}; \citealt{2013arXiv1309.7851B}).
The most prominent feature is a track of slowly rotating stars, with higher mass stars on the track having faster rotation rates than lower mass stars.
At 150~Myr, the track dominates for stars with masses above 0.8~M$_\odot$ and clearly extends down to stellar masses of 0.6~M$_\odot$.
At later ages, this track dominates even more as more quickly rotating stars spin down (\citealt{2003ApJ...586..464B}).
Eventually the track contains almost all stars (excluding tight binaries that are tidally locked and possibly stars below 0.35~M$_\odot$). 
Another feature that is clearly evident in Fig.~\ref{fig:supercluster100} is the track of rapid rotators, with a large distinct gap between the two tracks. 
The existence of the track of rapid rotators and the gap was seen in the Pleiades by \citet{1987A&AS...67..483V} and has been extensively discussed in the literature. 
The existence of these two tracks can be traced back to the pre-main sequence.
For example, two distinct tracks are clearly visible at all masses in the distribution of rotation rates in the $\sim$15~Myr old cluster h~Per (\citealt{2013A&A...560A..13M}).
Histograms showing the distributions of rotation periods in three different mass bins are shown in Fig.~\ref{fig:supercluster100hist}.

The next cluster in the age sequence that we use is M37, at an age of $\sim$550~Myr (\citealt{2008ApJ...675.1233H}). 
\citet{2009ApJ...691..342H} reported rotation periods for 575 stars in M37, of which 495 stars are in the mass range of interest to us. 
Given its age and the large number of measured rotation periods, this cluster is probably the most important one for our study since it gives us reliable constraints on how quickly the rapidly rotating stars of different masses spin down.
At approximately the same age is Praesepe, with an age of $\sim$570~Myr (\citealt{2011MNRAS.413.2218D}). 
Rotation periods for stars in Praesepe were reported by \citet{2011MNRAS.413.2218D}, \citet{2011MNRAS.413.2595S}, and \citet{2011ApJ...740..110A}, of which 108 are in the mass range of interest to us.
For our distribution of rotation rates at $\sim$550~Myr, we combine the rotation periods collected for M37 and Praesepe, and assume an age of 550~Myr for the resulting set of 603 stars.
The rotational distribution for stars in this aggregate cluster is shown in Fig.~\ref{fig:supercluster100}.

For solar mass stars, the slowly rotating track seen at 150~Myr has clearly spun down significantly and most of the rapid rotators have already converged onto the slowly rotating track. 
The 10th percentile for the 1.0~M$_\odot$ mass bin decreases from 4~$\Omega_\odot$ to 3~$\Omega_\odot$ between these two ages and the 90th percentile decreases from 50~$\Omega_\odot$ to 5~$\Omega_\odot$. 
At lower masses, although the rotational distribution has clearly evolved towards slower rotation, the rapidly rotating track seen at 150~Myr is still visible, and much less convergence of the two tracks has taken place. 
The 10th percentile for the 0.5~M$_\odot$ mass bin decreases from 4~$\Omega_\odot$ to 2~$\Omega_\odot$ between these two ages and the 90th percentile decreases from 86~$\Omega_\odot$ to 44~$\Omega_\odot$. 
Interestingly, the hole at intermediate masses and rotation rates seen at 150~Myr contains a large number of stars by 550~Myr.

The final cluster that we consider is NGC~6811, at an age of $\sim$1000~Myr. 
Rotation periods for 71 stars in NGC~6811 were reported by \citet{2011ApJ...733L...9M}, of which 51 stars are in the mass range of interest to us.
Unfortunately the sample of stars does not contain any stars below 0.8~M$_\odot$ and so cannot help us constrain the spin down timescales for the rapidly rotating low-mass stars.
All of the stars have converged onto the slowly rotating track, and no rapid rotators exist. 
There is almost no difference between the 10th and 90th percentiles.

Unfortunately, due to the difficulty of measuring the ages of individual stars, much less is known observationally about the evolution of rotation beyond 1~Gyr. 
Most of what is known is based on the current rotation rate and age of the Sun. 
It is often assumed in the literature that once all stars evolve onto the slowly rotating track seen in young clusters, this track spins down in a way that preserves its shape on the rotation-mass diagram, and therefore, the rotation period as a function of stellar mass and age can be represented as the product of two functions, i.e. $P_{\text{rot}} (M_\star, t) = f(M_\star) g(t)$ (\citealt{2003ApJ...586..464B}; \citealt{2007ApJ...669.1167B}).   
Recently, \citet{2008ApJ...687.1264M} derived an expression for the shape of this track as a function of stellar $(B-V)_0$ and age, such that 

\begin{equation} \label{eqn:mamajek}
P_{\text{rot}} = 0.407 \left[ (B-V)_0 - 0.495 \right]^{0.325} t^{0.566},
\end{equation}

\noindent where $t$ is in Myr, $P_{\text{rot}}$ is in days, and $(B-V)_0$ represents the mass dependence. 
The shape of the mass dependence is based on the shapes of the slowly rotating tracks in the Pleiades ($\sim$125~Myr) and Hyades ($\sim$625~Myr), and the dependence on time is influenced strongly by the current rotation rate of the Sun. 
In order to constrain our rotational evolution models at older ages, we assume the above equation is approximately true for stars beyond 1~Gyr. 
Using the above formula, we calculate the rotation period for stars in each of the six mass bins that we consider in our rotational evolution model at 5~Gyr. 
However, we warn that the shape of the mass dependence is not well constrained observationally beyond the age of Hyades.
This represents a significant uncertainty in the analysis of the next section.

\begin{figure}
\includegraphics[trim=5mm 5mm 5mm 5mm,width=0.47\textwidth]{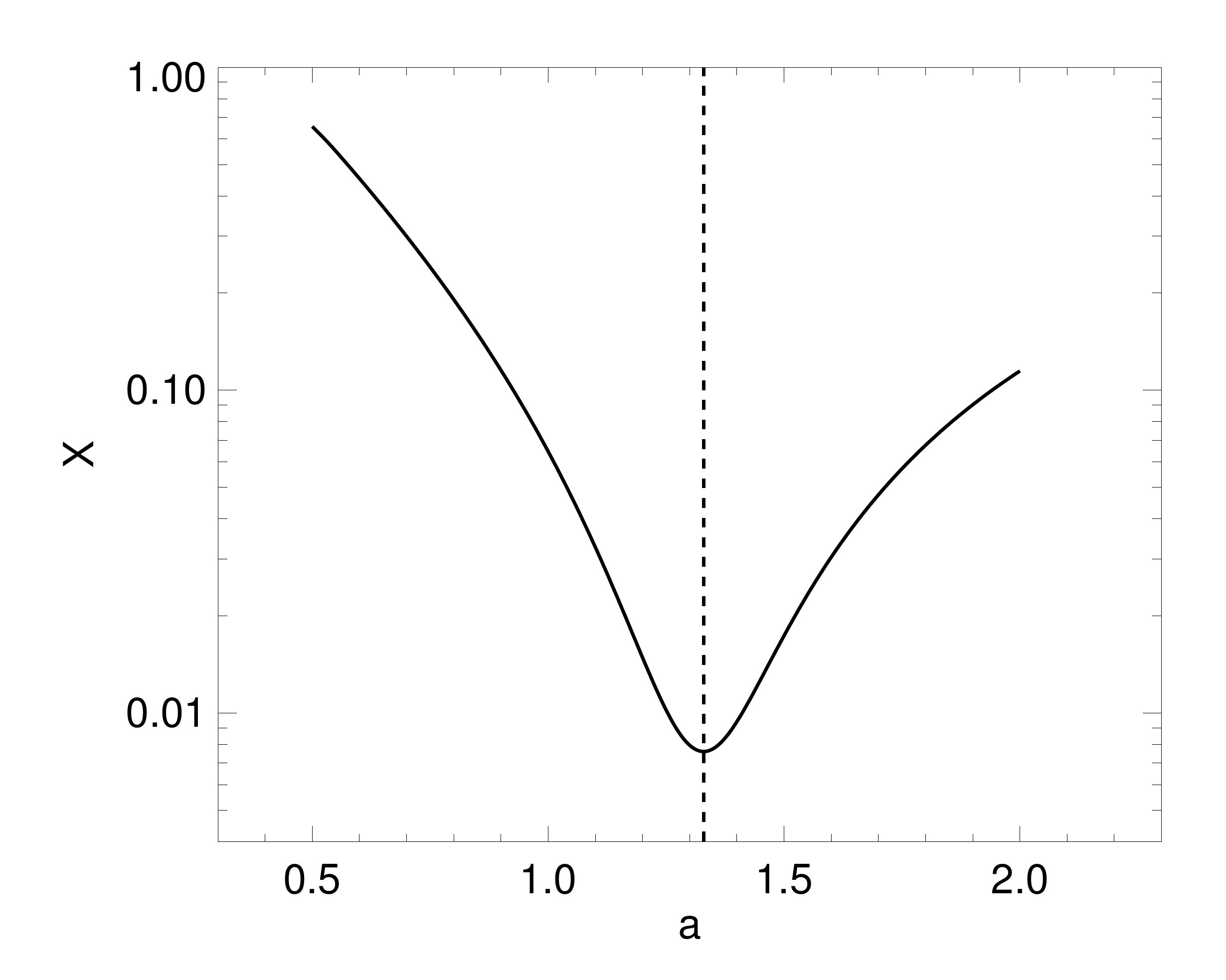}
\includegraphics[trim=5mm 5mm 5mm 5mm,width=0.47\textwidth]{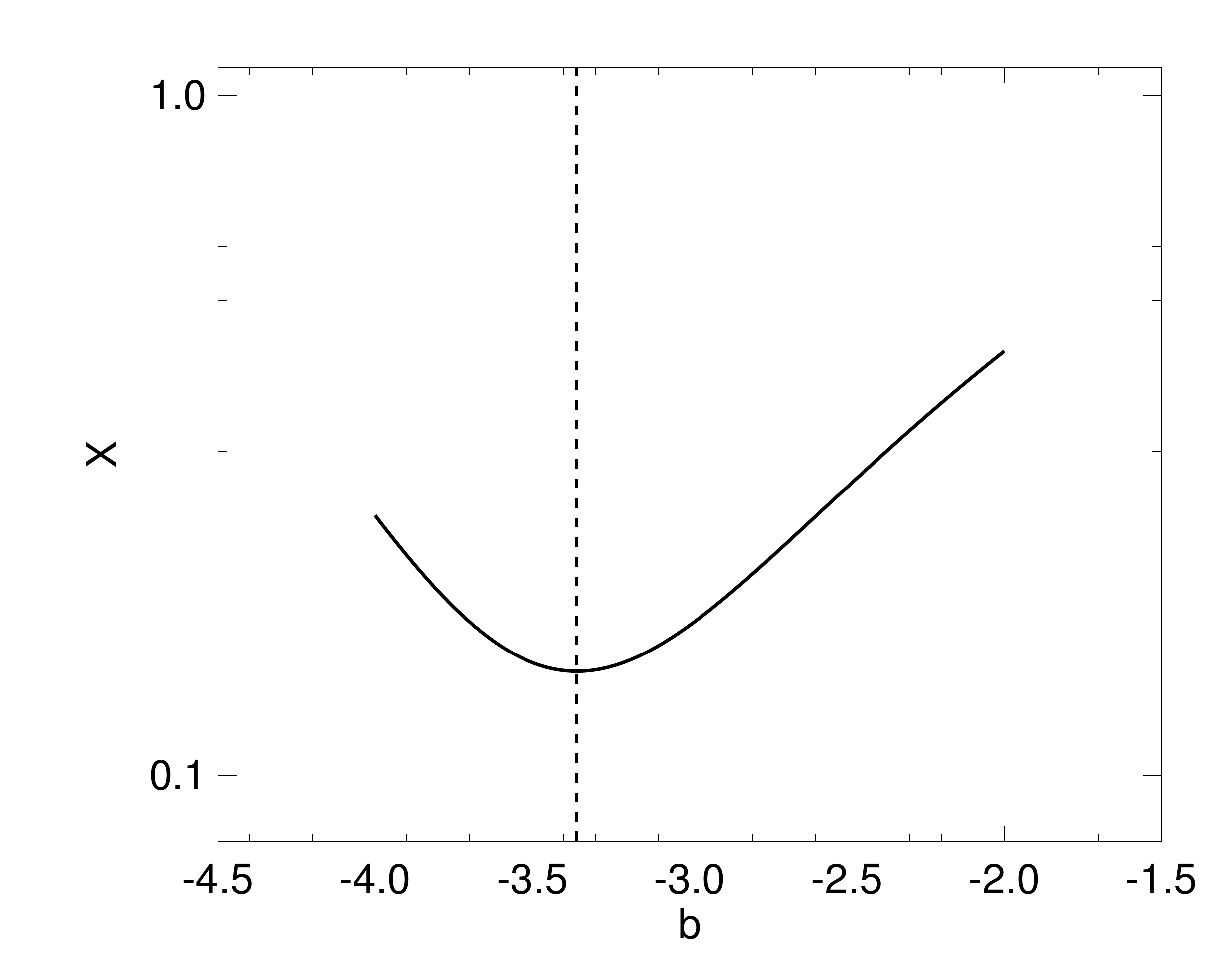}
\includegraphics[trim=5mm 5mm 5mm 5mm,width=0.47\textwidth]{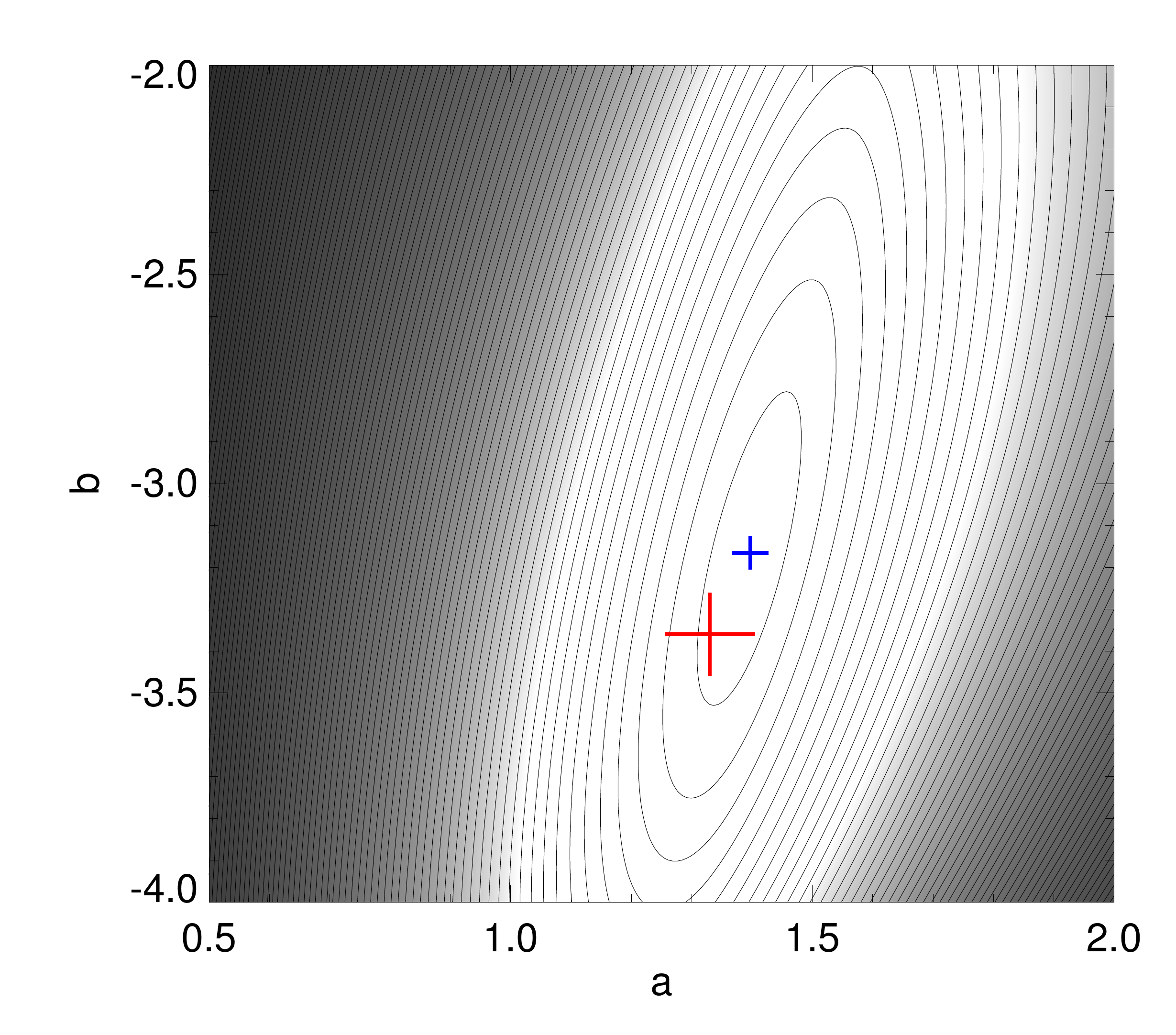}
\caption{
Plots showing how our goodness-of-fit parameter, $X$, depends on the free parameters $a$ and $b$. 
The upper panel shows $X$ against $a$ considering only the 1~M$_\odot$ mass bin.
The vertical dashed line shows our best fit value of $a \approx 1.33$.
The middle panel shows $X$ against $b$ assuming the above best fit value for $a$ and taking into account all three mass bins.
The vertical dashed line shows our best fit value of $b \approx -3.36$.
The lower panel shows a contour plot of $X$ against both $a$ and $b$, taking into account all mass bins.
The red cross shows the location of \mbox{$a=1.33$} and \mbox{$b=-3.36$}, and the blue cross shows the location of \mbox{$a=1.40$} and \mbox{$b=-3.17$}, corresponding to the minimum of $X$ for all values of $a$ and $b$.
}
 \label{fig:fita}
\end{figure}

\section{Rotational Evolution: Determination of the Free Parameters} \label{sect:rotevofitting}

\begin{figure*}
\centering
\includegraphics[trim=5mm 0mm 5mm 0mm,width=0.45\textwidth]{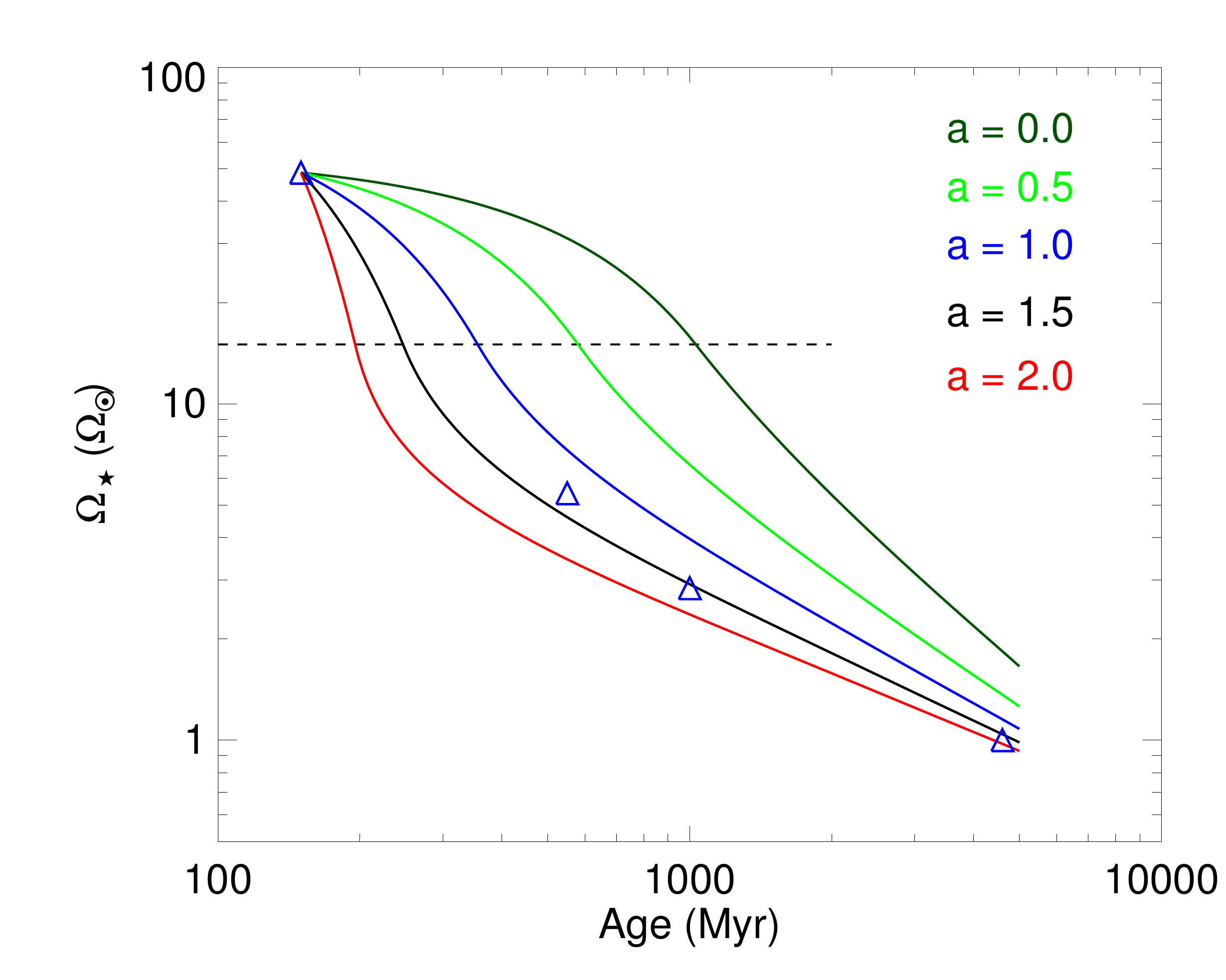}
\includegraphics[trim=5mm 0mm 5mm 0mm,width=0.45\textwidth]{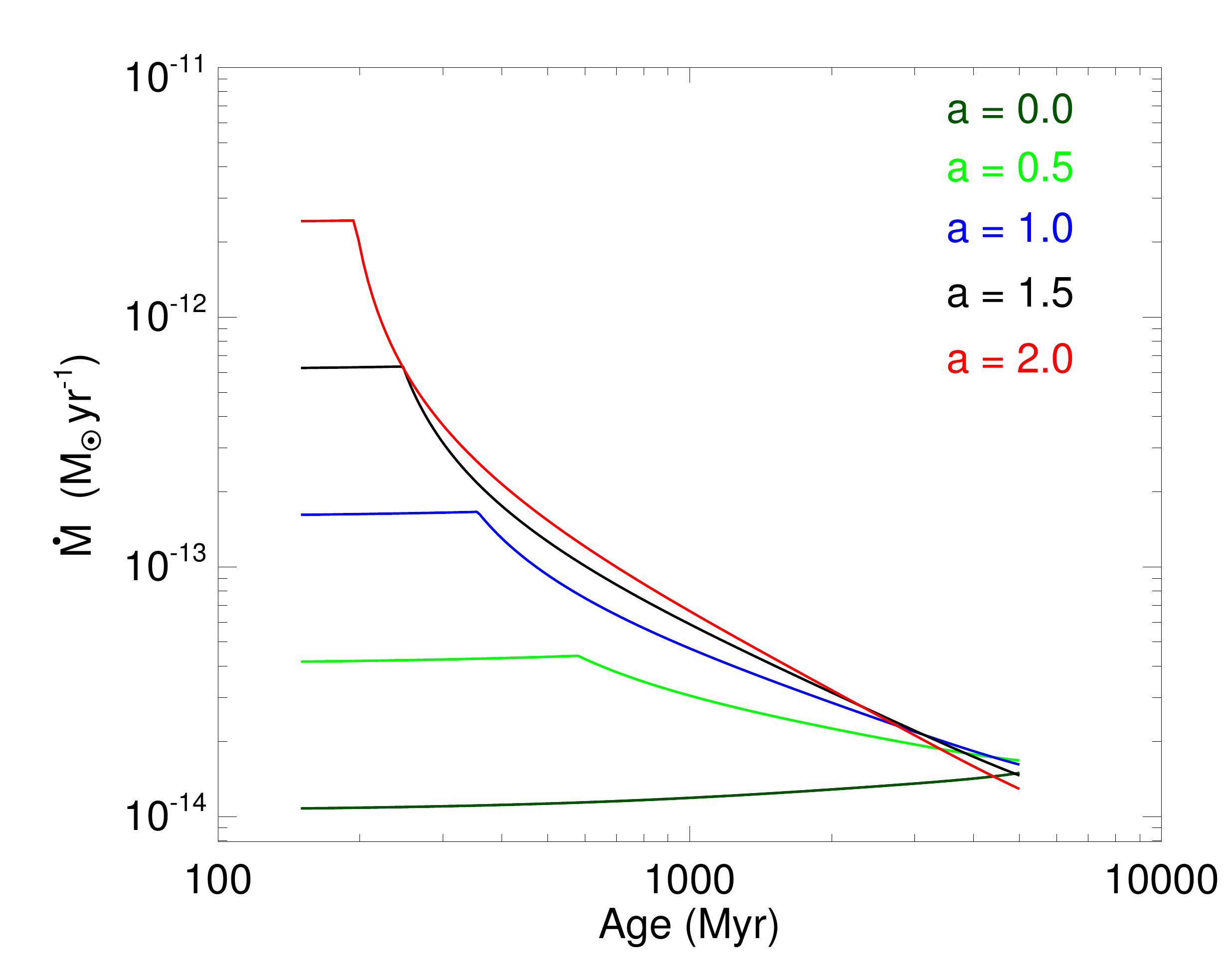}
\caption{
Plots showing the influence of the free parameter $a$ on the rotational evolution of solar mass stars starting at $\sim$150~Myr at the 90th percentile of the rotational distribution. 
The left panel shows angular velocity against stellar age for different values of $a$, where the triangles show the observational constraints on the 90th percentile track given in Table~\ref{tbl:percentiles}.
The horizontal dashed line shows the saturation threshold for both the mass loss rate and the magnetic field strength. 
The right panel shows corresponding wind mass loss rates along each track.
}
 \label{fig:influenceofa}
\end{figure*}

In the previous sections, we construct a rotational evolution model and collect rotation periods for more than 2000 stars in several young clusters.
In this section, we use the observational results to constrain the free parameters in the rotational evolution model. 
We evolve a star's initial rotation rate forward in time by integrating Eqn.~\ref{eqn:angularmomentumall} using the classical Runge-Kutta method and Eqn.~\ref{eqn:matttorque} to calculate the wind torque. 

The three free parameters that we fit later in this section are $K_\tau$, $a$, and $b$.
The parameter $c$ in Eqn.~\ref{eqn:saturation} is important since it controls the mass dependence of the saturation threshold.
We first fit $K_\tau$ and $a$ by considering the example of the solar wind and the spin down of solar mass stars.
We then fit $c$ by first setting $c = 2$; this is close to the value suggested from the saturation of \mbox{X-ray} emission but does not allow us to get acceptable fits to the spin down of 0.5~M$_\odot$ and 0.75~M$_\odot$ stars.
Using this value, we find the best fit value for $b$.
We then iteratively increase $c$, recalculating the best fit value of $b$ each time until we get a solution that we judge by eye as being an acceptable fit.
This allows us to get good fits to the rotational evolution models while keeping $c$ as close to our expectations from \mbox{X-ray} observations.

\subsection{The current solar wind and the value of $K_\tau$}

It is possible to constrain $K_\tau$  by considering the torque exerted on the Sun by the current solar wind. 
Observationally, the torque on the Sun averaged over long time periods can be easily estimated from observed spin down laws, such as Eqn.~\ref{eqn:mamajek}.
The angular velocity of solar mass stars beyond $\sim$1~Gyr is known to evolve with age, $t$, as $\Omega_\star \propto t^{-0.566}$. 
This implies that the rate at which the angular velocity of the Sun is changing is $d\Omega_\star / dt = -0.566 \Omega_\odot t_\odot^{-1}$, where $t_\odot = 4.6$~Gyr. 
Inserting this into Eqn.~\ref{eqn:angularmomentumall}, we get the torque from the current solar wind as

\begin{equation} \label{eqn:solartorque}
\tau_{\text{w},\odot} = \Omega_\odot \frac{dI_\odot}{dt} - 0.566 \Omega_\odot  I_\odot t_\odot^{-1},
\end{equation}

\noindent where \mbox{$I_\odot=6.87 \times 10^{53}$~g~cm$^{2}$}. 
The first term on the RHS of this equation, involving $dI_\odot / dt$, is negligible.
We get a torque exerted by the current solar wind on the Sun of $-7.15 \times 10^{30}$~erg~s$^{-1}$. 
This is the torque from the current solar wind averaged over time periods of hundreds of millions of years, and does not necessarily represent the solar wind torque at any instant in time.
However, to calculate $K_\tau$, we make the assumption that the current solar wind and magnetic field values are typical for the solar wind averaged over such long time periods.   
Inserting $B_{\text{dip}} = 1.35$~G and $\dot{M}_\star = 1.4 \times 10^{-14}$~M$_\odot$~yr$^{-1}$ into Eqn.~\ref{eqn:matttorque} gives a wind torque of $-6.46 \times 10^{30}$~erg~s$^{-1}$.  
Therefore, we find that $K_\tau=11$ is necessary for our predicted wind torque to match the current solar wind.

The fact that we need such a large value of $K_\tau$ could be interpreted in several ways.
This could be a result of the fact that we only consider the dipole component of the field and ignore the other field components, which could cause us to underestimate the wind torque, especially at cycle maximum when the dipole field is negligible.
Alternatively, it could be that we underestimate the values of $B_{\text{dip},\odot}$ and $\dot{M}_\odot$.
We derived these values using solar magnetic field and wind measurements that extend no more than a few decades into the past, whereas our determination of the solar wind torque using Eqn.~\ref{eqn:solartorque} is sensitive to the values averaged over hundred of millions of years.
If the Sun is currently at a long term level of low activity, then we would require a large value of $K_\tau$ to compensate for the low values of $B_{\text{dip},\odot}$ and $\dot{M}_\odot$.
This interpretation, while highly speculative, is supported by \citet{2011ApJ...743...48W} who showed that the average solar X-ray luminosity is a factor of 2-3 below what would be predicted given the Sun's mass and rotation rate.

\subsection{Fitting $a$ and $b$} \label{sect:rotevobestfit}

In order to find the best fit values of $a$ and $b$, we define a goodness-of-fit parameter that we call $X$.
To calculate this parameter for a given set of values for the free parameters, we first calculate rotation tracks starting at the 10th, 50th, and 90th percentiles of the 150~Myr rotational distribution in each mass bin considered.
For each of the observationally constrained percentiles listed in Table~\ref{tbl:percentiles}, we then calculate the square of the difference in the logarithm of the angular velocity between the observed value, $\Omega_{\text{obs}}$, and the value predicted by the model, $\Omega_{\text{model}}$.
We also include the predicted rotation rates at 5~Gyr from Eqn.~\ref{eqn:mamajek} for each mass bin, but only compare them to the 10th percentile tracks.
We do not attempt to encourage the 50th and 90th percentile tracks to spin down to these values.
Our goodness-of-fit parameter is then the sum of all of these values and is given by

\begin{equation}
X = \sum_i \gamma_i \left( \log \Omega_{\text{obs},i} - \log \Omega_{\text{model},i} \right)^2,
\end{equation}

\noindent where the sum is the sum over all percentiles for each age and each mass bin, plus also the 5~Gyr predicted rotation rates.
In this equation, $\gamma_i$ is a weighting parameter that we use to increase or decrease the importance of certain parts of the observational constraints. 
For most of the observed percentiles from young clusters, we set $\gamma_i = 1$, and for the comparison between the predictions of Eqn.~\ref{eqn:mamajek} for rotation at 5~Gyr and the 10th percentile models, we set $\gamma_i=5$.
This is done to force the models to spin down to the desired rotation rates at later ages.
Since one of the most difficult features of the observations to reproduce in the models is the slow spin down of rapidly rotating stars, especially at low masses, we set $\gamma_i = 2$ for the 90th percentiles of the distributions at 550~Myr.  

Other than $K_\tau$, which we have already determined by considering the current solar wind, the only free parameter that influences the rotational evolution of solar mass stars is~$a$.
We therefore determine $a$ by finding the value that minimises our goodness-of-fit parameter, $X$, considering only the 1.0~M$_\odot$ mass bin. 
The value of $a$ determines the dependence of the mass loss rate in the wind on stellar rotation. 
In the upper panel of Fig.~\ref{fig:fita}, we show our goodness-of-fit parameter, $X$, against $a$ and estimate that $a \approx 1.33$.
In Fig.~\ref{fig:influenceofa}, we show how the rotational evolution of the 90th percentile track, and the corresponding predicted mass loss rates, depends on our chosen value of $a$. 
Clearly both the predicted rotational evolution and the predicted mass loss rates are highly sensitive to the value of $a$. 
For low values of $a$, relative to the best fit value, a star starting at the 90th percentile at 150~Myr spins down too slowly and is well above the 90th percentiles at later ages.  
Similarly, high values of $a$ lead to the star spinning down too fast and reaching the saturation threshold too early. 
Interestingly, the rotation rate of the different tracks at 5~Gyr is not highly sensitive to the value of $a$, though there is some influence.

It is interesting to see if we can reproduce this value of $a$ using different reasoning.
At a given stellar mass, ignoring the change in stellar radius as the star ages, the mass loss rate is given by $\dot{M}_\star \propto \Omega_\star^a$.
Since we know that at ages beyond 1~Gyr, the angular velocity has a dependence on age of $\Omega_\star \propto t^{-0.566}$, we can estimate that

\begin{equation} \label{eqn:a_derivation1}
\frac{d \Omega_\star}{dt} \propto \Omega_\star^{2.77}.
\end{equation}

\noindent If we also ignore the evolution of the moment of inertia of the star, then $d\Omega_\star / dt$ is proportional to the wind torque, and therefore Eqn.~\ref{eqn:matttorque} implies 

\begin{equation} \label{eqn:a_derivation2}
\frac{d\Omega_\star}{dt} \propto B_{\text{dip}}^{0.87} \dot{M}_\star^{0.56} \Omega_\star.
\end{equation}

\noindent This is not exactly true since we have have ignored a $\Omega_\star$ term in the denominator of Eqn.~\ref{eqn:matttorque}, but it is a good approximation at these ages\footnotemark.
Inserting $\dot{M}_\star \propto \Omega_\star^a$ and $B_{\text{dip}} \propto \Omega_\star^{1.32}$, this becomes

\footnotetext{
Ignoring the denominator in Eqn.~\ref{eqn:matttorque} is reasonable for slow rotation.
For example, the current solar parameters give \mbox{$K_2^2 v_{\text{esc}}^2 = 9.8 \times 10^{12}$~cm$^{2}$~s$^{-2}$} and \mbox{$\Omega_\star^2 R_\star^2 = 3.5 \times 10^{10}$~cm$^{2}$~s$^{-2}$}, meaning that the rotational dependence is negligible. 
At fast rotation, the $\Omega_\star^2 R_\star^2$ term dominates the $K_2^2 v_{\text{esc}}^2$ term and so the approximation given in Eqn.~\ref{eqn:a_derivation2} is not reasonable.
For 0.5~M$_\odot$ and 1.0~M$_\odot$, the two terms become equal at rotation rates of approximately 30$\Omega_\odot$ and 17$\Omega_\odot$ respectively.
We take into account all terms in Eqn.~\ref{eqn:matttorque} in our derivation of $a \approx 1.33$ and $b \approx -3.36$. 
}

\begin{equation} \label{eqn:a_derivation3}
\frac{d\Omega_\star}{dt} \propto \Omega_\star^{0.56a + 2.15},
\end{equation}

\noindent which compared with Eqn.~\ref{eqn:a_derivation1} implies that \mbox{$a \sim 1.1$}.
This is similar to our determination of $a \approx 1.33$. 
Although this simple reasoning gives us a similar value of $a$ to the more sophisticated analysis, it is still necessary to determine $a$ by minimising our goodness-of-fit parameter because we are partly fitting $a$ to the rotational evolution of rapidly rotating stars in the first billion years when the stars do not follow a Skumanich style spin down law.
The fact that our determinations of $a$ are so similar shows that the early rotational evolution of rapidly rotating solar-mass stars is perfectly consistent with the Skumanich style spin down at later ages.

Now that we have derived the values of $K_\tau$ and $a$, we can find the best fit value of $b$.
To do this, we consider the rotational evolution of all three mass bins. 
In the middle panel of Fig.~\ref{fig:fita}, we show the dependence of $X$ on $b$, assuming that $a = 1.33$.
Our best fit model is the model where \mbox{$b \approx -3.36$}.

Once again, it is interesting to see if we can estimate a similar value of $b$ using different reasoning.
From Eqn.~\ref{eqn:matttorque}, the wind torque on a star is approximately given by

\begin{equation} \label{eqn:simplifiedMatt}
\tau_{\text{w}} \propto B_{\text{dip}}^{0.87} \dot{M}_\star^{0.56} R_\star^{2.87} \Omega_\star.
\end{equation}

\noindent When determining $K_\tau$, we showed that $d\Omega_\star / dt \propto \Omega_\star t_\star^{-1}$, which naturally follows from assuming a power-law dependence of $\Omega_\star$ on $t$.
Ignoring the time evolution of the moment of inertia, this implies that at a given age, $\tau_{\text{w}} \propto I_\star \Omega_\star$.
Since the mass loss rate is given by $\dot{M}_\star \propto R_\star^2 \Omega_\star^a M_\star^b$, this becomes 

\begin{equation} \label{eqn:bderivation2}
I_\star \Omega_\star \propto B_{\text{dip}}^{0.87} R_\star^{3.99} \Omega_\star^{1.74} M_\star^{0.56b},
\end{equation}

\noindent where we have inserted $a=1.33$.
We can see from this analysis that our best fit value of $b$ is likely to depend, at least weakly, on our best fit value of $a$.
We can estimate the value of $b$ by considering the ratio of the torques on two stars that have the same age.
If we consider two stars, Eqn.~\ref{eqn:bderivation2} implies that

\begin{equation}
\left( \frac{I'_{\star}}{I_{\star}} \right)
= 
\left( \frac{B'_{\text{dip}}}{B_{\text{dip}}} \right)^{0.87} \left( \frac{R'_\star}{R_\star} \right)^{3.99} \left( \frac{\Omega'_\star}{\Omega_\star} \right)^{0.74} \left( \frac{M'_\star}{M_\star} \right)^{0.56b},
\end{equation}

\noindent where quantities marked with a prime are for one star and
 quantities not marked with a prime are for the other star.
Rearranging for $b$ gives
 
\begin{equation} \label{eqn:eqnforb}
b = \frac{\log \left[ \left( \frac{I'_{\star}}{I_{\star}} \right) \left( \frac{B'_{\text{dip}}}{B_{\text{dip}}} \right)^{-0.87} \left( \frac{R'_\star}{R_\star} \right)^{-3.99} \left( \frac{\Omega'_\star}{\Omega_\star} \right)^{-0.74}   \right]}{ 0.56 \log \left( \frac{M'_\star}{M_\star} \right)}.
\end{equation}
 
\noindent Consider two 5~Gyr old stars with masses of 0.5~M$_\odot$ and 1.0~M$_\odot$. 
From Eqn.~\ref{eqn:dipoleRossby} and Eqn.~\ref{eqn:mamajek}, we can estimate the equatorial dipole field strengths and rotation periods, and from the stellar evolution models discussed in Section~\ref{sect:rotevo}, we can estimate the stellar radii and moments of inertia.
For the 0.5~M$_\odot$ star, we estimate a rotation period of $\sim$51~days, a dipole field strength of 1.85~G, a radius of 0.53~R$_\odot$, and a moment of inertia of \mbox{$2.1 \times 10^{53}$~g~cm$^{2}$}.
For the 1.0~M$_\odot$ star, we estimate a rotation period of $\sim$27~days, a dipole field strength of 1.25~G, a radius of 1.03~R$_\odot$, and a moment of inertia of \mbox{$6.9 \times 10^{53}$~g~cm$^{2}$}.
Inserting these quantities into Eqn.~\ref{eqn:eqnforb} gives $b \sim -4.1$.
This is similar to our derived value of $b \approx -3.36$.

To check that our estimates of $a$ and $b$ are robust, we fit the values once more by minimising the goodness-of-fit parameter, $X$.
Above, we fit the two parameters separately, but to check that this gives us approximately the best fit, we fit the two parameters simultaneously by producing a grid of rotational evolution models with ranges of values of $a$ and $b$. 
In the lower panel of Fig.~\ref{fig:fita}, we show a contour plot of $X$ from this grid.
Clearly there is some degeneracy between our values of $a$ and $b$, which can be understood from our above analysis.
The value of $a$ is well constrained, with 1.2 and 1.6 likely representing reasonable limits.
The constraints on $b$ are much looser, with reasonable lower and upper limits being approximately -4 and -2 respectively. 
The red cross shows our best fit of $a \approx 1.33$ and $b \approx -3.36$, which is clearly very close to the minimum of $X$, which lies at \mbox{$a \approx 1.40$} and \mbox{$b \approx -3.17$}.
We choose to use the first set of values of $a$ and $b$ since they give the best fit to the rotational evolution of solar mass stars, which is much better constrained beyond 1~Gyr observationally than the rotational evolution of lower mass stars.

One interesting result of these fits is the fact that at a given rotation rate in the unsaturated regime, low-mass stars have higher mass loss rates per unit surface area than high-mass stars. 
We predict that a 0.5~M$_\odot$ star has a mass loss rate per unit surface area that is a factor of ten larger than a 1.0~M$_\odot$ star with the same rotation rate.
This translates into a mass loss rate that is a factor of $\sim$3.4 higher for the 0.5~M$_\odot$ star. 
That low-mass stars have higher mass loss rates is an interesting result.
Assuming that $R_\star \propto M_\star^{0.8}$, this is inevitable if \mbox{$b < -1.6$}, which Fig.~\ref{fig:fita} shows clearly must be the case for our rotational evolution models to provide acceptable fits to the observational constraints\footnotemark.

\footnotetext{
Recently, \citet{2015arXiv150205801G} produced rotation models for stars with masses of 0.5~M$_\odot$ and 0.8~M$_\odot$.
To prescribe the mass loss rates, they used the model of \citet{2011ApJ...741...54C}, which predicts much lower values for low mass stars (for the 0.5~M$_\odot$ mass bin, the mass loss rate is between $10^{-17}$ and $10^{-16}$~M$_\odot$~yr$^{-1}$), in strong disagreement with our results.
However, to reproduce the observational constraints, \citet{2015arXiv150205801G} had to artificially increase the multiplicative constant in their formula for the wind torque for low mass stars, which has a similar effect as the $M^b$ term in our Eqn~4.
Our results correspond to the interpretation, discussed in Section~5.3 of \citet{2015arXiv150205801G}, that the \citet{2011ApJ...741...54C} model underestimates the mass loss rates of low mass stars.
\citet{2015arXiv150205801G} suggested that the mass loss rates of the 0.5~M$_\odot$ stars should be increased by a factor of 860 relative to what they would otherwise predict, which would be much more consistent with our results.
}

\citet{2014MNRAS.438.1162V} presented 3D MHD models of the winds from six M-dwarfs based on realistic non-dipolar magnetic field geometries measured using the Zeeman-Doppler Imaging technique. 
In Table~3 of \citet{2014MNRAS.438.1162V}, power-laws for the relations between different quantities were given based on fits to the results of their models.
They found that the wind torque is related to the mass loss rate as $\tau_{\text{w}} \propto \dot{M}_\star^{2.18 \pm 0.56}$. 
This might appear to be in contradiction to the $\dot{M}_\star^{0.56}$ dependence derived by \citet{2012ApJ...754L..26M} that we use in this paper, but it is important to recognise that the relation of \citet{2014MNRAS.438.1162V} is not the dependence of wind torque on mass loss rate, but is the correlation between wind torque and mass loss rate, and therefore also includes the correlations between mass loss rate and other quantities, such as magnetic field strength and rotation rate. 
At a given stellar mass and radius, the torque formula derived by \citet{2012ApJ...754L..26M} approximately reduces to $\tau_{\text{w}} \propto B_{\text{dip}}^{0.87} \dot{M}_\star^{0.56} \Omega_\star$.
Our derivation of the dependence of mass loss rate on rotation and Eqn.~\ref{eqn:dipoleRossby} imply that $\Omega_\star \propto \dot{M}_\star^{1/1.33}$ and $B_{\text{dip}} \propto \dot{M}_\star^{1.32/1.33}$.
Therefore, we find that $\tau_{\text{w}} \propto \dot{M}_\star^{2.18}$, in excellent agreement with the results of \citet{2014MNRAS.438.1162V} (the apparent perfect agreement is of course coincidental).
This agreement suggests that not considering non-dipolar magnetic field geometries in our rotational evolution model does not influence our results significantly.

It is possible with our results to construct simple scaling laws for the wind torque, $\tau_\text{w}$, as a function of stellar rotation and mass.
Assuming \mbox{$R_\star \propto M_\star^{0.8}$} means that the mass loss rates and dipole field strengths vary as \mbox{$\dot{M}_\star \propto \Omega_\star^{1.33} M_\star^{-1.76}$} and \mbox{$B_\text{dip} \propto \Omega_\star^{1.32} M_\star^{-1.43}$}, where the latter equation was derived by assuming a $M_\star^{-1.08}$ dependence to the convective turnover time (\mbox{\citealt{2014arXiv1408.6175R}}).
These are true until the saturation rotation rate given by \mbox{$\Omega_\text{sat} \propto M_\star^{2.3}$}.
Inserting these scaling laws into the simplified version of the wind torque formula given in Eqn.~\ref{eqn:simplifiedMatt} gives $\tau_\text{w}$ in the unsaturated regime as

\begin{equation}
\tau_\text{w} \approx \tau_{\text{w},\odot} \left( \frac{\Omega_\star}{\Omega_\odot} \right)^{2.89} ,
\end{equation}

\noindent where \mbox{$\tau_{\text{w},\odot} \approx -7.15 \times 10^{30}$~erg~s$^{-1}$} is the current solar wind torque.
In the saturated regime, we find instead that

\begin{equation}
\tau_\text{w} \approx \tau_{\text{w},\odot}  15^{1.89} \left( \frac{\Omega_\star}{\Omega_\odot} \right) \left(\frac{M_\star}{M_\odot} \right)^{4.42}.
\end{equation}

\noindent Given that Eqn.~\ref{eqn:simplifiedMatt} is only an accurate approximation at slow rotation, the rotation dependence in this formula is not reasonable for the fastest rotators. 
Interestingly, the wind torque has different dependences on stellar mass in the saturated and unsaturated regimes. 
This is due simply to the strong mass dependence of the saturation threshold.
This explains the interesting fact that rapidly rotating low-mass stars spin down slower than rapidly rotating high-mass stars, but slowly rotating low-mass stars spin down quicker than slowly rotating high-mass stars.

\subsection{Sources of uncertainty in the fit parameters} \label{sect:uncertainty}

We now have best fit values for the free parameters in our rotational evolution model. 
Constraining these free parameters is important because it not only allows us to predict the rotational evolution of stars on the main-sequence, but gives us an observationally constrained scaling law for mass loss rate as a function of stellar mass, radius, and rotation. 
In the stellar wind model developed in Paper~I, the largest unknown was how to scale the base temperature and density to other stars.
This scaling law for the mass loss rate gives us a large part of the solution to this problem. 
However, before we analyse what these results mean for the evolution of wind properties on the main-sequence, we should discuss sources of uncertainty in our results.
We concentrate mostly on the parameters $a$ and $b$ since they are the most important for our wind model. 
Probably the main source of uncertainty in our predicted values of $a$ and $b$ comes from how we relate stellar and wind properties to wind torques. 
Although our results are also based on stellar evolution models and observed rotation rates of stars at different ages, these are much better constrained.
We emphasise that we are only speculating in this section about reasons why our results \emph{might} be inaccurate and have no clear reason to think that there are such inaccuracies.

The driver of our rotational evolution model is the torque formula (Eqn.~\ref{eqn:matttorque}) derived by \citet{2012ApJ...754L..26M}, which allows us to calculate wind torques from the stellar mass, radius, rotation rate, dipole magnetic field strength, and mass loss rate. 
This equation is derived from 2D MHD wind models. 
\citet{2012ApJ...754L..26M} assumed a polytropic equation of state, with $\alpha=1.05$ everywhere, and a base temperature and density of the wind that is uniform over the stellar surface.
Such models produce realistic distributions of wind speeds for a given magnetic field geometry, but are inaccurate in that the contrast between the slow and fast wind is underestimated (for example, see Fig.~5 and Fig.~6 of \citealt{2009ApJ...699..441V}).
\citet{2012ApJ...754L..26M} also set the temperature in all models by assuming that the sound speed is a fixed fraction of the escape velocity at the base of the wind. 
Finally, they assumed that the stellar magnetic fields are all dipolar.
In real winds, these assumptions might not be fulfilled, which could lead to the dependences of torque on the various parameters being different to the predictions of \citet{2012ApJ...754L..26M}.

We have needed to add an extra free parameter, $K_\tau$, into our model for calculating the torque (i.e. Eqn.~\ref{eqn:modifyMatttorque}). 
The free parameter $K_\tau$ is not a meaningless fudge factor, but represents real physical processes that we do not understand, such as the details of the wind driving mechanisms and non-dipolar field geometries.
Also, $K_\tau$ is influenced by our measurements of the current solar mass loss rate and dipole field strength, which might not be fully representative of the values averaged over long time periods.
We assume that $K_\tau$ is the same for all stars; however, it could be that this parameter should also have a dependence on stellar mass and rotation.
We could take this into account by assuming $K_\tau \propto \Omega_\star^e M_\star^f$ and using $e$ and $f$ as free parameters in the rotational evolution models.
However, $e$ and $f$ would influence the models in a similar way as $a$ and $b$, i.e. both assumptions add power law dependences on $\Omega_\star$ and $M_\star$ into the wind torque.

Another source of uncertainty is in how we calculate the strength of the dipole component of the field for each star given its mass and rotation rate. 
We do this using Eqn.~\ref{eqn:dipoleRossby}, which is based on the correlation between the magnetic field strength and rotation derived by \citet{2014MNRAS.441.2361V}.
The form of this relation is $B_{\text{dip}} \propto Ro^d$. 
The exact value of $d$, however, is uncertain since the relation of \citet{2014MNRAS.441.2361V} is based on measurements of the large scale components of the magnetic field strength, which is not exactly the same as the dipole component of the field, and since the relation between the magnetic field strength and the Rossby number contains a lot of scatter, from which a range of values of $d$ can provide acceptable fits.
The parameter $d$ has the same effect on the rotational evolution models as $a$, since they both influence the power law dependence of the wind torque on angular velocity.
A change in $d$ would lead to a change in the opposite direction in the best fit value of $a$. 
In addition, it is not completely clear that the dipole component of the field should be scaled with Rossby number instead of some other parameter.
Using the Rossby number adds a mass dependence into the dipole component of the field, such that at a given rotation rate, lower mass stars have larger magnetic field strengths than higher mass stars.
This dependence has a similar effect as the $M_\star^b$ dependence in our assumption about the mass loss rate. 
If instead, we scaled $B_{\text{dip}}$ with $\Omega_\star$, we would then need a stronger mass dependence for the mass loss rate to compensate in the rotational evolution models.

Looking at the above discussion in a more general way, by making the assumption that $\dot{M}_\star \propto R_\star^2 \Omega_\star^a M_\star^b$, we are adding extra power law dependences on rotation and mass into the wind torque which we then interpret as being entirely due to variations in the mass loss rate.
If there were other sources of such dependences then we would confuse them with the dependence of mass loss rate on stellar rotation and mass.

In the previous section, we are able to estimate values of $a$ and $b$ that are similar to our best fit values using cruder analytic reasoning.
This is both encouraging and discouraging.
It is encouraging because one of the major sources of uncertainty in our model is the angular velocity at which saturation occurs for the mass loss rate and dipole field strength.
Since the cruder reasoning only considers rotation at later ages, we can see that this uncertainty does not significantly influence our result.
It is discouraging because it means that our best fit value of $b$ is mostly sensitive to the spin down of low mass stars at later ages, which is constrained by Eqn.~\ref{eqn:mamajek}.
However, the spin down of stars at later ages, as predicted by Eqn.~\ref{eqn:mamajek}, is constrained almost entirely by the Sun, and so additional observational confirmation of the spin down of low-mass stars is desirable.

Another source of uncertainty in our model comes from the possibility that young stars arrive at the ZAMs with inner cores that are rotating faster than the stellar surfaces.
Core-envelope decoupling would lead to a spin up torque acting on the stellar surface that would work against the wind torque, meaning that in order to reproduce the same rotation track, a larger spin down torque, and therefore a larger mass loss rate, would be needed from the wind.
For solar mass stars, this would lead to our predicted value of $a$ being larger. 
However, \citet{2013A&A...556A..36G} predicted from rotational evolution models of solar mass stars that around the ZAMS, the core would be rotating a factor of $\sim$2 faster than the envelope, and that by the age of the Sun, the core and the envelope would have the same angular velocities.
In this scenario, our wind would need to remove twice as much angular momentum from the star between the ZAMS and 5~Gyr, which is unlikely to require a much larger value of~$a$.

In our models, we consider only the quiet winds of stars.
It has recently been suggested that the winds of more active stars could be dominated by coronal mass ejections (CMEs; \citealt{2007AsBio...7..167K}; \citealt{2012ApJ...760....9A}; \citealt{2013ApJ...764..170D}). 
Lacking more detailed physical knowledge of CME ejections from more active stars, the assumption that the angular momentum removal per unit mass loss is the same for hypothetical CMEs as for quiet winds implies that our scaling law for the mass loss rate remains valid.
Should a CME remove more (less) angular momentum per unit mass loss than the quiet wind, we would overestimate (underestimate) the wind mass loss rate when fitting our scaling law for $\dot{M}$ to the observations (if CMEs do indeed carry away a significant amount of mass and angular momentum).
Given the fundamental uncertainties of these questions and the lack of observational support for strong CME dominated winds in active stars (\citealt{2014MNRAS.443..898L}), we do not consider the question of CME angular momentum transport further.

\begin{figure}[!htp]
\includegraphics[trim=10mm 5mm 5mm 5mm,width=0.49\textwidth]{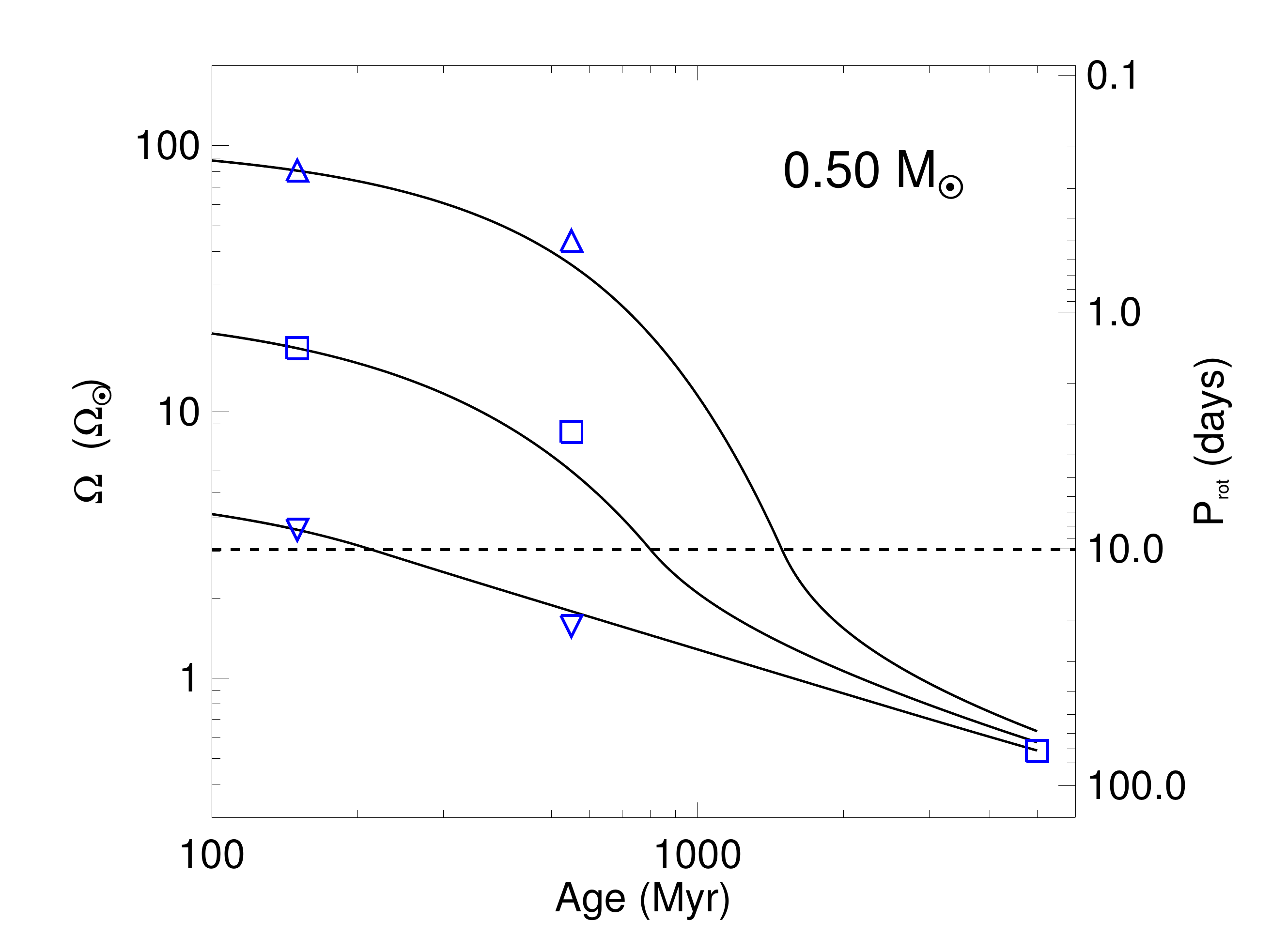}
\includegraphics[trim=10mm 5mm 5mm 5mm,width=0.49\textwidth]{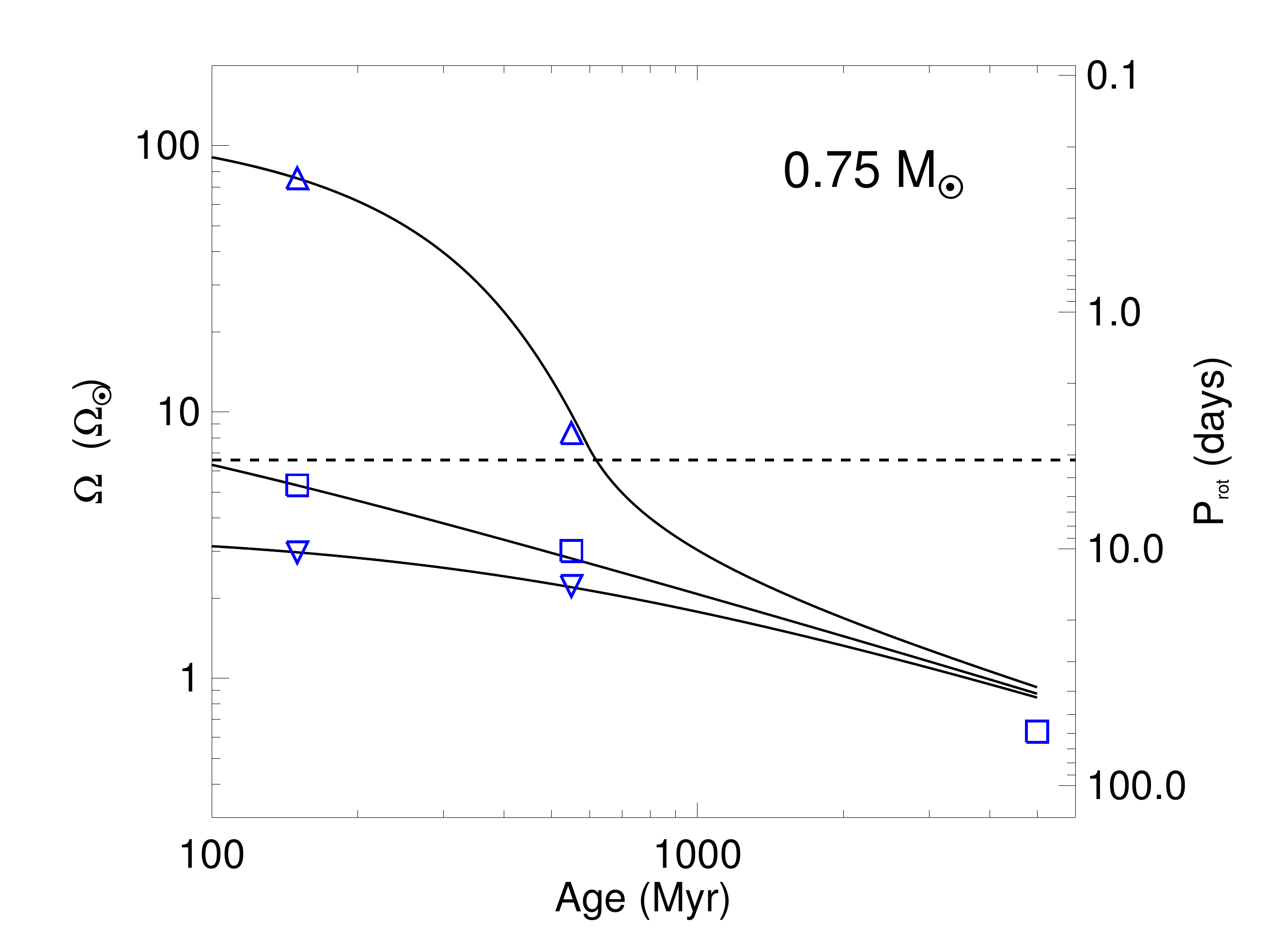}
\includegraphics[trim=10mm 5mm 5mm 5mm,width=0.49\textwidth]{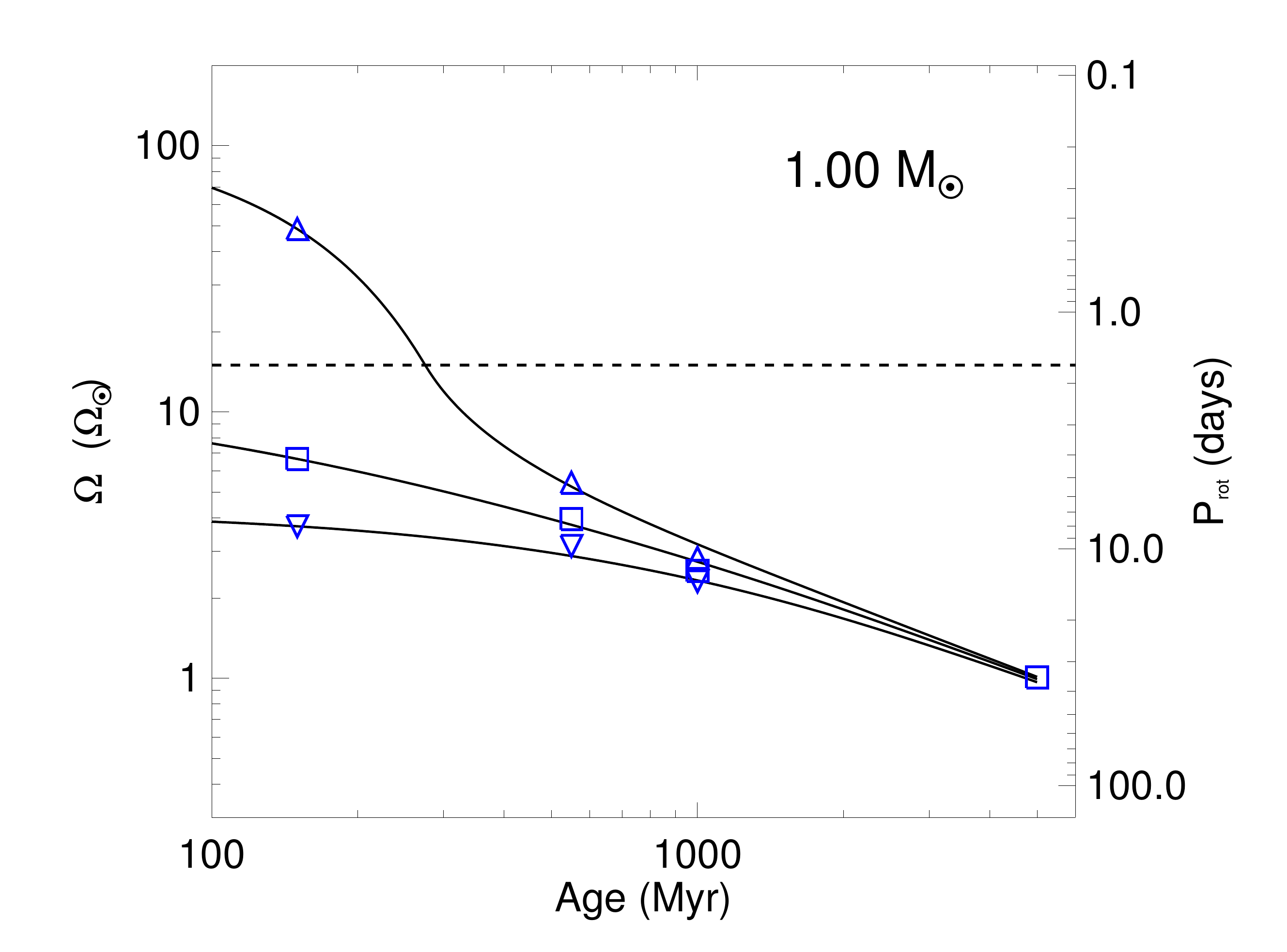}
\caption{
Plots showing the rotational evolution of stars at the 10th, 50th, and 90th percentiles of the rotational distribution from 100~Myr to 5~Gyr.
The three panels show the rotation tracks for stars with masses of  0.50~M$_\odot$ (\emph{upper panel}), 0.75~M$_\odot$ (\emph{middle panel}), and 1.00~M$_\odot$ (\emph{lower panel}).
The observational constraints on the rotational evolution of stars at the 10th percentile (\emph{downward pointing triangles}), 50th percentile (\emph{squares}), and 90th percentile (\emph{upward pointing triangles}) of the rotation distribution. 
The horizontal dashed lines show the saturation thresholds for the mass loss rates and dipole field strengths.
}
 \label{fig:rotevotracks}
\end{figure}


\section{Rotational Evolution: Results} \label{sect:rotevoresults}

In Fig.~\ref{fig:rotevotracks}, we show the rotational evolution tracks for our best fit models.
The three panels show the evolution of stars with masses of 0.50~M$_\odot$, 0.75~M$_\odot$, and 1.00~M$_\odot$.
The observational constraints are shown as triangles for the 10th and 90th percentiles and squares for the 50th percentiles which are summarised in Section~\ref{sect:obsconstraints} and Table~\ref{tbl:percentiles}. 
The squares at older ages are predictions of the 5~Gyr rotation rates from the gyrochronological relation derived by \citet{2008ApJ...687.1264M}. 
Our model clearly fits the observational constraints very well, especially in the 0.5~M$_\odot$ and 1.0~M$_\odot$ mass bins.

For solar mass stars, we are able to explain the rotational evolution of stars from 100~Myr to 5~Gyr well.
This is done with only two free parameters, $K_\tau$ and $a$.
According to our model, a star at the 90th percentile of the rotational distribution at 150~Myr will remain in the saturated regime until about 300~Myr. 
By 500~Myr, the tracks have mostly converged, and by 1~Gyr, there is very little difference between the tracks.
Our tracks probably overestimate slightly the spread in rotation rates at 1~Gyr compared to the observational constraints, though the difference is very small.

\begin{figure*}[!htbp]
\centering
\includegraphics[trim=5mm 5mm 5mm 5mm,width=0.49\textwidth]{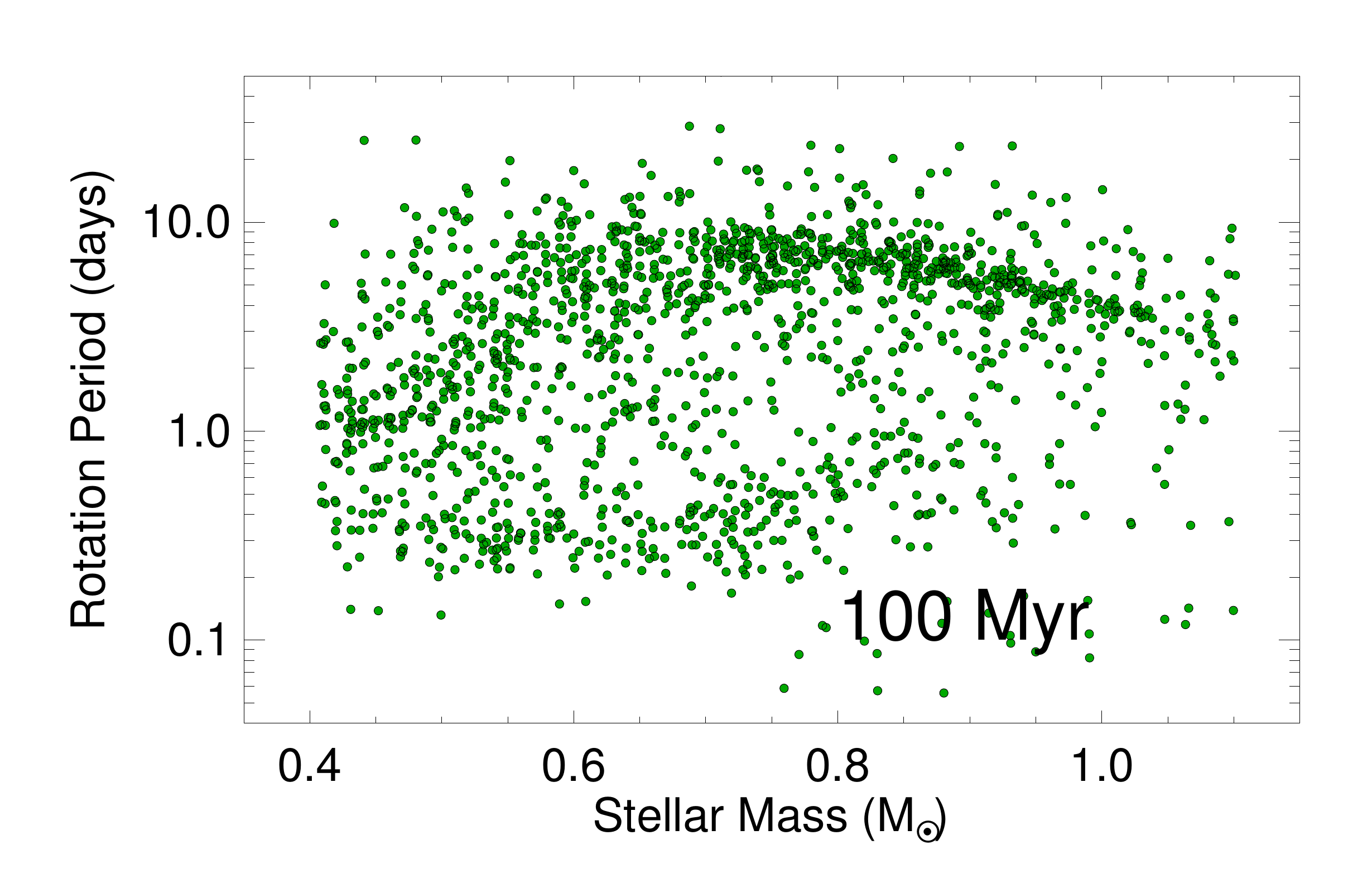}
\includegraphics[trim=5mm 5mm 5mm 5mm,width=0.49\textwidth]{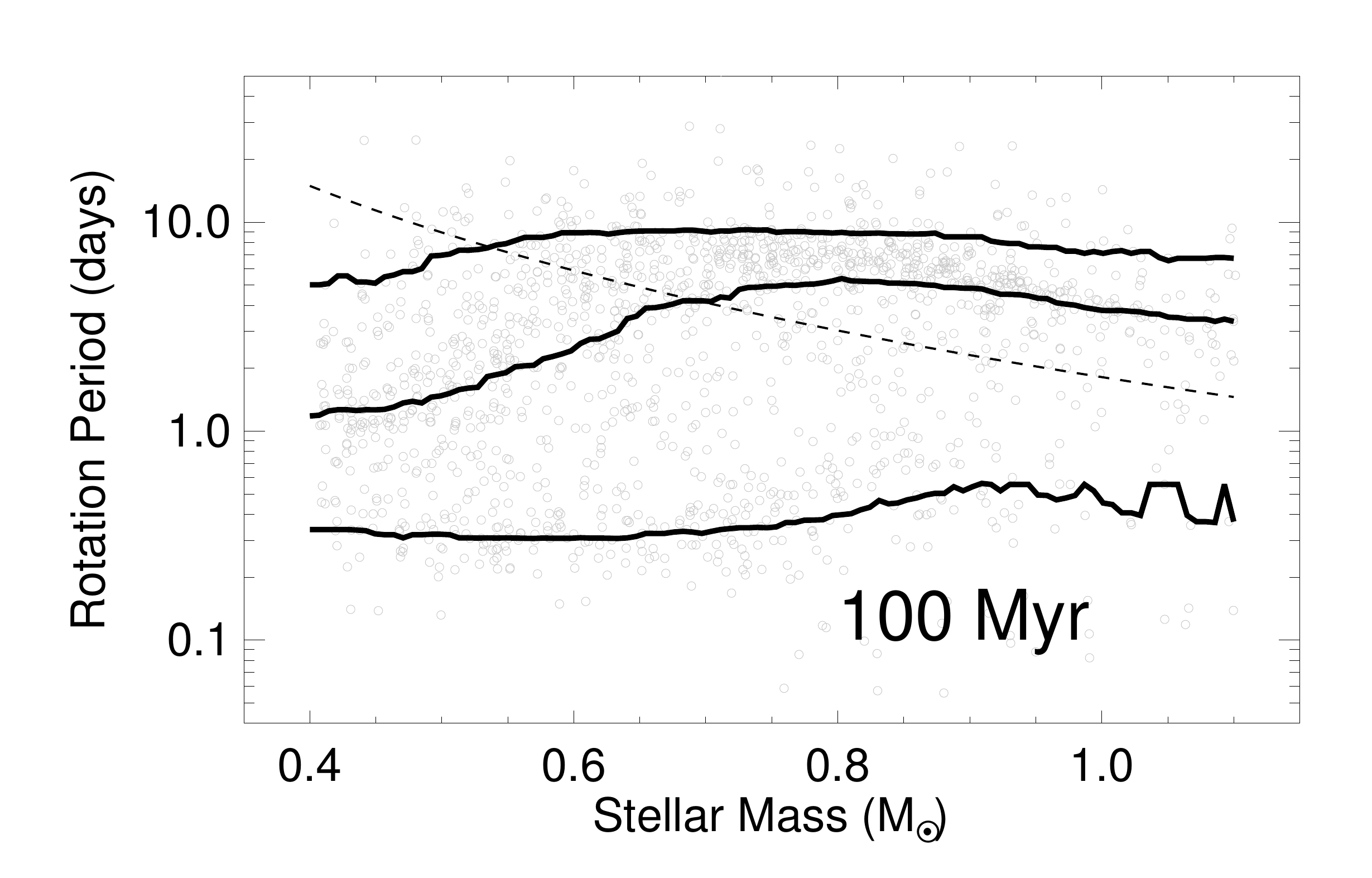}
\includegraphics[trim=5mm 5mm 5mm 5mm,width=0.49\textwidth]{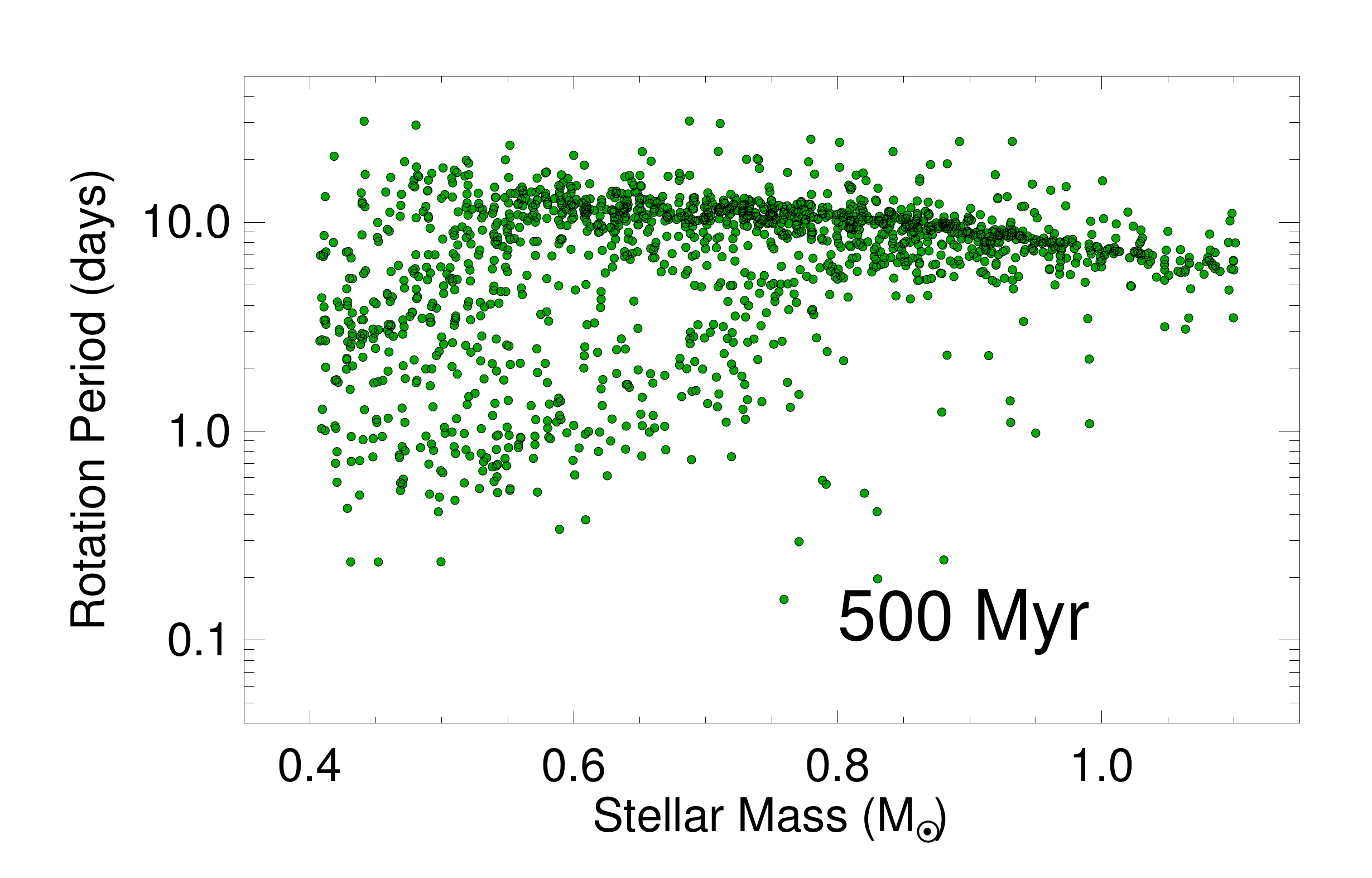}
\includegraphics[trim=5mm 5mm 5mm 5mm,width=0.49\textwidth]{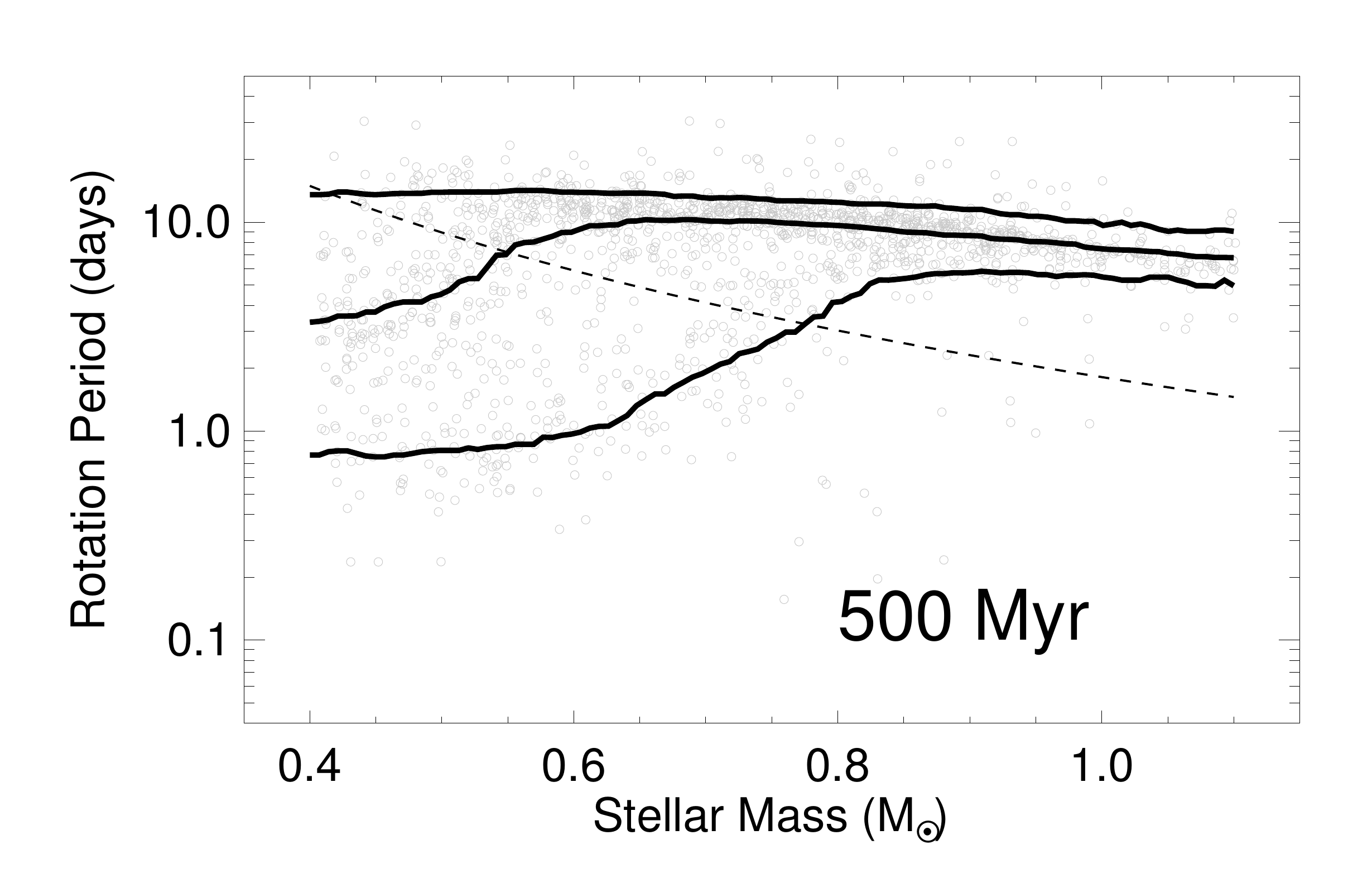}
\includegraphics[trim=5mm 5mm 5mm 5mm,width=0.49\textwidth]{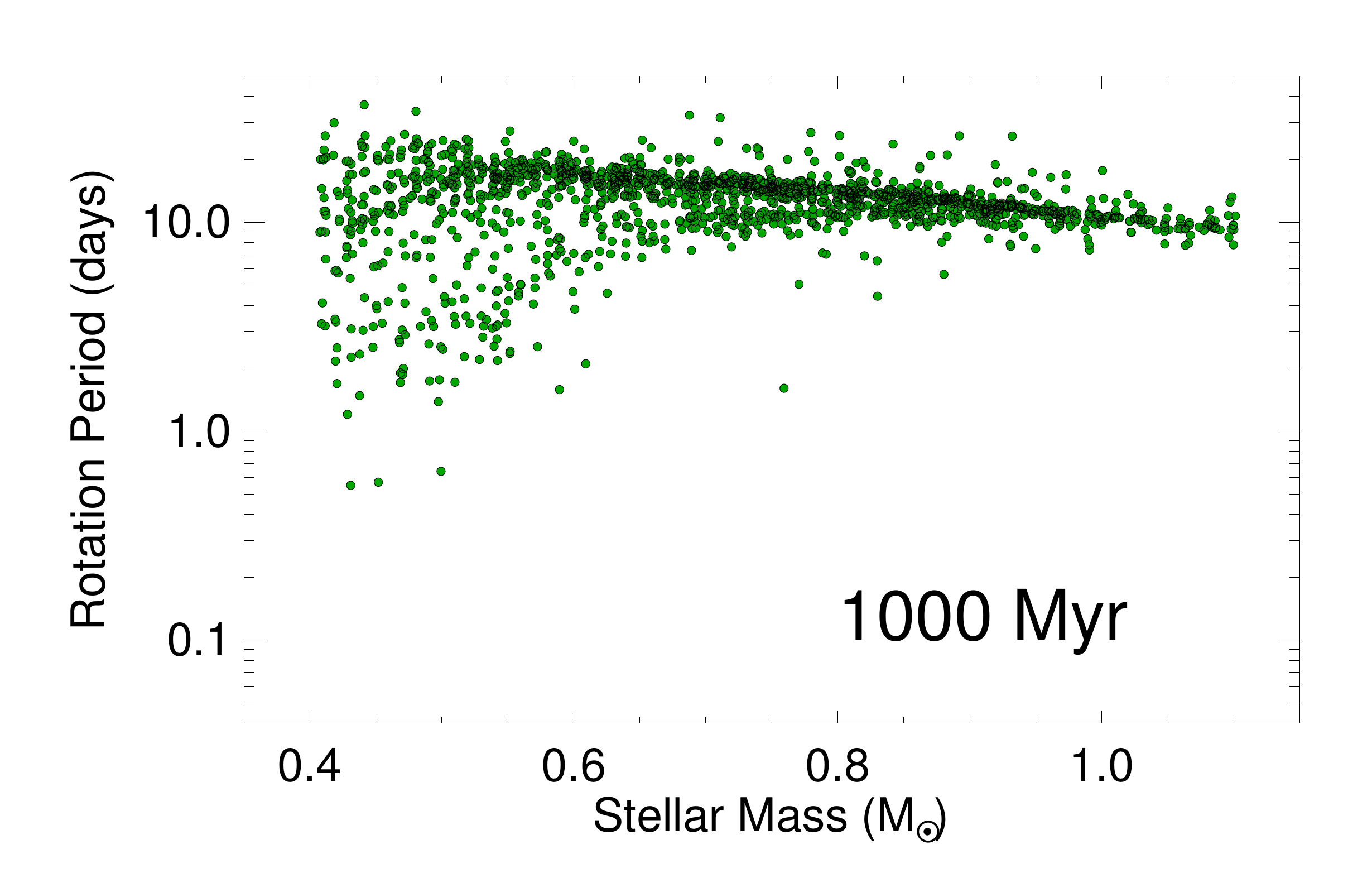}
\includegraphics[trim=5mm 5mm 5mm 5mm,width=0.49\textwidth]{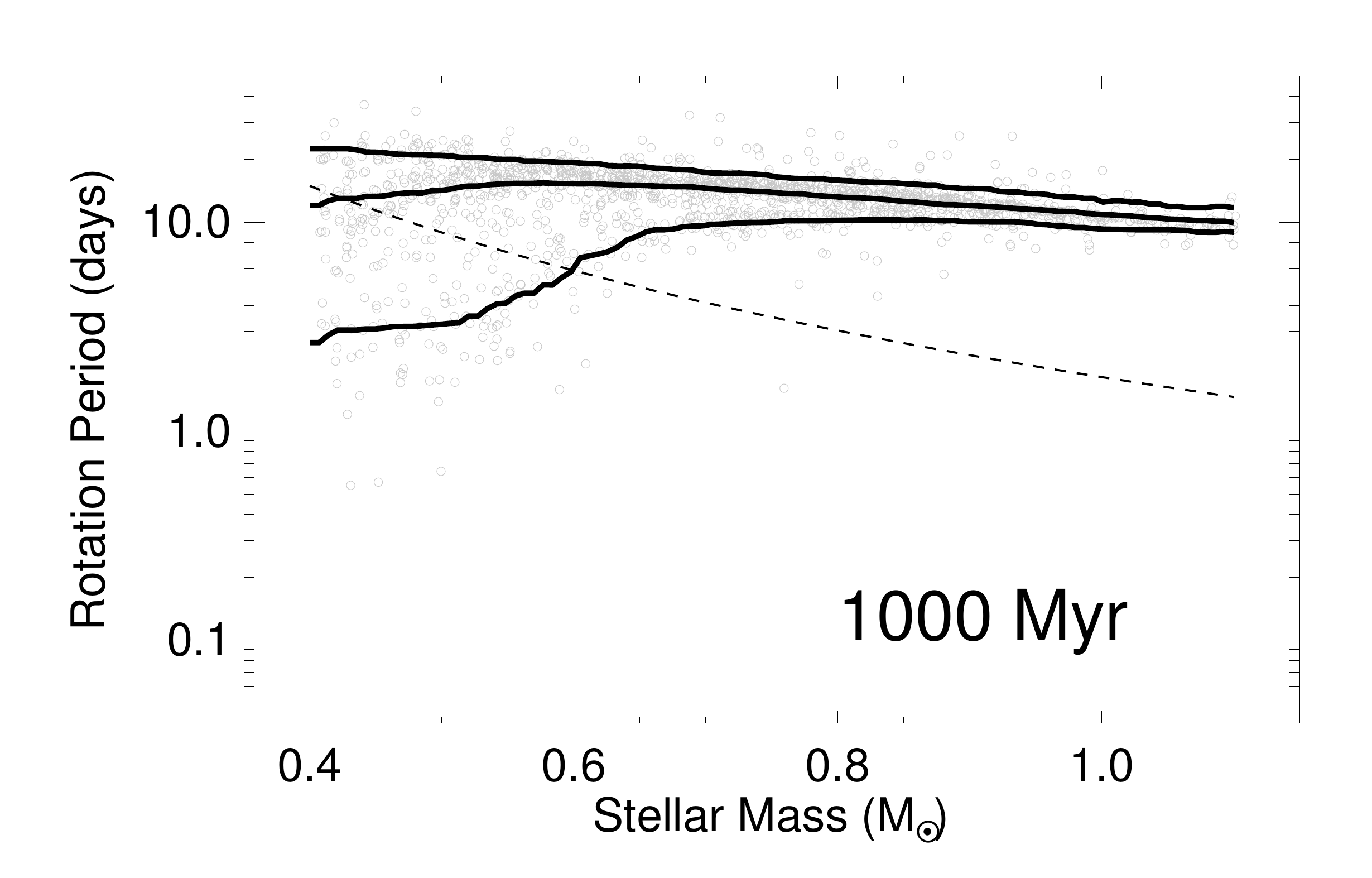}
\includegraphics[trim=5mm 5mm 5mm 5mm,width=0.49\textwidth]{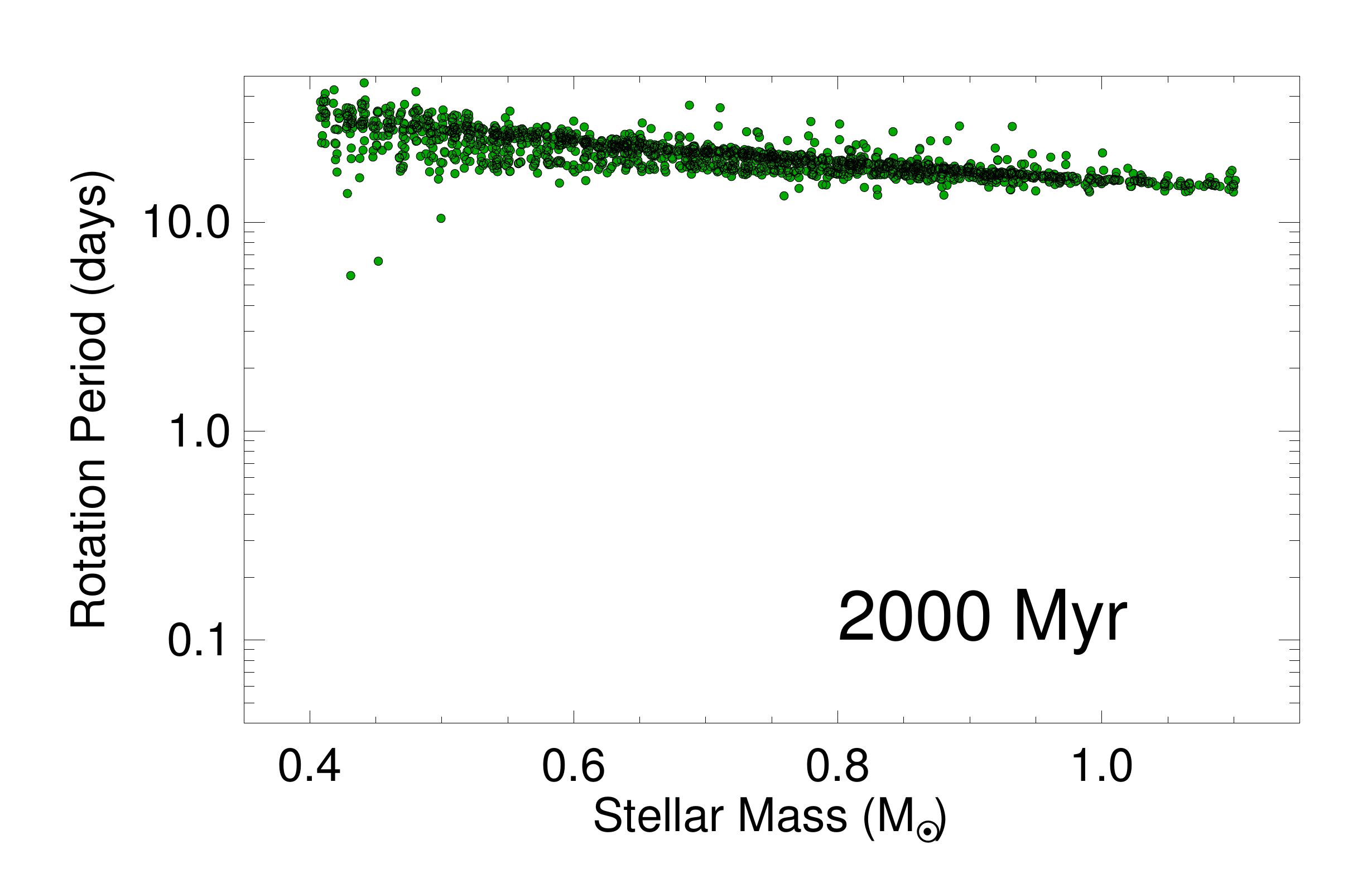}
\includegraphics[trim=5mm 5mm 5mm 5mm,width=0.49\textwidth]{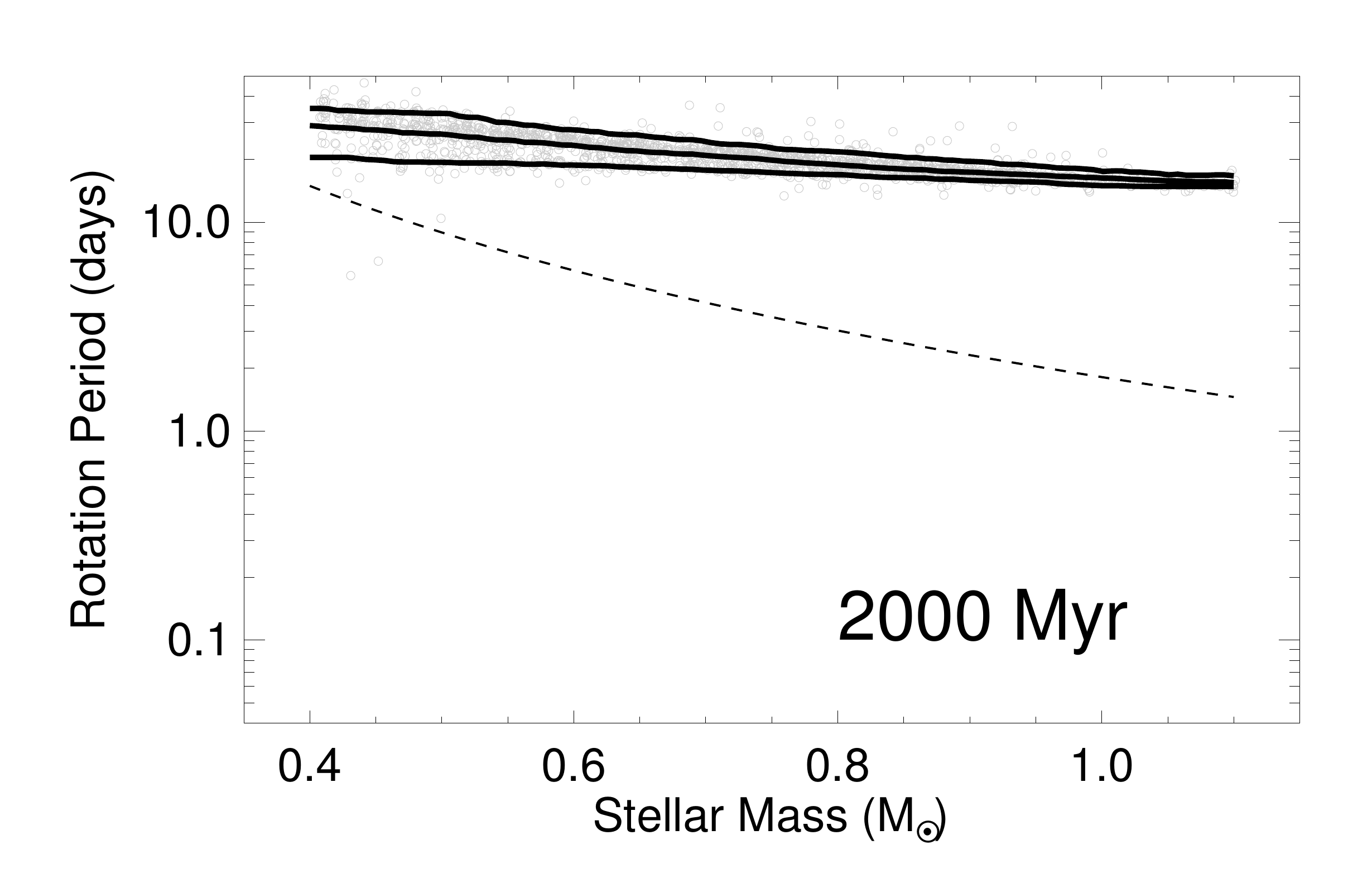}
\caption{
Figure showing how rotation in our composite cluster evolves between 100~Myr and 2~Gyr according to our rotational evolution model.
The cluster is made up of $\sim$1500 stars in the four young clusters Pleiades, M50, M35, and NGC~2516, all of which have ages $\sim$150~Myr.
The upper row shows the distribution predicted by evolving this composite cluster back to 100~Myr, and the other three rows show the distribution evolved forward to 500~Myr, 1~Gyr, and 2~Gyr. 
The left column shows the position of each star in the distribution and the right column shows the 10th, 50th, and 90th percentiles of the distribution as a function of mass for each age.
The dashed lines show the saturation threshold determined by Eqn.~\ref{eqn:saturation}.
}
 \label{fig:superclusterevolution}
\end{figure*}

\begin{table*}
\centering 
\begin{tabular}{ccccccccc}
Stellar Mass (M$_\odot$) & $\bar{\Omega}_\star$ ($\Omega_\odot$) & $\sigma_\Omega$ ($\Omega_\odot$) & $\Omega_{10}$ ($\Omega_\odot$) & $\Omega_{50}$ ($\Omega_\odot$) & $\Omega_{90}$ ($\Omega_\odot$) & $\dot{M}_{10}$ ($\dot{\text{M}}_\odot$) & $\dot{M}_{50}$ ($\dot{\text{M}}_\odot$) & $\dot{M}_{90}$ ($\dot{\text{M}}_\odot$) \\
\hline
\hline
100 Myr \\
\hline
All & 27.43 & 44.62 & 3.24 & 8.49 & 78.79 & - & - & -\\
0.5 & 32.39 & 35.09 & 3.88 & 18.05 & 85.25 & 11.84 & 11.84 & 11.84\\
0.6 & 29.04 & 35.28 & 3.12 & 11.13 & 88.65 & 8.49 & 14.36 & 14.36\\
0.7 & 27.22 & 42.93 & 3.02 & 6.52 & 82.76 & 6.26 & 17.42 & 17.72\\
0.8 & 25.44 & 52.11 & 3.07 & 5.20 & 69.82 & 5.08 & 10.21 & 21.13\\
0.9 & 22.65 & 54.66 & 3.23 & 5.64 & 53.11 & 4.50 & 9.48 & 25.20\\
1.0 & 23.71 & 51.17 & 3.76 & 7.15 & 60.10 & 4.45 & 10.46 & 28.02\\
\hline
250 Myr \\
\hline
All & 16.43 & 28.08 & 2.67 & 5.19 & 50.29 & - & - & -\\
0.5 & 23.00 & 26.70 & 2.73 & 12.11 & 62.18 & 10.29 & 11.88 & 11.88\\
0.6 & 19.07 & 24.65 & 2.51 & 6.35 & 60.91 & 6.37 & 14.38 & 14.38\\
0.7 & 15.62 & 27.11 & 2.56 & 3.89 & 45.81 & 5.02 & 8.78 & 17.76\\
0.8 & 13.56 & 32.32 & 2.62 & 3.77 & 34.54 & 4.11 & 6.67 & 21.18\\
0.9 & 11.07 & 30.92 & 2.82 & 4.24 & 16.39 & 3.78 & 6.52 & 25.30\\
1.0 & 10.74 & 23.39 & 3.25 & 5.03 & 14.74 & 3.74 & 6.69 & 27.94\\
\hline
500 Myr \\
\hline
All & 7.56 & 12.68 & 2.11 & 3.28 & 19.24 & - & - & -\\
0.5 & 12.47 & 15.90 & 1.96 & 5.77 & 33.83 & 6.63 & 11.94 & 11.94\\
0.6 & 9.11 & 12.26 & 1.96 & 3.04 & 28.13 & 4.58 & 8.21 & 14.41\\
0.7 & 6.05 & 11.03 & 2.08 & 2.69 & 13.93 & 3.84 & 5.39 & 17.82\\
0.8 & 5.23 & 12.41 & 2.19 & 2.82 & 6.82 & 3.25 & 4.57 & 14.75\\
0.9 & 4.47 & 9.67 & 2.37 & 3.16 & 4.77 & 3.02 & 4.43 & 7.66\\
1.0 & 4.12 & 2.88 & 2.69 & 3.64 & 4.98 & 3.00 & 4.49 & 6.81\\
\hline
750 Myr \\
\hline
All & 4.10 & 5.82 & 1.74 & 2.52 & 7.43 & - & - & -\\
0.5 & 6.60 & 8.87 & 1.57 & 2.84 & 17.12 & 4.99 & 10.96 & 12.00\\
0.6 & 4.48 & 5.45 & 1.61 & 2.16 & 11.89 & 3.55 & 5.22 & 14.43\\
0.7 & 3.00 & 3.73 & 1.79 & 2.17 & 4.34 & 3.15 & 4.07 & 10.22\\
0.8 & 2.84 & 3.56 & 1.91 & 2.37 & 3.55 & 2.73 & 3.62 & 6.21\\
0.9 & 2.82 & 2.00 & 2.08 & 2.62 & 3.34 & 2.55 & 3.48 & 4.80\\
1.0 & 2.98 & 0.61 & 2.36 & 2.93 & 3.57 & 2.59 & 3.45 & 4.50\\
\hline
1000 Myr \\
\hline
All & 2.66 & 2.82 & 1.48 & 2.08 & 3.49 & - & - & -\\
0.5 & 3.60 & 4.74 & 1.32 & 1.89 & 8.33 & 3.96 & 6.41 & 12.06\\
0.6 & 2.57 & 2.19 & 1.41 & 1.78 & 4.70 & 2.97 & 4.05 & 14.46\\
0.7 & 2.11 & 0.99 & 1.58 & 1.88 & 2.82 & 2.67 & 3.36 & 5.77\\
0.8 & 2.16 & 0.81 & 1.71 & 2.06 & 2.67 & 2.36 & 3.01 & 4.28\\
0.9 & 2.29 & 0.45 & 1.88 & 2.27 & 2.71 & 2.25 & 2.88 & 3.66\\
1.0 & 2.51 & 0.37 & 2.10 & 2.50 & 2.93 & 2.28 & 2.86 & 3.55\\
\hline
1500 Myr \\
\hline
All & 1.66 & 0.72 & 1.16 & 1.60 & 2.08 & - & - & -\\
0.5 & 1.56 & 1.21 & 1.00 & 1.27 & 2.22 & 2.75 & 3.79 & 7.98\\
0.6 & 1.48 & 0.36 & 1.15 & 1.38 & 1.96 & 2.28 & 2.90 & 4.63\\
0.7 & 1.56 & 0.26 & 1.31 & 1.53 & 1.91 & 2.09 & 2.57 & 3.45\\
0.8 & 1.68 & 0.24 & 1.43 & 1.67 & 1.95 & 1.87 & 2.30 & 2.83\\
0.9 & 1.82 & 0.22 & 1.59 & 1.83 & 2.07 & 1.82 & 2.20 & 2.58\\
1.0 & 1.98 & 0.20 & 1.76 & 1.98 & 2.22 & 1.88 & 2.19 & 2.56\\
\hline
2000 Myr \\
\hline
All & 1.33 & 0.30 & 0.96 & 1.34 & 1.67 & - & - & -\\
0.5 & 1.09 & 0.32 & 0.82 & 1.04 & 1.41 & 2.15 & 2.93 & 4.39\\
0.6 & 1.19 & 0.18 & 0.98 & 1.16 & 1.45 & 1.85 & 2.32 & 3.10\\
0.7 & 1.32 & 0.17 & 1.13 & 1.31 & 1.53 & 1.73 & 2.10 & 2.60\\
0.8 & 1.43 & 0.16 & 1.25 & 1.44 & 1.61 & 1.58 & 1.90 & 2.21\\
0.9 & 1.56 & 0.15 & 1.39 & 1.57 & 1.72 & 1.54 & 1.81 & 2.04\\
1.0 & 1.67 & 0.13 & 1.54 & 1.67 & 1.82 & 1.62 & 1.81 & 2.03\\
\hline
\end{tabular}
\caption{
Table giving statistical properties of our composite cluster at different ages.
The cluster is composed of $\sim$1500 stars with known rotation periods and masses from Pleiades, M50, M35, and NGC~2516, all of which have ages $\sim$150~Myr.
We have evolved the cluster backwards to 100~Myr and forwards to 2~Gyr using our rotational evolution model. 
The distributions of rotation rates at ages of 100~Myr, 500~Myr, 1~Gyr, and 2~Gyr for this cluster are shown in Fig.~\ref{fig:superclusterevolution}. 
In this table, we give, from left to right, the average angular velocity, $\bar{\Omega}_\star$, the standard deviation in the angular velocity, $\sigma_\Omega$, the 10th, 50th, and 90th percentiles in the angular velocity distributions, and our predicted wind mass loss rates at these percentiles. 
For each age, we give the values for all of the stars between 0.4~M$_\odot$ and 1.1~M$_\odot$, and for masses of 0.4~M$_\odot$, 0.5~M$_\odot$, 0.6~M$_\odot$, 0.7~M$_\odot$, 0.8~M$_\odot$, 0.9~M$_\odot$, and 1.0~M$_\odot$.
At each specific mass, we calculate the values of each parameter using all stars within 0.1~M$_\odot$ of the given mass. 
}
\label{tbl:percentiles2}
\end{table*}

It is interesting to compare our models to those of \citet{2013A&A...556A..36G} for solar mass stars. 
The main difference between our model and the model of \citet{2013A&A...556A..36G} on the main-sequence is the use of core-envelope decoupling. 
\citet{2013A&A...556A..36G} found that without the extra spin-up torque from the core on the envelope due to core-envelope decoupling, rapidly rotating stars would spin down too quickly by the age of $\sim$650~Myr.
As we show in the next section, our model for $\dot{M}_\star$ provides similar results to the model of \cite{2011ApJ...741...54C} used in their model, so the difference is unlikely due to difference values of $\dot{M}_\star$.
The difference is likely that we use a weaker scaling for $B_{\text{dip}}$ with rotation rate.
In our model $B_{\text{dip}} \propto \Omega_\star^{1.32}$, whereas in the model of \citet{2013A&A...556A..36G}, a stronger scaling closer to $B_{\text{dip}} \propto \Omega_\star^{2.6}$ was used.
This means that we predict weaker torques for rapidly rotating stars, allowing them to spin down slower without core-envelope decoupling.
We warn however that we have only shown that the evolution of solar mass stars on the main-sequence can be explained without the use of core-envelope decoupling.
However, core-envelope decoupling is also an important factor on the PMS, which is a phase of rotational evolution that we do not consider.

The rotation tracks for the lowest mass stars that we consider are shown in the upper panel of Fig.~\ref{fig:rotevotracks}.
Our models clearly provide a good fit to the observational constraints, though there is some indication that we slightly underestimate the spread in rotation rates at 550~Myr.
The difference, however, is very small.
There are two main differences between the rotational evolution of 1.0~M$_\odot$ stars and 0.5~M$_\odot$ stars.
Firstly, in the first Gyr of evolution just after the ZAMS, much less spin down of the rapidly rotating stars takes place.
Secondly, low-mass slowly rotating stars spin down faster than higher mass stars.
This results in the slowly rotating tracks in young clusters having a shape such that higher mass stars rotate faster than lower mass stars, with a similar trend at later ages.

The most noticeable difference between the models and the observations are in the 0.75~M$_\odot$ mass bin.
All three tracks spin down to approximately the same rotation rate by 5~Gyr, but this value is slightly larger than the predicted value. 
It is unclear if this is a result of the model underestimating the wind torques at later ages for 0.75~M$_\odot$ stars, or if Eqn.~\ref{eqn:mamajek} overestimates the rotation periods for stars of this mass.
This latter suggestion is plausible given that the exact form of the rotation period dependence on stellar mass at a given age beyond 1~Gyr is not well constrained observationally, but is the result of assuming that the shape of the slowly rotating tracks seen in young clusters remains the same as the stars evolve to later ages. 
The nearby $\sim$0.7~M$_\odot$~star 61~Cyg~A provides a test for which of the two predictions is likely to be more realistic. 
\citet{1996ApJ...466..384D} estimated that 61~Cyg~A has a rotation period of $\sim$35~days and \citet{2008A&A...488..667K} estimated an age for the star of \mbox{$6.0 \pm 1.0$~Gyr}.
This suggests that our prediction for the spin down of stars with this mass are more likely.
However, \citet{1996ApJ...466..384D} also estimated a similar rotation period of $\sim$38~days for the lower mass companion, 61~Cyg~B, which is difficult to reconcile with both our predictions and the gyrochronological relation of \citet{2008ApJ...687.1264M} if the age of 6~Gyr is reasonable.

In Section~\ref{sect:obsconstraints}, we combine the rotation period measurements for the Pleiades, M50, M35, and NGC~2516.
These are four young clusters with ages of $\sim$150~Myr, and we assume that the combination of the rotation periods for the four clusters represents the rotational distribution at 150~Myr. 
To illustrate the results of our rotational evolution model, we evolve this cluster back to 100 Myr, and forward to 2~Gyr. 
In Fig.~\ref{fig:superclusterevolution}, we show the distribution at 100~Myr, 500~Myr, 1~Gyr, and 2~Gyr.
The left column shows the locations of each star in the distribution at each age and the right column shows the 10th, 50th, and 90th percentiles of the distributions as a function of mass. 
The percentiles were calculated at each mass by considering all stars within 0.1~M$_\odot$ of that mass. 
In Table~\ref{tbl:percentiles2}, we give these percentiles for masses of 0.5~M$_\odot$, 0.6~M$_\odot$, 0.7~M$_\odot$, 0.8~M$_\odot$, 0.9~M$_\odot$, and 1.0~M$_\odot$ at ages of 100~Myr, 250~Myr, 500~Myr, 750~Myr, 1~Gyr, 1.5~Gyr, and 2~Gyr. 
As before, we assume mass bins with widths of 0.2~M$_\odot$.
The dashed lines in Fig.~\ref{fig:superclusterevolution} show the mass-dependent saturation threshold.

\begin{figure}
\includegraphics[trim=10mm 5mm 5mm 5mm,width=0.49\textwidth]{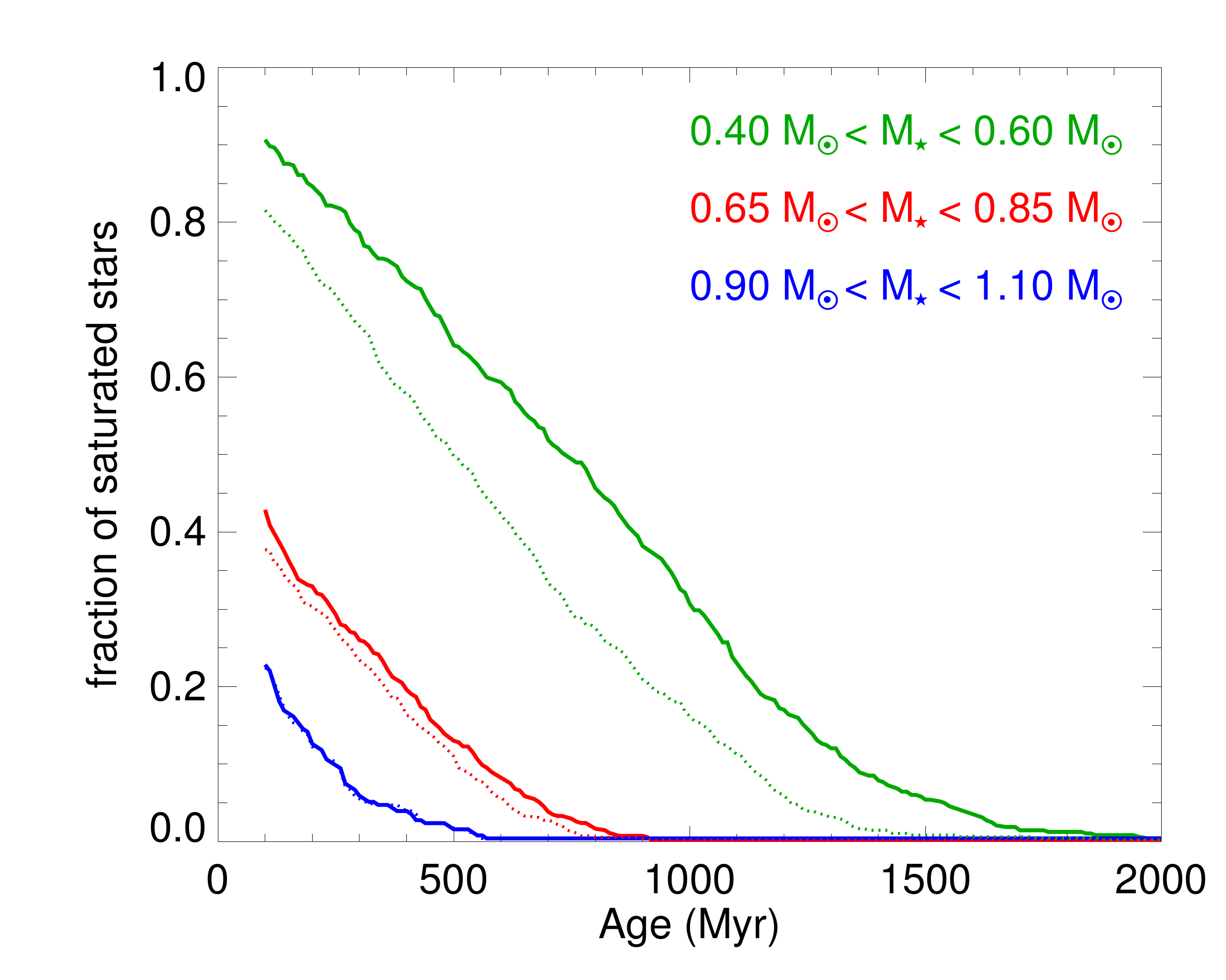}
\caption{
Plot showing the fraction of stars in the saturated regime as a function of age based on the results shown in Fig.~\ref{fig:superclusterevolution} for different mass bins.
The solid lines correspond to the values calculated using our constraints on the mass dependence of the saturation threshold given by Eqn.~\ref{eqn:saturation} with $c=2.3$, and the dotted lines show the values calculated using the empirical saturation threshold for \mbox{X-ray} emission derived by \citet{2011ApJ...743...48W}.
}
 \label{fig:fracsat}
\end{figure}

Fig.~\ref{fig:superclusterevolution} shows clearly how the rotation rates of stars in each mass bin converge as they age. 
At 100~Myr, the 10th and 90th percentiles are almost independent of mass, indicating that the slowest and fastest rotators at each mass have similar rotation rates.
However, the 50th percentile is strongly mass dependent.
At all masses, a significant fraction of stars are in the saturated regime, especially for low mass stars where almost all stars are saturated.
By 500~Myr, the 90th percentile has converged for solar mass stars, but is still at very fast rotation for low-mass stars.
Almost all stars of solar mass have evolved out of the saturated regime, though a significant fraction of stars at low masses remain saturated.
A similar distribution is seen at 1~Gyr, though the 90th percentile track has made more progress in converging with the other tracks at low masses and very few stars at low masses are saturated. 
Between 1~Gyr and 2~Gyr, the remaining rapid rotators converge, and at 2~Gyr, all stars lie on one track, with lower mass stars rotating slower than higher mass stars.
At 2~Gyr, there are no remaining saturated stars in the distribution.
The age dependence of the fraction of stars in the saturated regime for the different mass bins is shown in Fig.~\ref{fig:fracsat}.



\section{The Evolution of Stellar Winds on the Main-Sequence} \label{sect:windevo}

It is now possible to couple the stellar wind model developed in Paper~I with the rotational evolution model that we develop in this paper to predict how stellar wind properties evolve on the main-sequence for a range of stellar masses.
We review our wind model in Section~\ref{sect:windmodel}, and then apply it to the evolution of the solar wind in Section~\ref{sect:solarwindintime} and to stars with a range of masses in Section~\ref{sect:windsallmasses}.

\subsection{Stellar wind model} \label{sect:windmodel}

In Paper~I, we construct a stellar wind model based on scaling the solar wind to other stars.
Our wind model was calculated using a 1D numerical hydrodynamic model of the solar wind run using the \emph{Versatile Advection Code} (\citealt{1996ApL&C..34..245T}; \citealt{1997JCoPh.138..981T}). 
The fundamental driving mechanism of our wind is thermal pressure gradients. 
In order for a thermal pressure driven wind to gain enough energy to accelerate to the correct wind speeds, it is necessary to heat the wind as it expands.
To do this, we assume that pressure and density are related by a polytropic equation of \mbox{state, $p \propto \rho^\alpha$.}
In order to accurately describe the wind, we choose a value of $\alpha$ that varies with radial distance from the solar surface. 
Once $\alpha$ is set, the two free parameters that determine the wind properties are the base temperature, $T_0$, which determines the wind acceleration and strongly influences the wind densities, and the base density, $n_0$, which strongly influences the wind densities only.
 
\begin{figure}
\includegraphics[trim = 10mm 0mm 5mm 10mm,clip=true,width=0.49\textwidth]{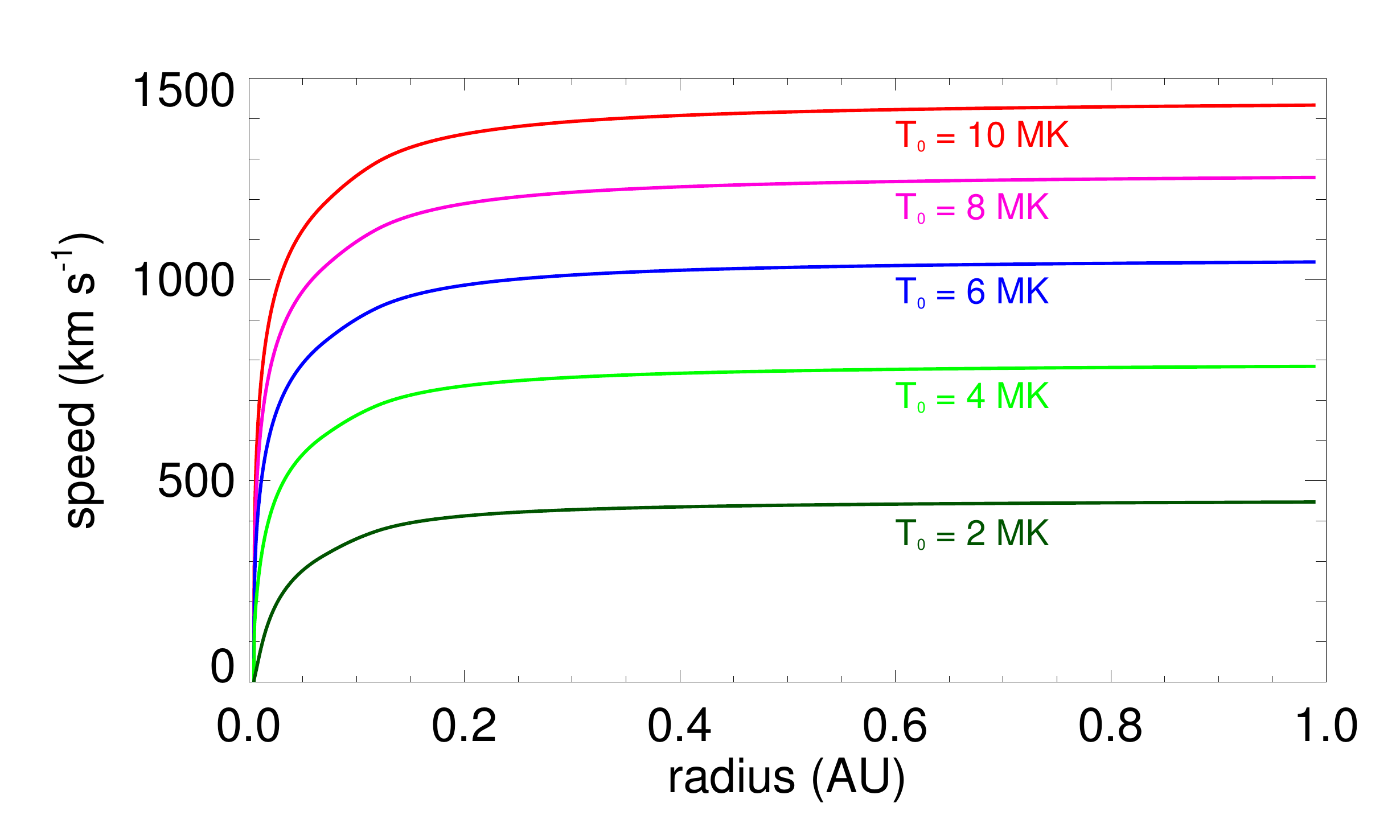}
\includegraphics[trim = 10mm 0mm 5mm 10mm,clip=true,width=0.49\textwidth]{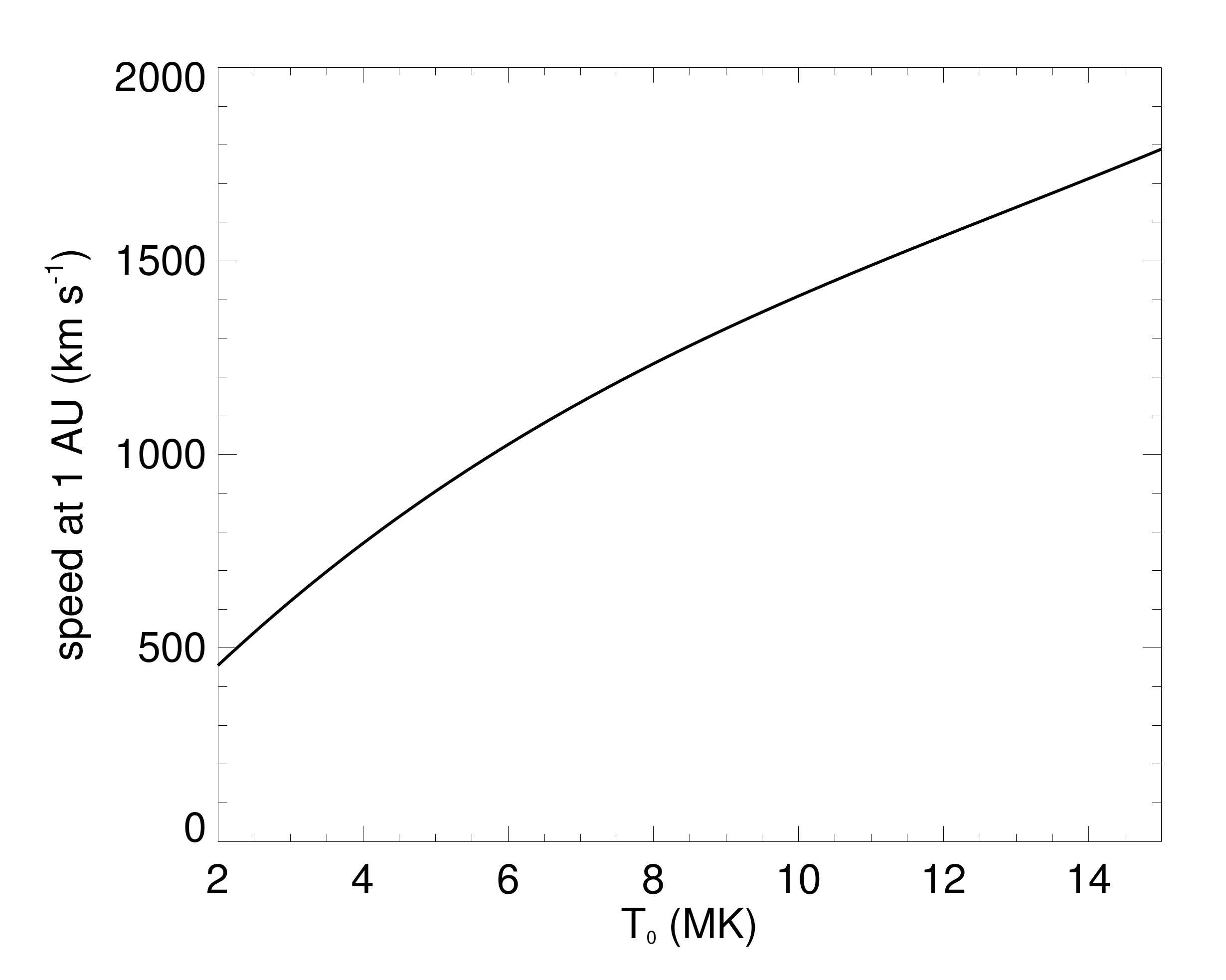}
\caption{
Figure demonstrating the dependence of wind speed on the wind base temperature in our model.
The upper panel shows wind speed against radial distance from the stellar surface for a solar mass and radius star for five different base temperatures.
The lower panel shows the dependence of wind speed at 1~AU on $T_0$, as given by Eqn.~\ref{eqn:windspeed}.
}
 \label{fig:windspeedmodel}
\end{figure}

In Paper~I, we run a grid of 1200 numerical wind models with a range of stellar masses, stellar radii, and wind temperatures.
We show that the wind speed can be predicted from $T_0$ only with the equation

\begin{equation} \label{eqn:windspeed}
v_{1\text{AU}} \approx 73.39 + 224.14 T_0 - 11.28 T_0^2 + 0.28 T_0^3,
\end{equation}

\noindent where $v_{1\text{AU}}$ is the wind speed at 1~AU in km~s$^{-1}$ and $T_0$ is in MK. 
The dependence of wind speed on $T_0$ is shown in Fig.~\ref{fig:windspeedmodel}.
Once $T_0$ is known, the base density can be calculated in order to give the required densities far from the star.

For the solar wind, we constrain $T_0$ and $n_0$ using spacecraft measurements.
The solar wind breaks down quite clearly into two components based on the speed of propagation. 
The spatial distribution of slow and fast wind is closely associated with the structure of the solar magnetic field (\citealt{1990ApJ...355..726W}; \citealt{2000JGR...10510465A}).
The slow wind travels approximately 400~\kms\hspace{0mm} and at 1~AU has typical proton densities and temperatures of $\sim5$~cm$^{-3}$ and \mbox{$7.5 \times 10^4$~K}. 
The fast wind travels approximately 760~\kms\hspace{0mm} and at 1~AU has typical proton densities and temperatures of $\sim2$~cm$^{-3}$ and \mbox{$1.8 \times 10^5$~K}. 
Fitting our models to the spacecraft measurements, we found base temperatures of 1.8~MK and 3.8~MK for the slow and fast components of the solar wind respectively.
Our models provide an excellent description of the real solar wind far from the solar surface, but are unrealistic within the solar corona. 
In our model, we assume that the winds of all low-mass main-sequence stars break down into slow and fast wind components, and that the two components scale to other stars in the same way.

For stars other than the Sun, spacecraft measurements are not available.
We use instead the constraints on the mass loss rates from our rotational evolution model to derive the wind mass fluxes, $\rho v$.
However, disentangling $\rho$ and $v$ is difficult and requires that we know the wind temperature. 
In Paper~I, we develop two models for scaling the base temperature.

\begin{itemize}

\item
In Model~A, we follow \citet{2007A&A...463...11H} and assume that the base temperature is proportional to the coronal temperature, such that

\begin{equation}
T_0 = T_{0,\odot} \left( \frac{\bar{T}_{\text{cor}}}{\bar{T}_{\text{cor},\odot}} \right),
\end{equation}

\noindent where $\bar{T}_{\text{cor}}$ is the plasma temperature averaged over the entire corona. 
In order to do this, we use the result of Johnstone \& G\"{u}del (2015) that there exists one universal relation between coronal average temperature, $\bar{T}_{\text{cor}}$, and \mbox{X-ray} surface flux, $F_\text{X} = L_\text{X} / (4 \pi R_\star^2)$, for all low-mass main-sequence stars, such that \mbox{$\bar{T}_{\text{cor}} \approx 0.11 F_\text{X}^{0.26}$}, where $\bar{T}_{\text{cor}}$ is in MK and $F_\text{X}$ is in \mbox{erg s$^{-1}$ cm$^{-2}$}.
We use the empirical relations of \citet{2011ApJ...743...48W} to estimate $F_\text{X}$ as a function of stellar mass and rotation. 

\vspace{2mm}

\item
In Model~B, we follow \citet{2012ApJ...754L..26M} and assume that the sound speed at the base of the wind is a fixed fraction of the escape velocity, such that 

\begin{equation}
T_0 = \frac{2 G  \mu m_\text{p} }{\gamma k_\text{B}} \left( \frac{c_s}{v_{\text{esc}}} \right)^2 \left( \frac{M_\star}{R_\star} \right),
\end{equation}

\noindent where the fraction $c_s/v_{\text{esc}}$ is a constant that we constrain from the solar wind models
For the slow and fast winds respectively, we find $c_s/v_{\text{esc}}=0.329$ and $c_s/v_{\text{esc}}=0.478$.

\end{itemize}


\noindent The two models are by definition equivalent for the Sun, but for other stars they can lead to different results.

The final effect that we consider in our wind model is the acceleration of the wind from magneto-rotational effects. 
As an ionised wind expands away from a rotating magnetised star, it gains kinetic energy and angular momentum from the magnetic field, which can lead to increased wind speeds far from the star and, in extreme cases, increased mass loss rates. 
Due to this acceleration, for rapidly rotating stars, the wind structure becomes highly latitude dependent and properly taking this into account would require 2D or 3D MHD models which would be inappropriate in this paper.  
On the other hand, the influence of magneto-rotational effects on the wind speeds is likely to be so significant for the most rapidly rotating stars that it would also be inappropriate to ignore them entirely.
For simplicity, we assume the influence of the rotation of the star can be charactarised by the Michel velocity, given by

\begin{equation} \label{eqn:MichelVel}
v_{\text{M}} = \left( \frac{R_\star^4 B_\star^2 \Omega_\star^2}{\dot{M}_\star} \right)^{\frac{1}{3}}.
\end{equation}

\noindent As we justify in Paper~I, $B_\star = 0.2 B_\text{dip}$, where $B_\text{dip}$ is the polar strength of the dipole component of the magnetic field. 
In the following section, we discuss our model for calculating $B_{\text{dip}}$ for other stars. 
We follow \citet{1976ApJ...210..498B} and consider two regimes: these are the \emph{slow magnetic rotator}~(SMR) and the \emph{fast magnetic rotator}~(FMR) regimes.
A star is in the SMR regime when the speed that the wind would have in the absence of the star's rotation is much larger than $v_\text{M}$, in which case, magneto-rotational effects can be ignored.
A star is in the FMR regime when $v_\text{M}$ is larger than the speed that the wind would have in the absence of rotation, in which case, the terminal velocity of the wind is approximately $v_\text{M}$. 
The current solar wind has a Michel velocity of $\sim$40~\kms, and is therefore in the SMR regime.
While this model is crude, it gives us a simple way to estimate the influence of magneto-rotational acceleration of the wind in the equatorial plane.

\begin{figure*}
\centering
\includegraphics[trim=5mm 5mm 5mm 5mm,width=0.49\textwidth]{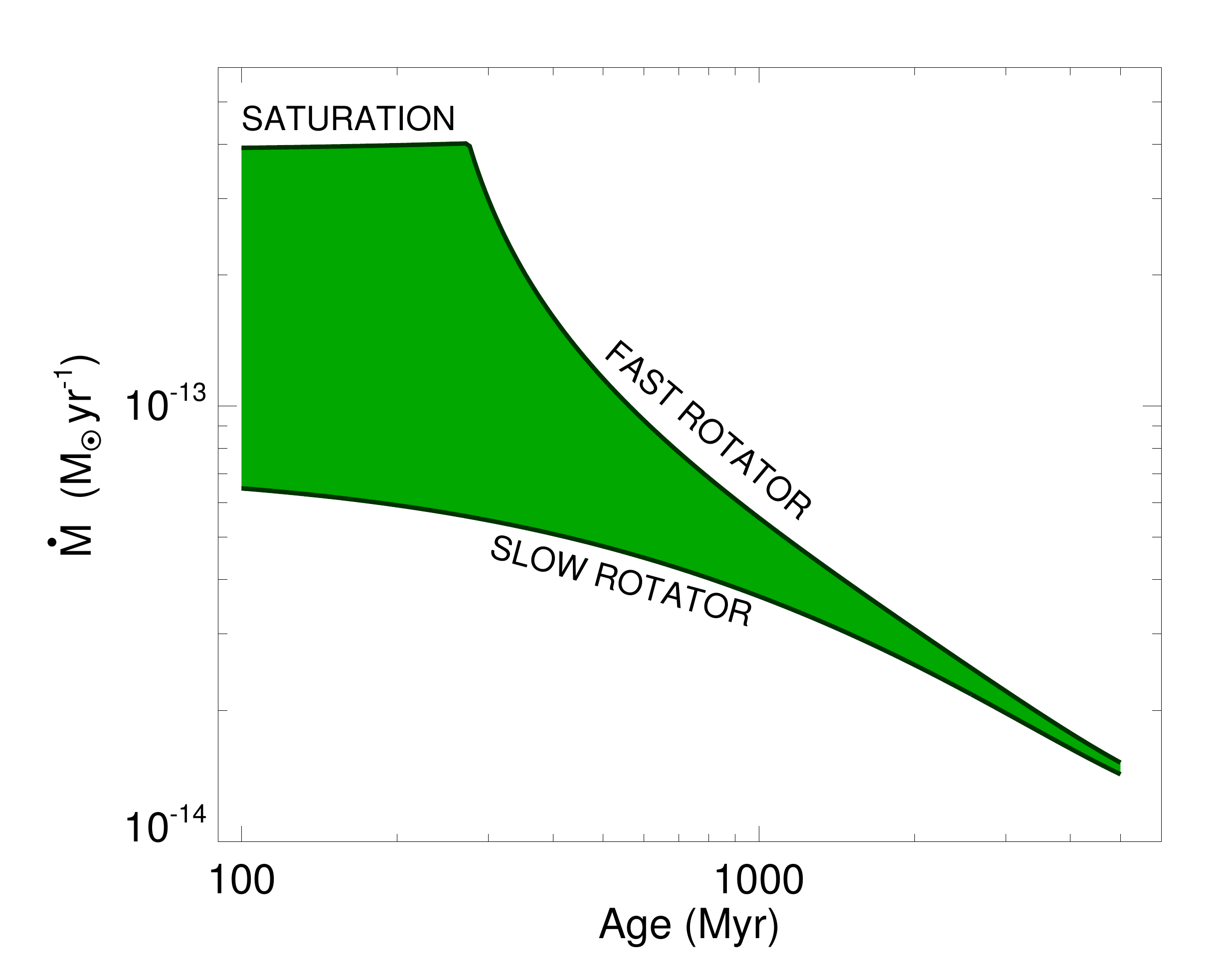}
\includegraphics[trim=5mm 5mm 5mm 5mm,width=0.49\textwidth]{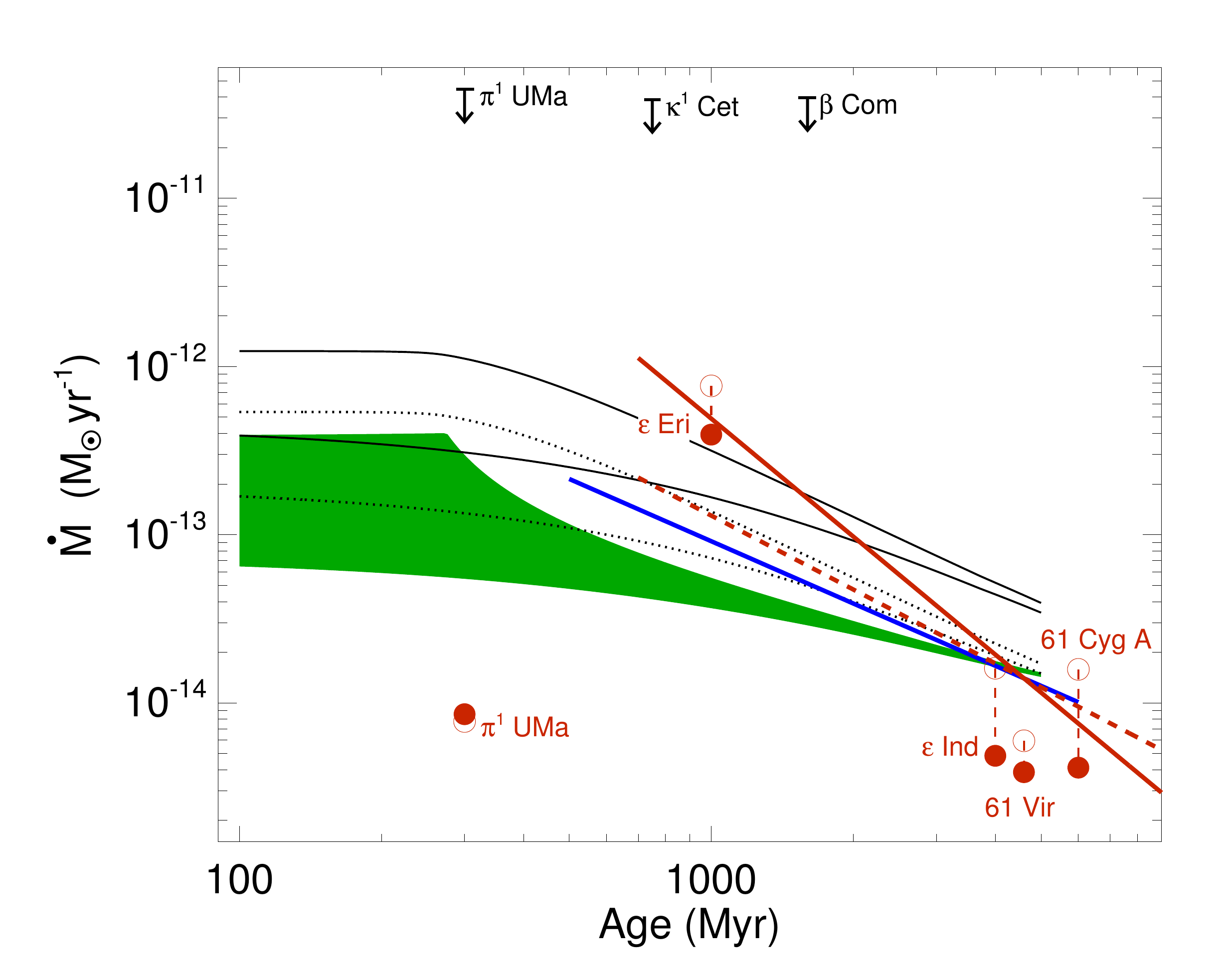}
\caption{
Figures showing the evolution of the solar wind mass loss rate with age on the main-sequence.
The left panel shows the mass loss rates that we derive from our rotational evolution models, such that $\dot{M}_\star \propto \Omega_\star^{1.33}$ for a given stellar radius and mass. 
The lower and upper lines show $\dot{M}_\star$ along the 10th and 90th percentile rotation tracks respectively.
The right panel compares these estimates with other constraints on the mass loss rates of solar mass stars from the literature.
The green shaded region shows the results of our wind model, as shown in the left panel. 
The solid black lines show the predictions of the model developed by \citet{2011ApJ...741...54C}, calculated using the BOREAS code, along our 10th and 90th percentile rotation tracks.
The dotted black lines show the same thing, but with all of the mass loss rates scaled down by a factor of 2.3 in order to make the prediction for the mass loss rate of the current solar wind match our value. 
The solid blue line shows the current solar wind mass loss rate extrapolated into the past by assuming a time dependence of $t^{-1.23}$, as predicted by \citet{2013PASJ...65...98S}.
The three black downward pointing arrows show the upper limits on the mass loss rates for three young solar analogues derived from non-detections of radio emission by \citet{2000GeoRL..27..501G}.
The solid red line shows the solar wind extrapolated into the past assuming a time dependence of $t^{-2.33}$, as suggested by \citet{2005ApJ...628L.143W} from measurements of astrospheric Ly$\alpha$ absorption, and the dashed red line shows the $t^{-1.46}$ dependence that we derive in Section~\ref{sect:solarwindintime} using the results of \citet{2005ApJ...628L.143W}.
The red circles show measurements of the mass loss rates of several stars using this technique by \citet{2005ApJ...628L.143W}, and \citet{2014ApJ...781L..33W}. 
The empty circles correspond to the reported mass loss rates normalised by the stellar surface areas, $\dot{M}_\star R_\star^{-2}$, and the solid circles correspond to $\dot{M}_\star R_\star^{-2} M_\star^{-3.36}$, where $R_\star$ and $M_\star$ are in solar units.
We have chosen an age of 4~Gyr for $\epsilon$~Ind based on \citet{2010A&A...510A..99K}, though significant uncertainty exists in the age determination of this star (see Section~7.3 of \citealt{2010A&A...510A..99K}).
}
 \label{fig:solarMdotintime}
\end{figure*}

\begin{figure*}
\centering
\includegraphics[trim=5mm 0mm 5mm 0mm,width=0.49\textwidth]{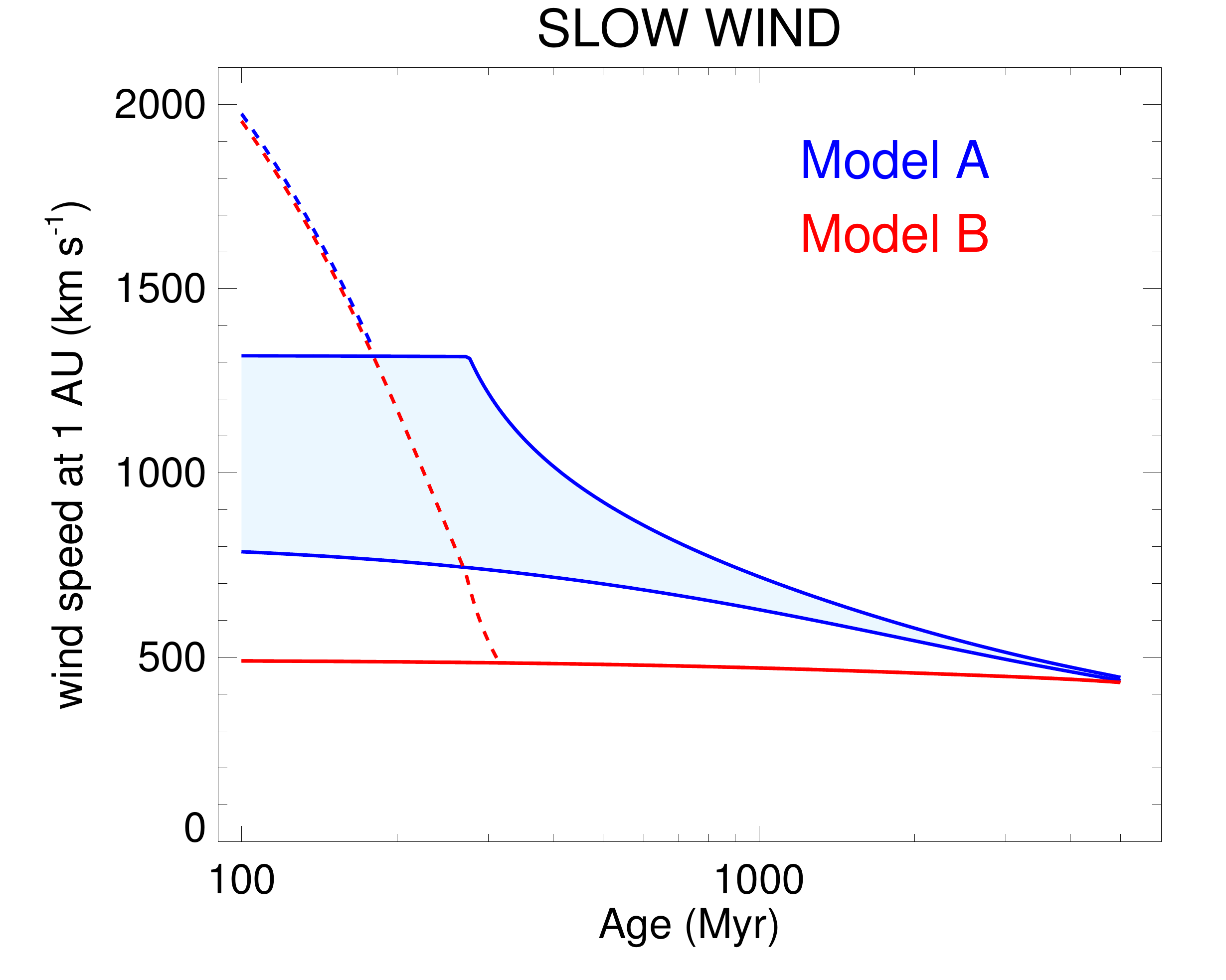}
\includegraphics[trim=5mm 0mm 5mm 0mm,width=0.49\textwidth]{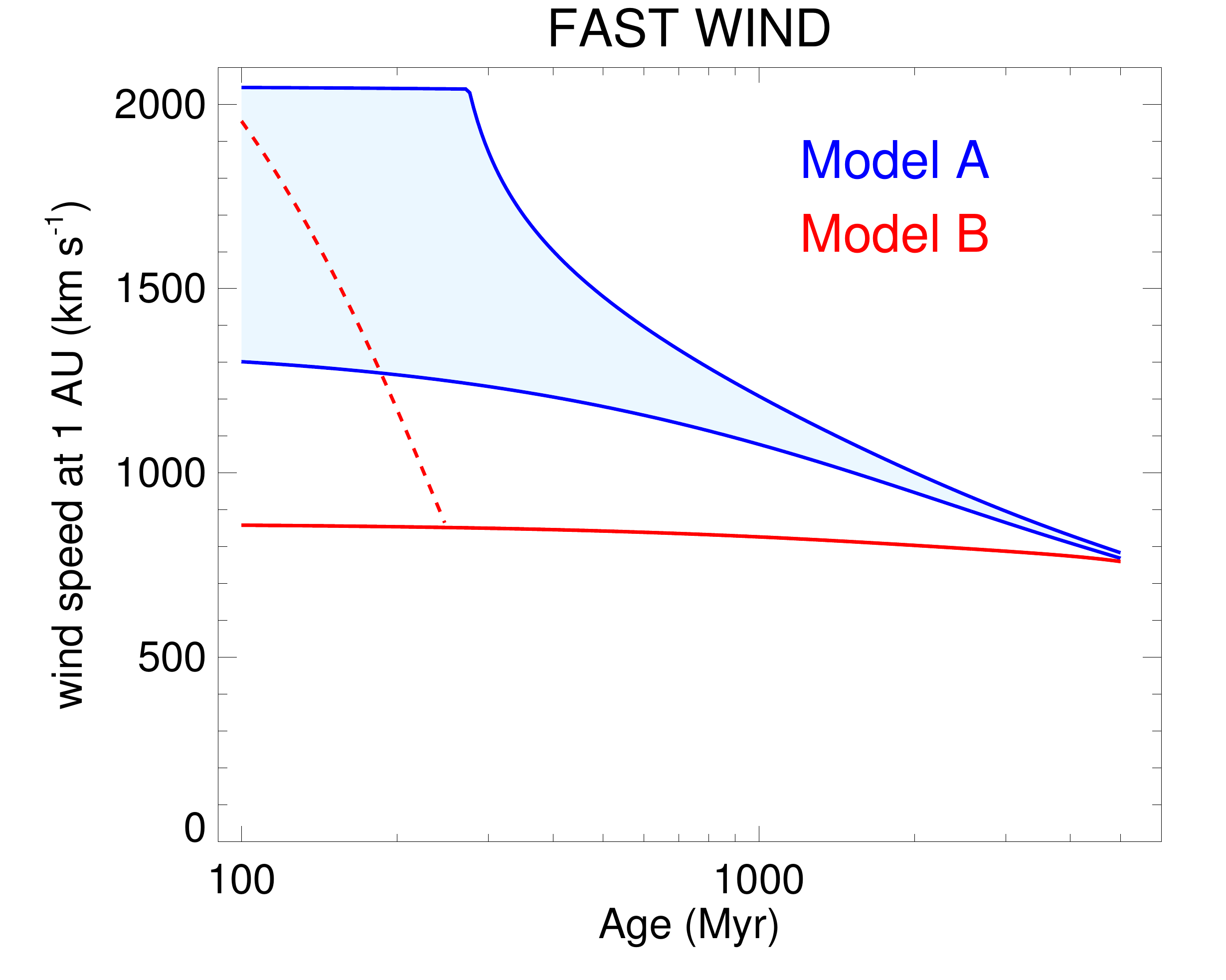}
\includegraphics[trim=5mm 0mm 5mm 0mm,width=0.49\textwidth]{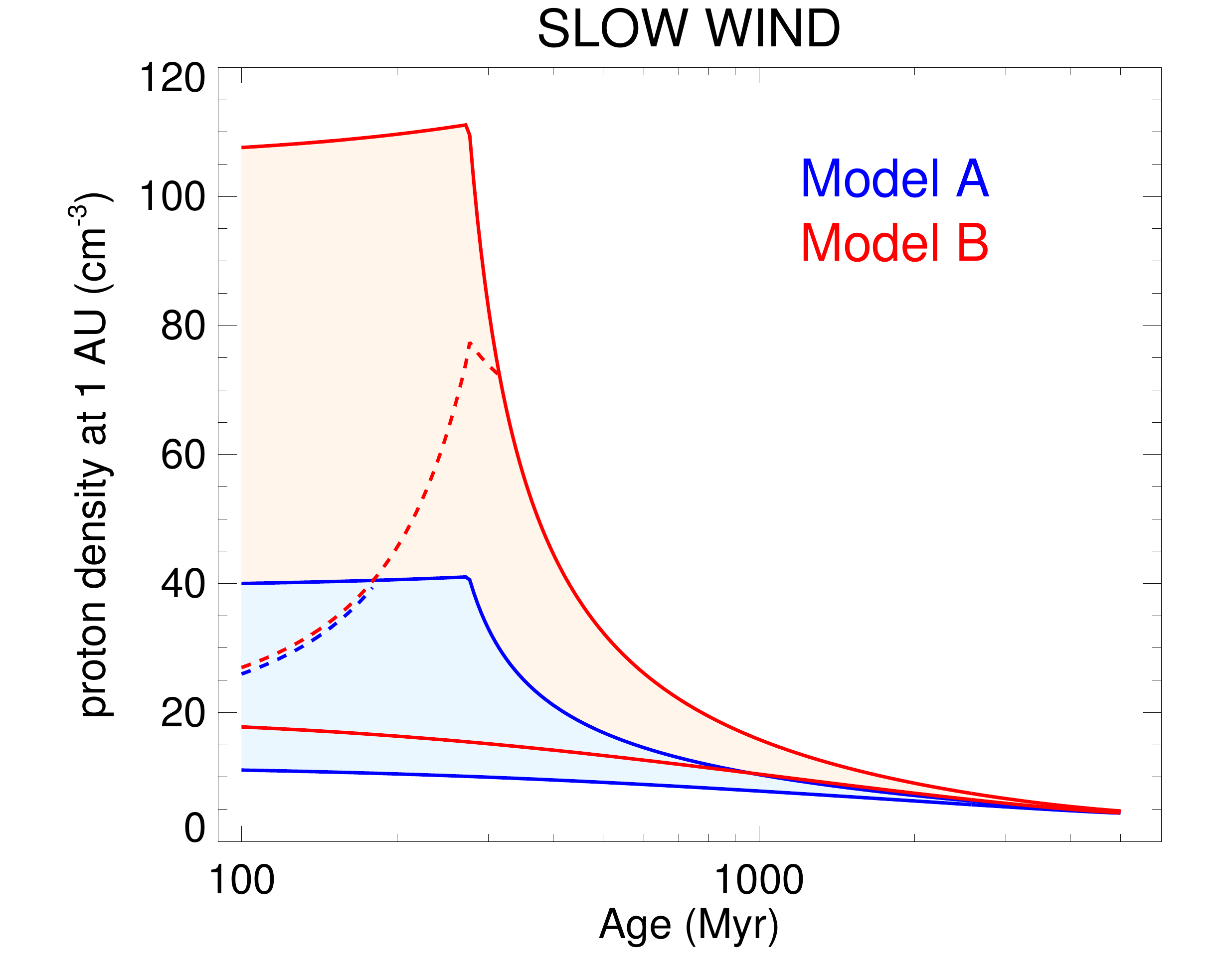}
\includegraphics[trim=5mm 0mm 5mm 0mm,width=0.49\textwidth]{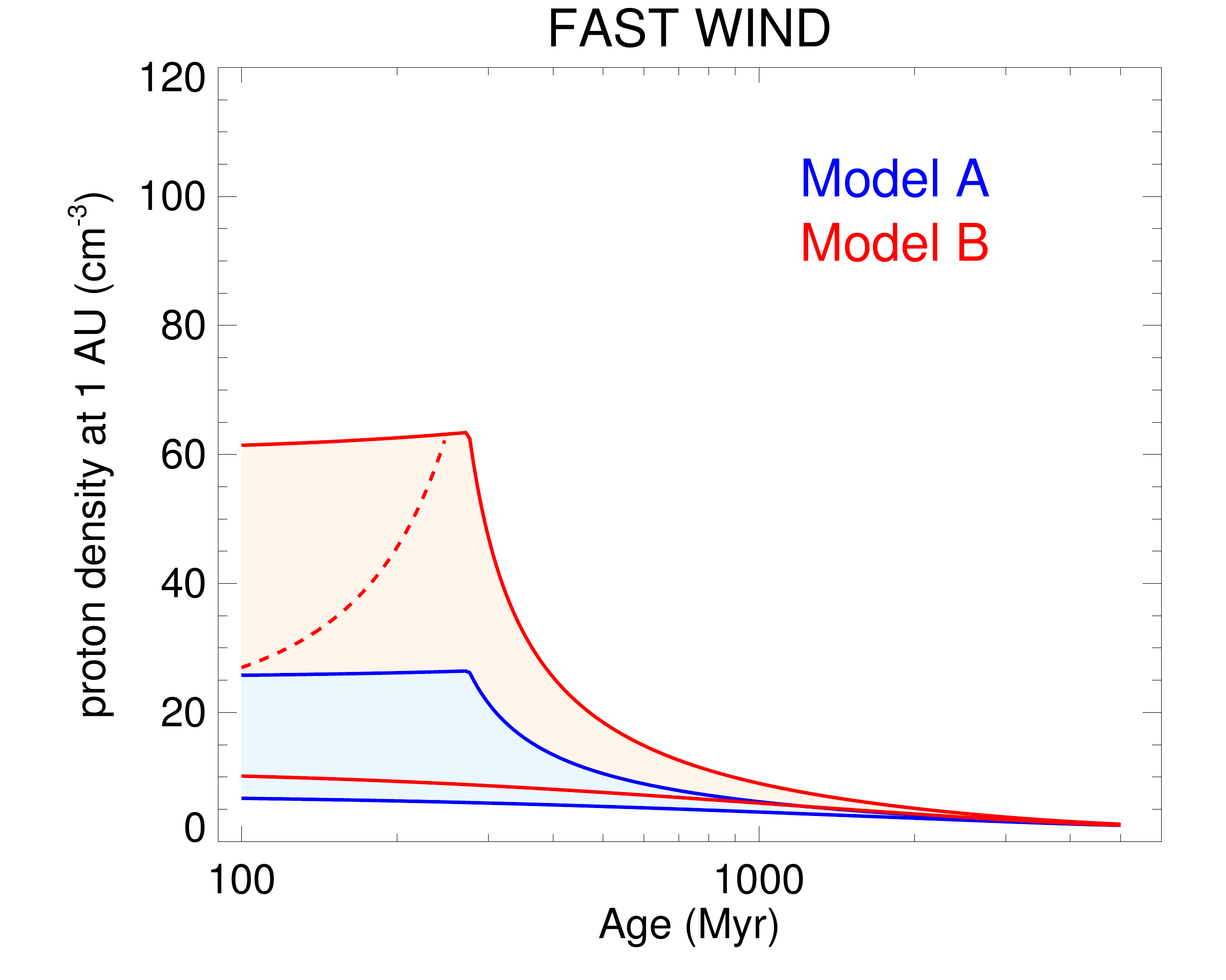}
\caption{
Figure showing the evolution of the solar wind speed (\emph{upper~row}) and density (\emph{lower~row}) at 1~AU on the main-sequence.
In all panels, the lower and upper lines show the properties along the 10th and 90th percentile rotation tracks respectively.
We show the evolution of the slow (\emph{left~column}) and fast (\emph{right~column}) components for both Model~A (\emph{blue}) and Model~B (\emph{red}). 
In all panels, the solid lines show the wind properties assuming only thermal pressure driving of the wind and the dashed lines show the results for the 90th percentile track at 1~AU in the equatorial plane taking into account magneto-rotational acceleration of the wind, as described in Section~\ref{sect:windmodel}. 
}
 \label{fig:solarwind1AUproperties}
\end{figure*}

\subsection{The solar wind in time} \label{sect:solarwindintime}

Our predictions of the properties of the winds of solar mass stars are likely to be more reliable than for lower-mass stars given that our wind model is based on the solar wind, and that our rotational evolution models are based on better observational constraints at solar masses. 
Although the discussion in this section is about the time evolution of the solar wind, our results apply equally to the winds of other solar mass stars.

In Fig.~\ref{fig:solarMdotintime}, we show our predicted mass loss rates for the solar wind as a function of time along the 10th and 90th percentile rotation tracks. 
At 100~Myrs, we predict that the solar wind mass loss rate would have been \mbox{$\sim7 \times 10^{-14}$~M$_\odot$~yr$^{-1}$} if the Sun had been close to the 10th percentile of the rotation distribution, and  \mbox{$\sim4 \times 10^{-13}$~M$_\odot$~yr$^{-1}$}, if the Sun had been close to or above the saturation threshold.
An interesting result is that the spread in possible mass loss rates for the Sun at 100~Myr due to the uncertainties in how rapidly the Sun was rotating is only about a factor of six. 
This is due to the fact that the mass loss rates saturate at around 15$\Omega_\odot$, such that there is no difference between the mass loss rates for stars with rotation rates of 15$\Omega_\odot$ and 100$\Omega_\odot$.
For every track shown in Fig.~\ref{fig:solarMdotintime}, the total mass losses from the solar wind integrated from 100~Myr to 5~Gyr are \mbox{$< 3 \times 10^{-4}$~M$_\odot$}, indicating that the winds of solar mass stars are unable to significantly influence their masses during their main-sequence lives.

If we consider just the ages beyond 700~Myr when the rotation rates of solar mass stars have converged and follow a dependence on age given by Eqn.~\ref{eqn:mamajek}, our model predicts that 
\begin{equation} \label{eqn:oursolarMdotintime}
\dot{M}_\star \propto t^{-0.75}.
\end{equation}
\noindent In the right panel of Fig.~\ref{fig:solarMdotintime}, we show how our predictions for the evolution of the solar wind mass loss rate compare with other constraints from the literature.
The solid area shows the range of wind mass loss rates from our model shown in the left panel of Fig.~\ref{fig:solarMdotintime}.
The solid black lines show predictions of the mass loss rate along our 10th and 90th percentile rotation tracks from the model of \citet{2011ApJ...741...54C}\footnotemark.
The \citet{2011ApJ...741...54C} model predicts mass loss rates that are a factor of a few higher than the mass loss rates that we predict. 
This is mostly because the mass loss rate that we used for the current solar wind is approximately a factor of 2.3 lower than the mass loss rate predicted by the \citet{2011ApJ...741...54C} model, which is slightly higher than \mbox{$3 \times 10^{-14}$~\mdot}. 
As we show in Fig.~5 of Paper~I, based on spacecraft measurements of the solar wind mass flux, this is likely to be too large.
The black dotted lines in Fig.~\ref{fig:solarMdotintime} show the predictions of the \citet{2011ApJ...741...54C} model scaled down by a factor of 2.3. 
These predictions match our model very well, with the \citet{2011ApJ...741...54C} model predicting a slightly stronger decrease in solar wind mass loss rate with age, and therefore a slightly higher mass loss rate at 100~Myr.

The solid blue line in Fig.~\ref{fig:solarMdotintime} shows the solar wind mass loss rate scaled backwards in time by assuming $\dot{M}_\star \propto t^{-1.23}$, as predicted by \citet{2013PASJ...65...98S}.  
The age dependence is slightly steeper than our prediction, but very similar, and is in good agreement with the predictions of the \citet{2011ApJ...741...54C} model.
Our model could very easily be made to agree with the models of \citet{2011ApJ...741...54C} and \citet{2013PASJ...65...98S} if we assumed a slightly weaker scaling between the dipole field strength and rotation, which is certainly plausible.
\citet{2014A&A...570A..99S} assumed the wind temperature is a constant and scaled the wind base density with coronal electron densities.
Since the wind mass loss rate in their model is proportional to the base density, for stars with a given radius, their Eqn.~5 implies that $\dot{M}_\star \propto L_\text{X}^{1/2}$.
Combined with the dependence between rotation and \mbox{X-ray} emission from \citet{2011ApJ...743...48W} and Eqn.~\ref{eqn:mamajek}, the model of \citet{2014A&A...570A..99S} implies $\dot{M}_\star \propto t^{-0.62}$, which is very similar to our determination.

\footnotetext{
We calculate the mass loss rates from the model of \citet{2011ApJ...741...54C} using the BOREAS code (\citealt{2011ascl.soft08019C}), assuming solar metallicity and taking into account the evolution of the solar radius and luminosity.
}

In Fig.~\ref{fig:solarMdotintime}, we also compare our predictions to observational constraints. 
The downward pointing arrows at the top of the figure show upper limits on the mass loss rates for three young solar analogues derived from non-detections of stellar winds in radio by \citet{2000GeoRL..27..501G}.
Clearly all other wind mass loss rate predictions shown in Fig.~\ref{fig:solarMdotintime} are consistent with these upper limits.
More important are the measurements of the mass loss rates of several solar analogues by \citet{2002ApJ...574..412W}, \citet{2005ApJ...628L.143W}, and \citet{2014ApJ...781L..33W}. 
The mass loss rates were derived by measurements of excess absorption of the stellar Ly$\alpha$ emission line, which is thought to be a result of neutral hydrogen walls at the edges of the stellar astrospheres. 
The solid red line in Fig.~\ref{fig:solarMdotintime} shows the relation between mass loss rate and age $\dot{M}_\star \propto t^{-2.33}$ suggested by \citet{2005ApJ...628L.143W}.

The history of the solar wind suggested by Ly$\alpha$ mass loss rate measurements is in stark contradiction to our predictions for the mass loss rates. 
Our results suggest a much weaker dependence of mass loss rate on stellar age than the $t^{-2.33}$ dependence suggested by \citet{2005ApJ...628L.143W}, and we do not find any break down of the winds at ages younger than 700~Myr.
Our mass loss rate predictions are a result of considering the rotational evolution of solar mass stars, and can be calculated based on the dependence of rotational evolution on wind mass loss rates.
It is interesting to consider what implications the scenario of \citet{2005ApJ...628L.143W} would have for the rotational evolution of solar mass stars.
By 700~Myr, the rotation rates of solar mass stars have mostly converged onto the slowly rotating track, and the stars spin down approximately according to Eqn.~\ref{eqn:mamajek}, such that $\Omega_\star \propto t^{-0.566}$, which implies that $d \Omega_\star / dt \propto \Omega_\star^{2.8}$.
Assuming $\dot{M}_\star \propto t^{-2.33}$ implies that $\dot{M}_\star \propto \Omega_\star^{4.12}$.
Inserting $B_{\text{dip}} \propto \Omega_\star^{1.32}$ and this dependence into Eqn.~\ref{eqn:a_derivation2} gives $d \Omega_\star / dt \propto \Omega_\star^{4.5}$, which contradicts the Skumanich style spin down. 

A possible reason why the age dependence of the mass loss rate derived by \mbox{\citet{2005ApJ...628L.143W}} might be overestimated is that they combined stars of different masses to derive a single relation between $F_\text{X}$ and \mbox{$\dot{M}/R_\star^2$}.
However, our results suggest a further $M_\star^{-3.36}$ dependence for the mass loss rate, and so the results could have been biased by the fact that the intermediate activity stars in the sample of \mbox{\citet{2005ApJ...628L.143W}} were mostly K~stars, whereas the low activity stars were mostly G~stars.
The four most Sun-like single stars in the sample of \mbox{\citet{2005ApJ...628L.143W}} are $\epsilon$~Eri, 61~Cyg~A, $\epsilon$~Ind, and 61~Vir.
Including the Sun in this sample, we find that a power-law fit between $F_\text{X}$ and $\dot{M}/R_\star^2$ gives 
\begin{equation}
\frac{\dot{M}}{R_\star^2} \propto F_\text{X}^{1.33},
\end{equation}
\noindent in agreement with the results of \mbox{\citet{2005ApJ...628L.143W}}.
If we instead include the $M_\star^{-3.36}$ dependence, we find
\begin{equation} \label{eqn:newWoodfit}
\frac{\dot{M}}{R_\star^2 M_\star^{-3.36}} \propto F_\text{X}^{1.18},
\end{equation}
\noindent which is indeed shallower\footnotemark.
However, combining this with the relation \mbox{$F_\text{X} \propto t^{-1.74}$} used by \mbox{\citet{2005ApJ...628L.143W}} leads to an age dependence of $\dot{M}$ that is still significantly steeper than our prediction. 
A large part of this contradiction can be further resolved if we consider that the $F_\text{X} \propto t^{-1.74}$ relation is likely too strong.
Combining $\Omega_\star \propto t^{-0.566}$ and $R_\text{X} \propto \Omega_\star^{2.18}$ from \citet{2011ApJ...743...48W} gives $F_\text{X} \propto t^{-1.23}$, which combined with Eqn.~\ref{eqn:newWoodfit} gives us a relation for solar mass stars of
\begin{equation} \label{eqn:newWooddependence}
\dot{M} \propto t^{-1.46},
\end{equation} 
\noindent which is much closer to the \mbox{$\dot{M} \propto t^{-0.75}$} that we derive.
If we instead assume that \mbox{$\Omega_\star \propto t^{-0.5}$} (\citealt{1972ApJ...171..565S}) and \mbox{$R_\text{X} \propto \Omega_\star^{2}$} (\citealt{2014arXiv1408.6175R}), we find \mbox{$\dot{M} \propto t^{-1.18}$}.
In Fig.~\ref{fig:solarMdotintime}, we compare Eqn.~\ref{eqn:newWooddependence} with the other predictions.
The previous $t^{-2.33}$ determination appears to fit $\epsilon$~Eri much better than the new determination, though it could simply be that the age of $\sim$1~Gyr that we use is overestimated; the gyrochronological age derived by \citet{2007ApJ...669.1167B} of 460~Myr fits Eqn.~\ref{eqn:newWooddependence} much better.

\footnotetext{
The $F_\text{X}^{1.18}$ dependence is highly sensitive to the physical parameters for $\epsilon$~Eri, since it is the only star at intermediate activity in the sample.
The radius of 0.74~R$_\odot$ that we use was derived by \citet{2012ApJ...744..138B} by comparing their measured interferometric radius with the star's parallax.
They used their results to derive a mass of 0.82~M$_\odot$ and age $\sim$1~Gyr from the Yonsei-Yale isochrones.
}

The suggestion that the relation between $\dot{M}_\star$ and age breaks down at an age of 700~Myr, with younger stars having low mass loss rates, is also difficult to reconcile with the observed rotational evolution of stars at these ages. 
Assuming that young solar analogues have mass loss rates similar to that of the current solar wind is similar to assuming in our rotational evolution model that $a \sim 0$ at these ages, where $\dot{M}_\star \propto \Omega_\star^a$.
The rotation track that we would expect for a rapidly rotating star in this scenario is shown as the dark green line in the left panel of Fig.~\ref{fig:influenceofa}.
In the first few hundred million years after the ZAMS, there is almost no spin down, in contradiction to the observational constraints.
For example, observations indicate that a star at the 90th percentile of the distribution of rotation periods at 150~Myr would have a rotation rate of $\sim$50~$\Omega_\odot$, and that it would spin down to $\sim$5~$\Omega_\odot$ by 550~Myr.
However, the rotation track in Fig.~\ref{fig:influenceofa} indicates that if we assume the mass loss rates are similar to that of the current solar wind, then the star would instead only spin down to $\sim$30~$\Omega_\odot$.

\begin{figure}
\centering
\includegraphics[trim=5mm 5mm 5mm 5mm,width=0.49\textwidth]{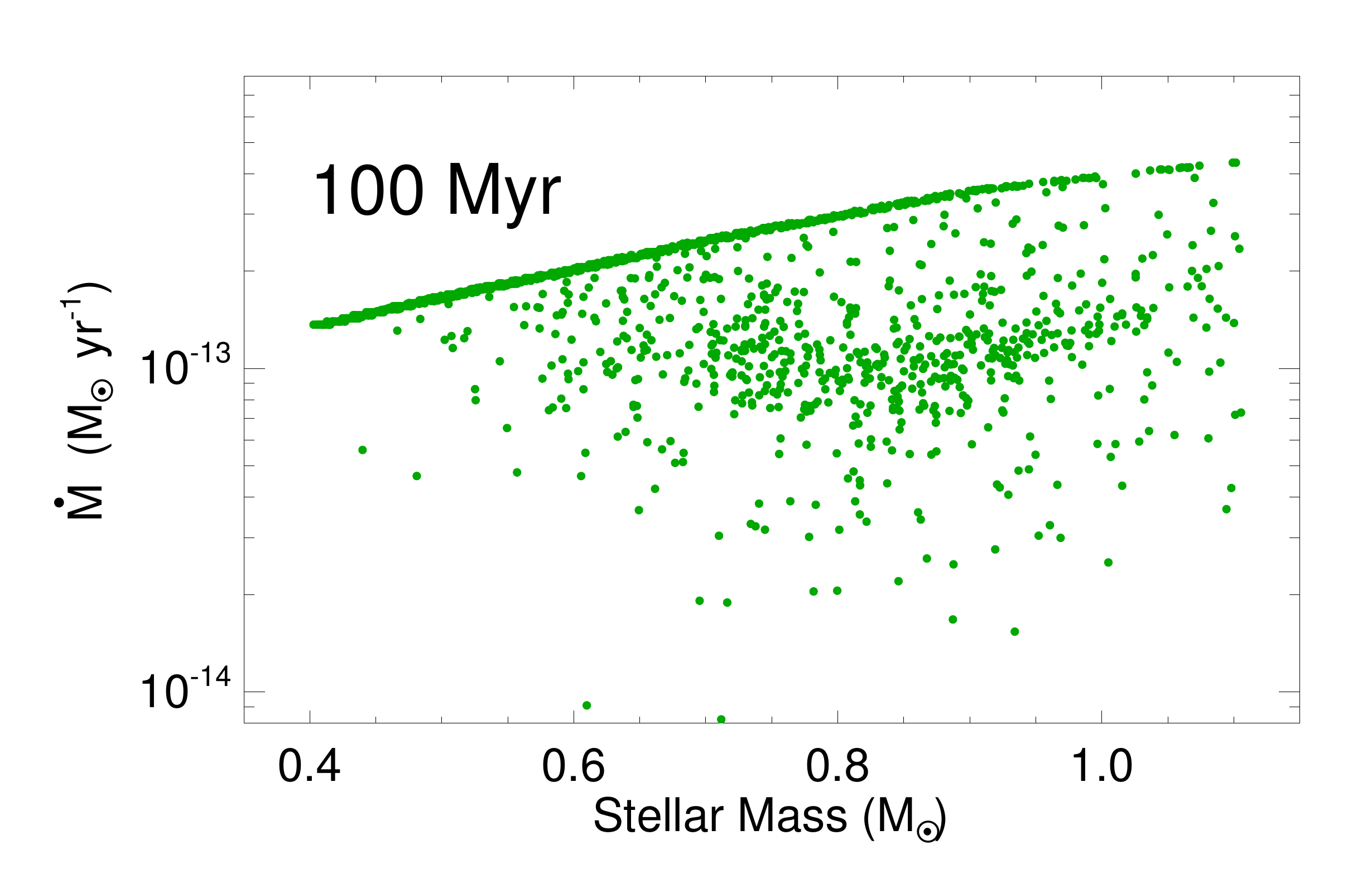}
\includegraphics[trim=5mm 5mm 5mm 5mm,width=0.49\textwidth]{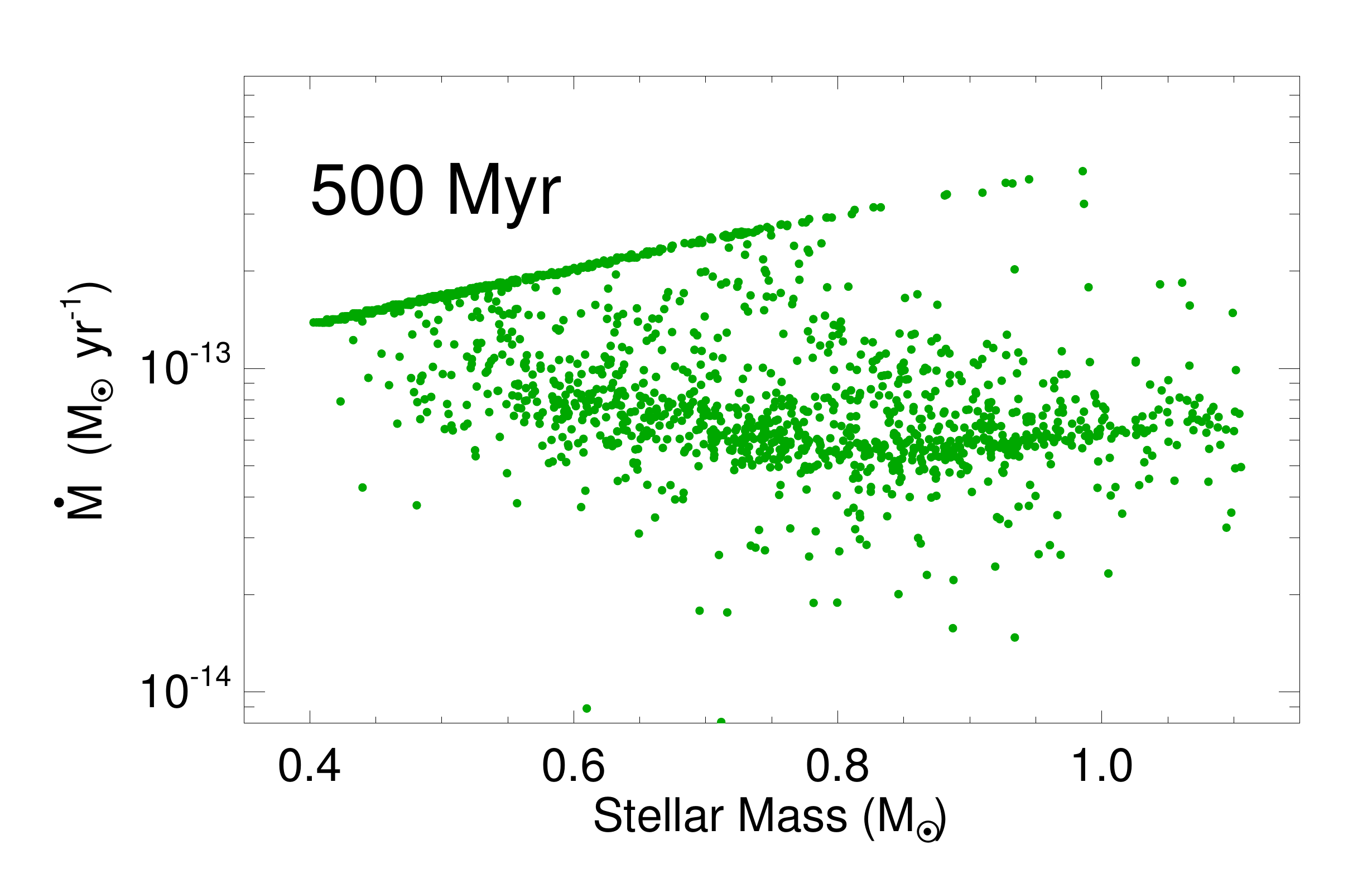}
\includegraphics[trim=5mm 5mm 5mm 5mm,width=0.49\textwidth]{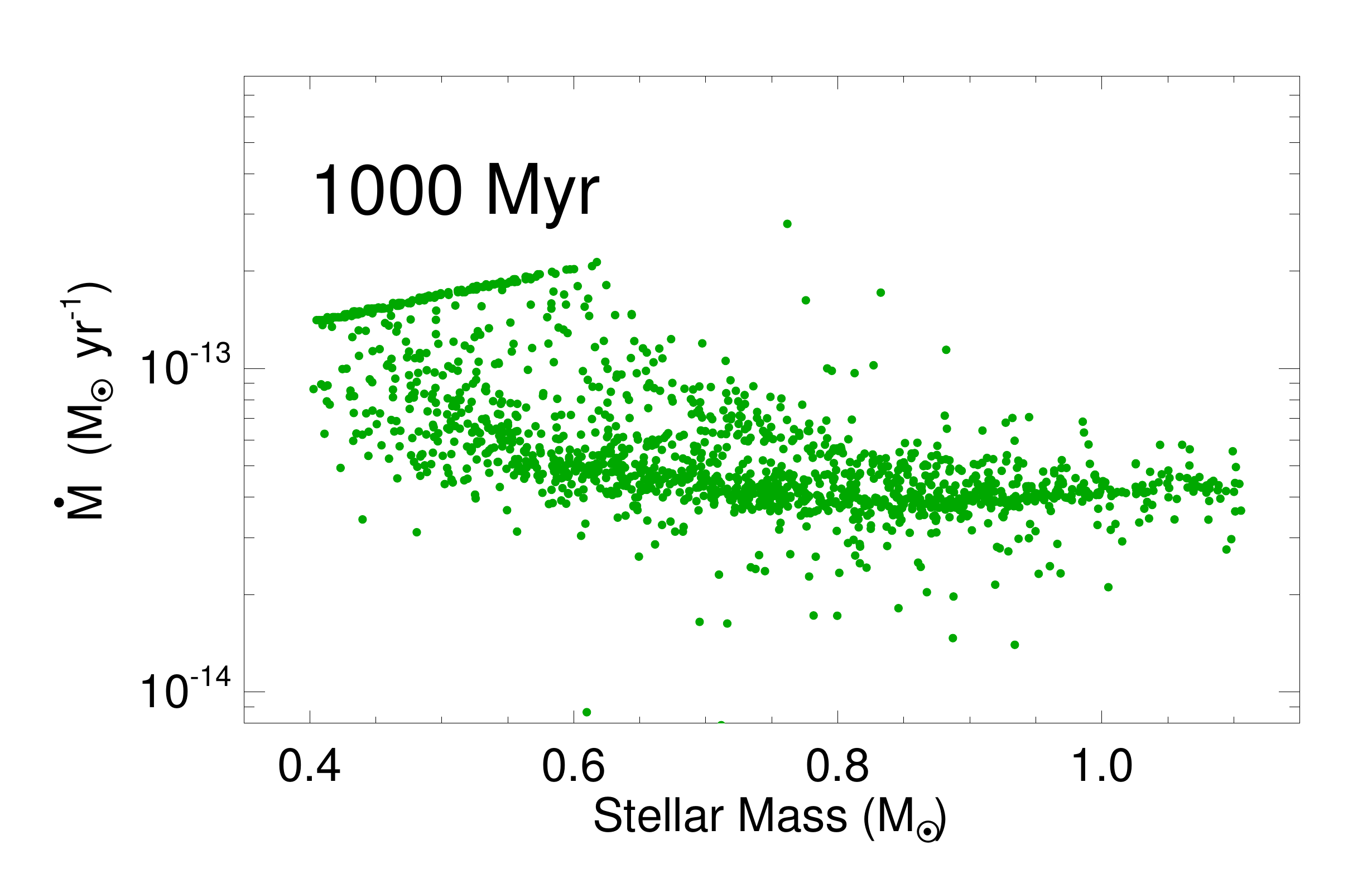}
\includegraphics[trim=5mm 5mm 5mm 5mm,width=0.49\textwidth]{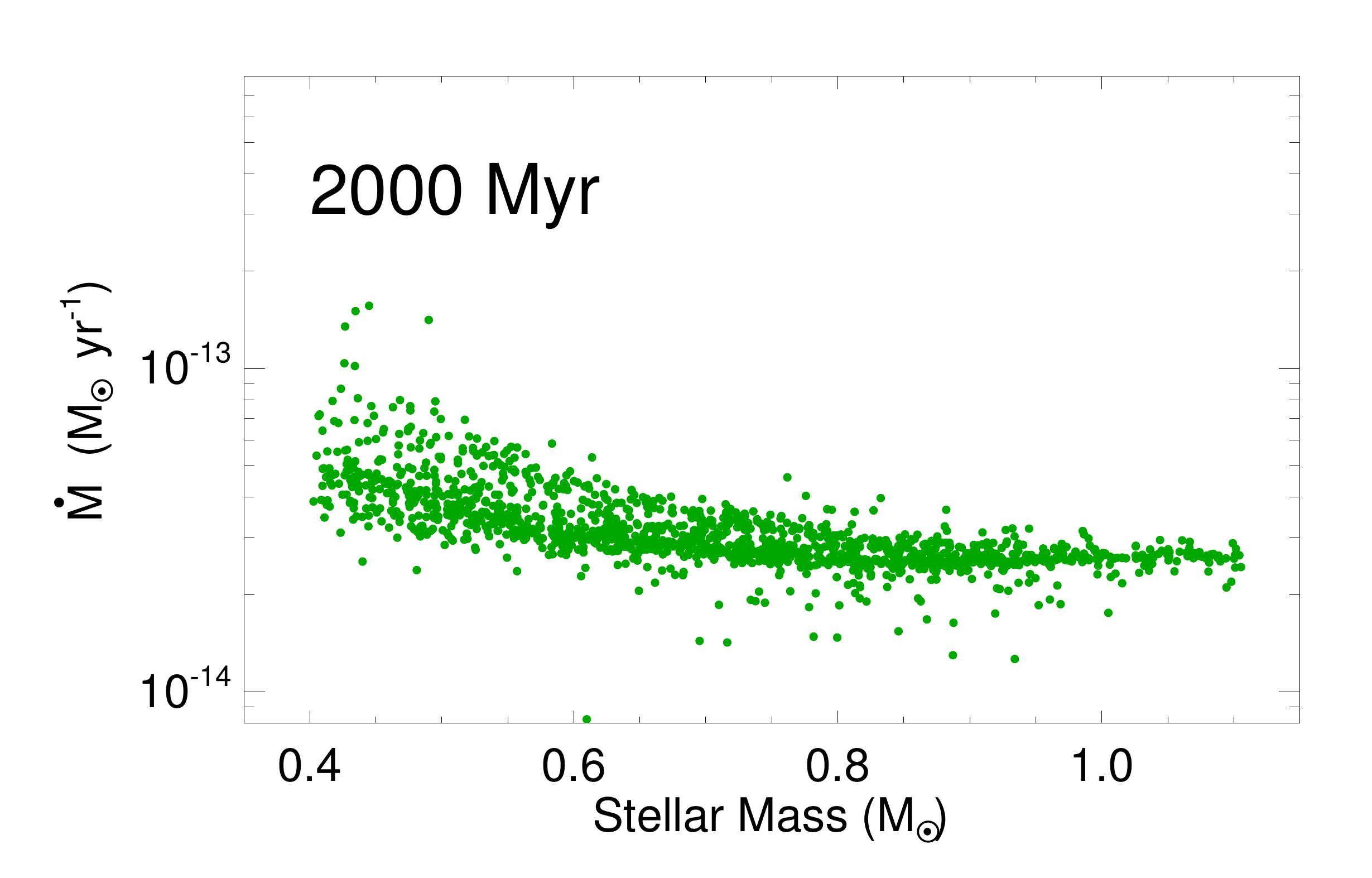}
\caption{
Figure showing the distribution of stellar mass loss rates at 100~Myr, 500~Myr, 1000~Myr, and 2000~Myr based on our scaling law for mass loss rate and the distributions of rotation rates shown in Fig.~\ref{fig:superclusterevolution}.
}
 \label{fig:superclusterMdot}
\end{figure}

In Fig.~\ref{fig:solarwind1AUproperties}, we show the wind speeds and densities for the slow and fast winds calculated from both Model~A and Model~B as a function of age.
Although the mass fluxes in Model~A and Model~B are the same, the two models can lead to significantly different predictions for the density and speed of the wind.  
In Model~A, since we scale the wind temperature with coronal temperatures, rapidly rotating stars have much higher wind temperatures than slowly rotating stars, and therefore much higher wind speeds. 
On the other hand, the wind temperature for Model~B is determined by the surface escape velocity, and is therefore independent of rotation. 
For Model~B, the solar wind speeds are approximately constant in time, with small decreases of about 70~\kms\hspace{0mm} between 100~Myr and 5~Gyr.
The decrease in the wind speed is due to the expansion of the Sun which causes the wind base temperature to decrease.
For Model~A, the wind speeds are much larger at young ages, with values as high as 2000~\kms\hspace{0mm} for the fast wind in the saturated regime. 
In both models, large changes in the proton densities at 1~AU are seen between the different rotation tracks, and between 100~Myr and 1~Gyr.
However, the variations in the densities in Model~A are a factor of a few smaller than the variations in the densities in Model~B.
The differences in the densities between Model~A and Model~B cancel out the differences in wind speeds, leading to the same mass loss rates in both models. 
An important difference between the two models is therefore the wind ram pressure, which is given by $\rho v^2$.
Since the wind ram pressure is more sensitive to the wind speed than the density, in Model~A the ram pressures for the young solar wind will be a factor of a few higher than in Model~B.


We should however be cautious when interpreting the results for the 90th percentile track shown in Fig.~\ref{fig:solarwind1AUproperties}.
For the most rapidly rotating stars, magneto-rotational acceleration of the wind is likely to dominate over thermal acceleration. 
As we show in Paper~I, the rotation rates at which solar mass stars transition into the FMR regime depend on which model we use to calculate the wind temperature: for Model~A and Model~B respectively, the transition from the SMR regime to the FMR regime happens at \mbox{$\sim60 \Omega_\odot$} and \mbox{$\sim15 \Omega_\odot$}.
Since most solar mass stars never reach such high rotation rates, we conclude that the Sun probably did not make it into the FMR regime, though there is still a significant possibility that it did. 
The dashed lines in the upper panel of Fig.~\ref{fig:solarwind1AUproperties} show the Michel velocities along the 90th percentile rotation track where they become larger than the thermal speeds.
For a solar mass star at the 90th percentile of the rotational distribution at 100~Myr, the terminal wind speed in the equatorial plane is likely to be $\sim$2000~\kms, regardless of the wind temperature. 
The winds therefore have a much lower density in the equatorial plane than we would predict from thermal acceleration alone.

There is clearly a lot of uncertainty in the properties of the solar wind at young ages.
This is not just due to the lack of theoretical understanding of stellar winds, but also because we do not know how fast the Sun was rotating in its first few hundred million years.
In the following section, we extend our discussion to the winds of lower-mass stars.

\begin{figure}
\centering
\includegraphics[trim=5mm 5mm 5mm 5mm,width=0.43\textwidth]{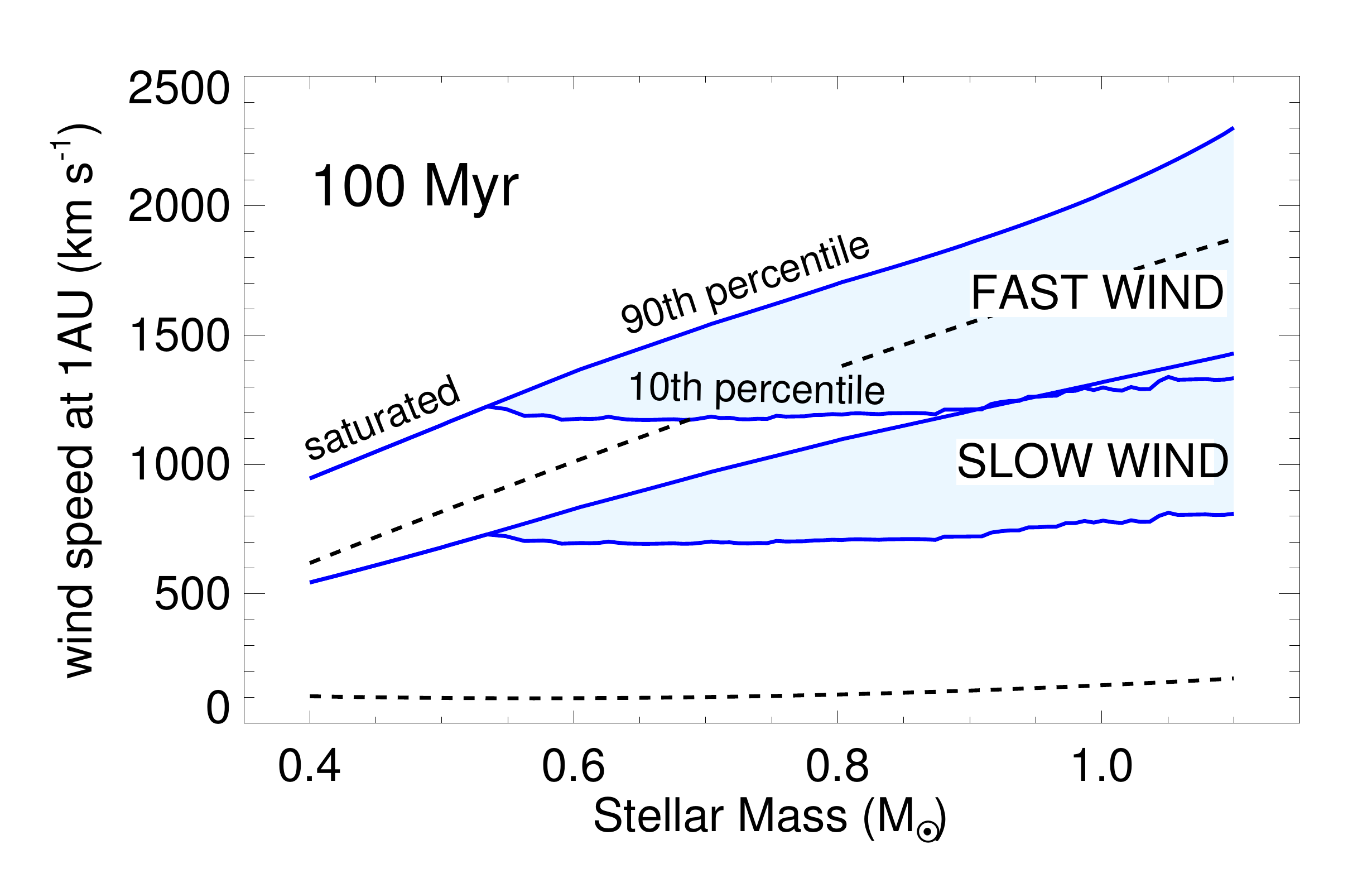}
\includegraphics[trim=5mm 5mm 5mm 5mm,width=0.43\textwidth]{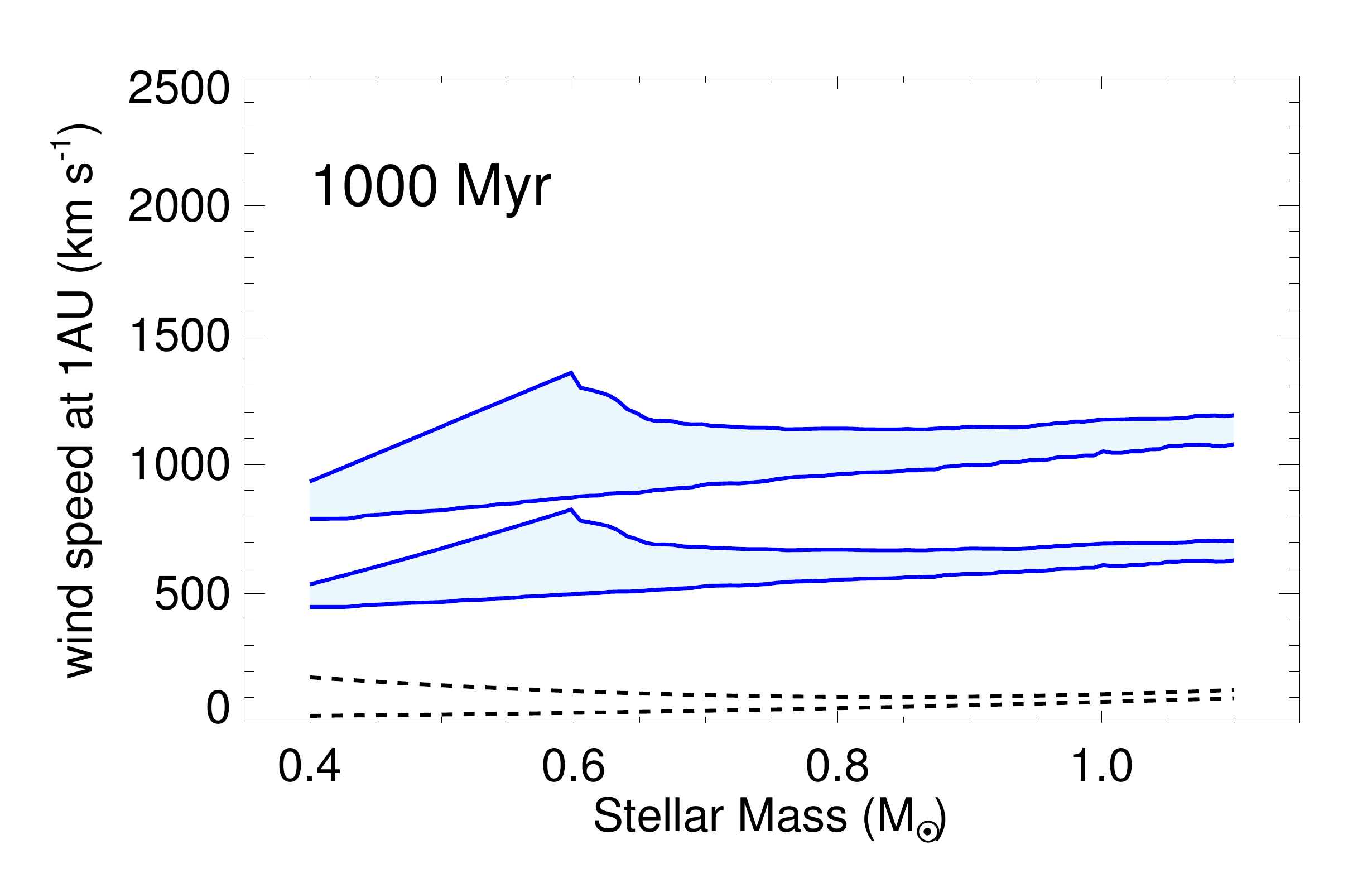}
\includegraphics[trim=5mm 5mm 5mm 5mm,width=0.43\textwidth]{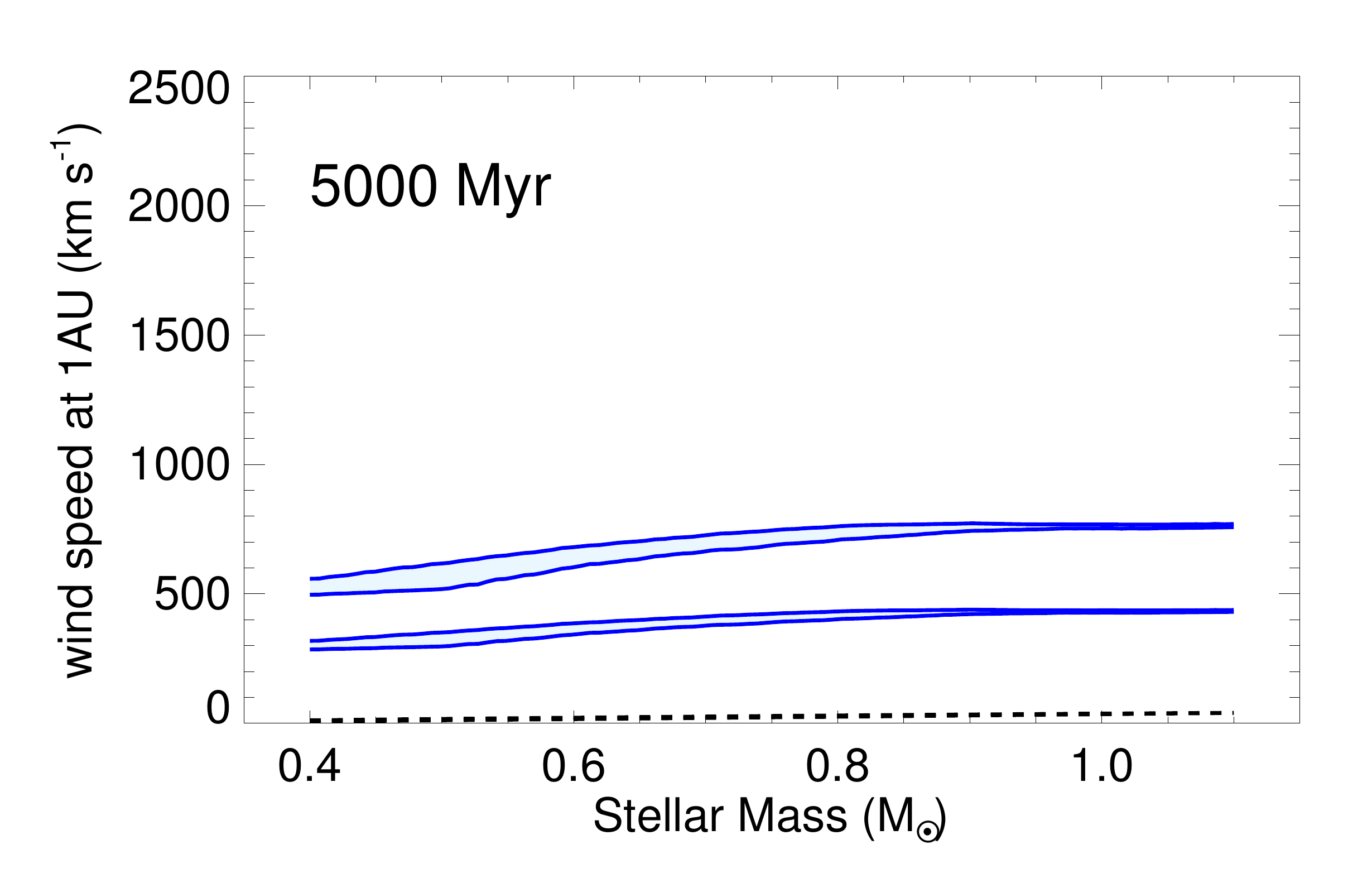}
\caption{
Figure showing the evolution of the slow and fast wind speeds at 1~AU for different stellar masses assuming Model~A for determining the wind temperature. 
In this model, we assume that the wind temperature can be scaled with the coronal temperature.
We calculate the wind speed at 1~AU using Eqn.~\ref{eqn:windspeed}.
The three panels correspond to different ages, and in each panel, the lower and upper lines correspond to the wind speeds for stars at the 10th and 90th percentiles of the rotational distributions respectively. 
In the case of the low-mass stars in the upper panel, there is only one line because both percentiles lie above the saturation threshold, and therefore have the same wind temperatures.
The lower and upper dashed black lines show the Michel velocity along the 10th and 90th percentile tracks, which is an estimate of the terminal wind speed in the equatorial plane due to magneto-rotational acceleration, as discussed in Section~\ref{sect:windmodel}.
}
 \label{fig:windvelpercentiles}
\end{figure}

\subsection{The winds of stars from 0.4~M$_\odot$ to 1.1~M$_\odot$} \label{sect:windsallmasses}

\begin{figure*}
\centering
\includegraphics[trim=5mm 5mm 5mm 5mm,width=0.44\textwidth]{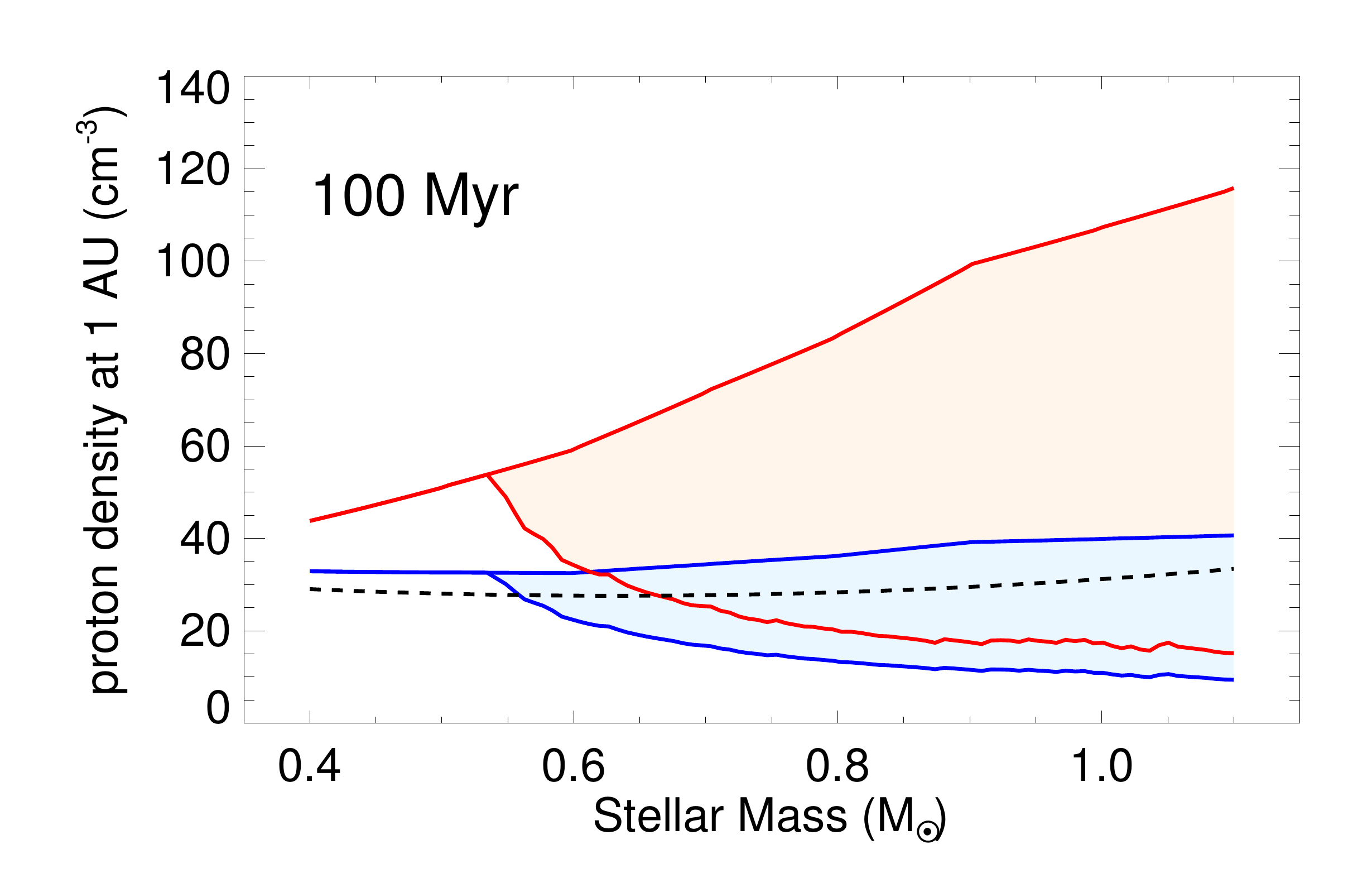}
\includegraphics[trim=5mm 5mm 5mm 5mm,width=0.44\textwidth]{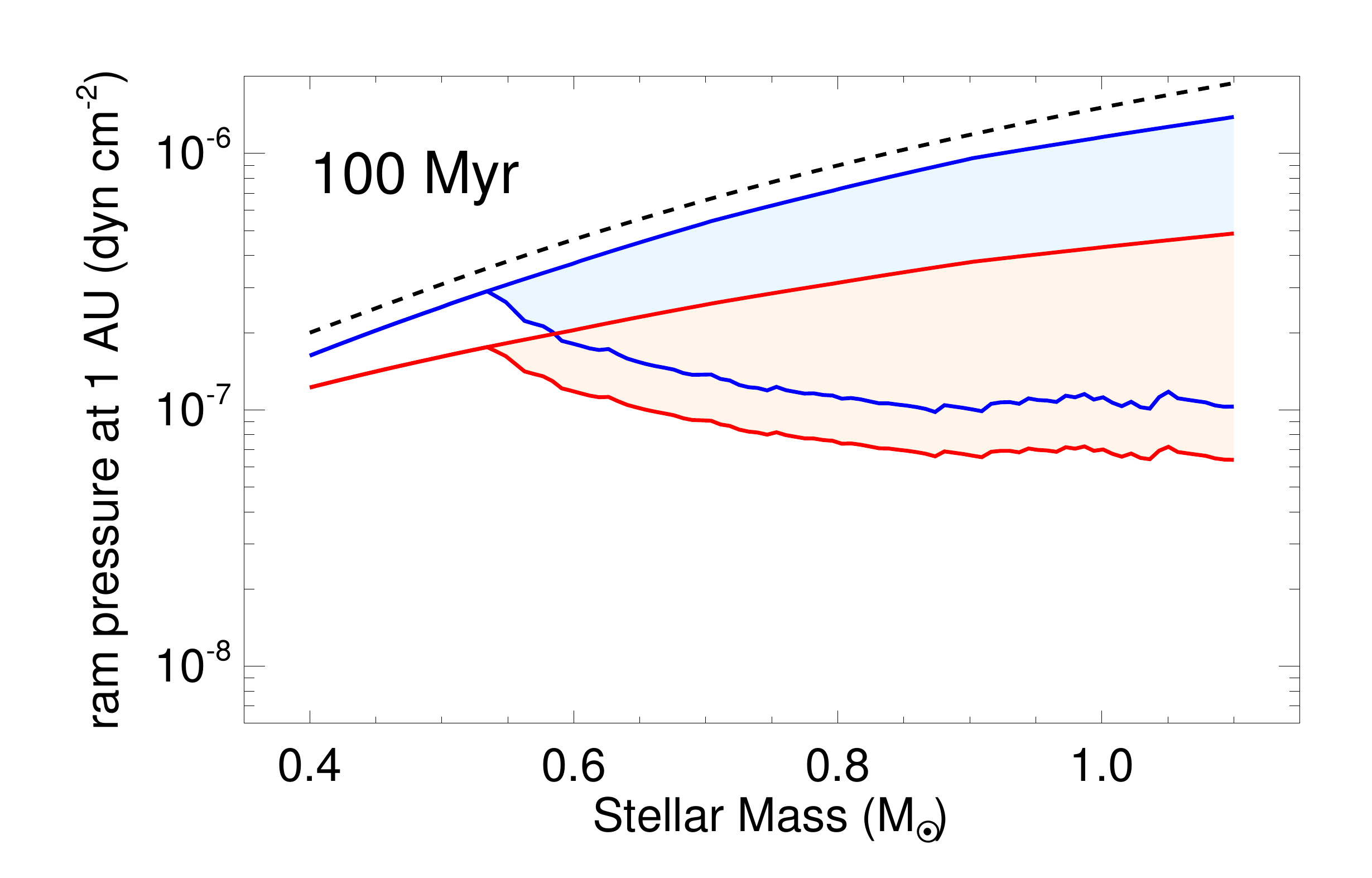}
\includegraphics[trim=5mm 5mm 5mm 5mm,width=0.44\textwidth]{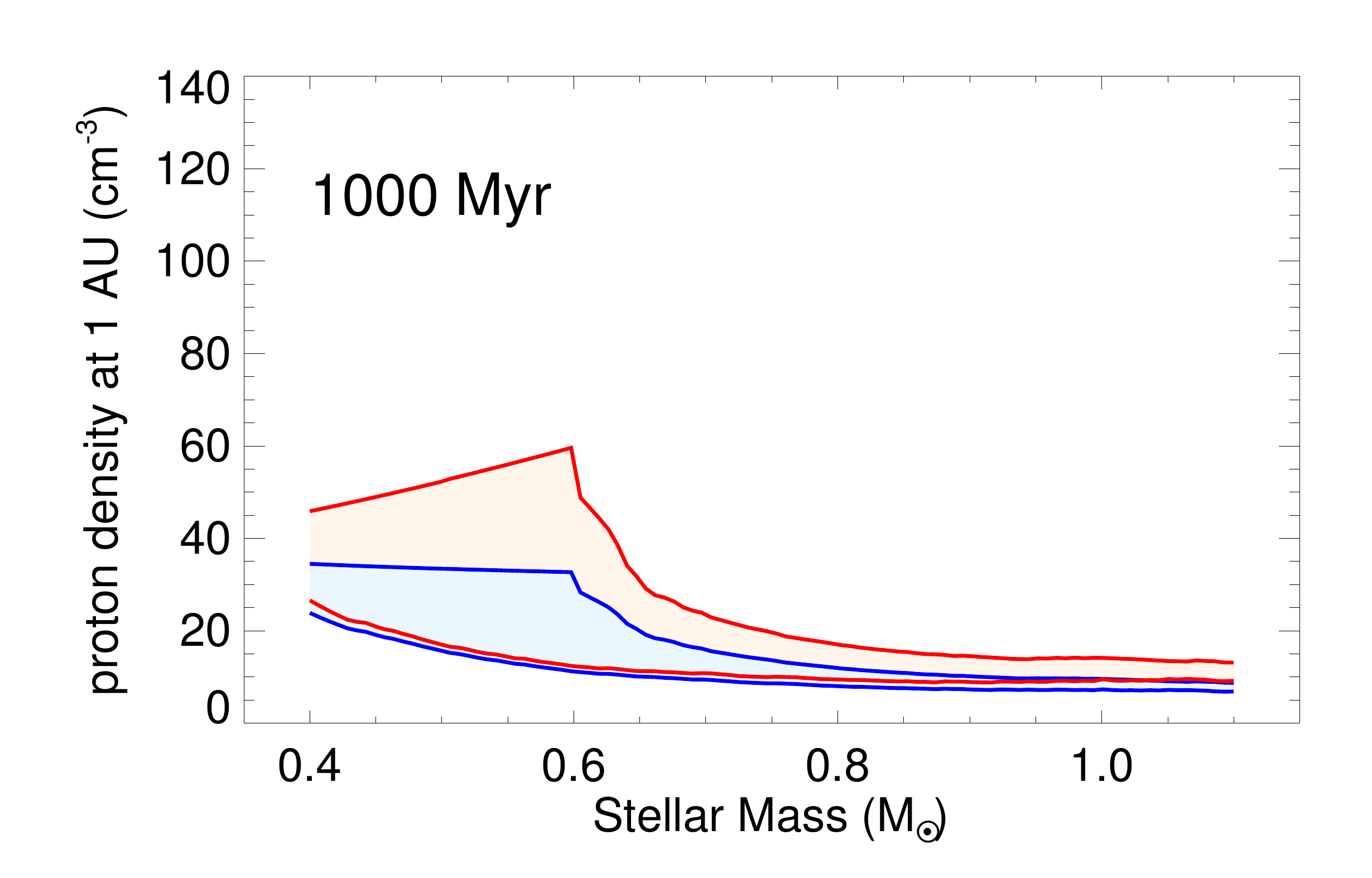}
\includegraphics[trim=5mm 5mm 5mm 5mm,width=0.44\textwidth]{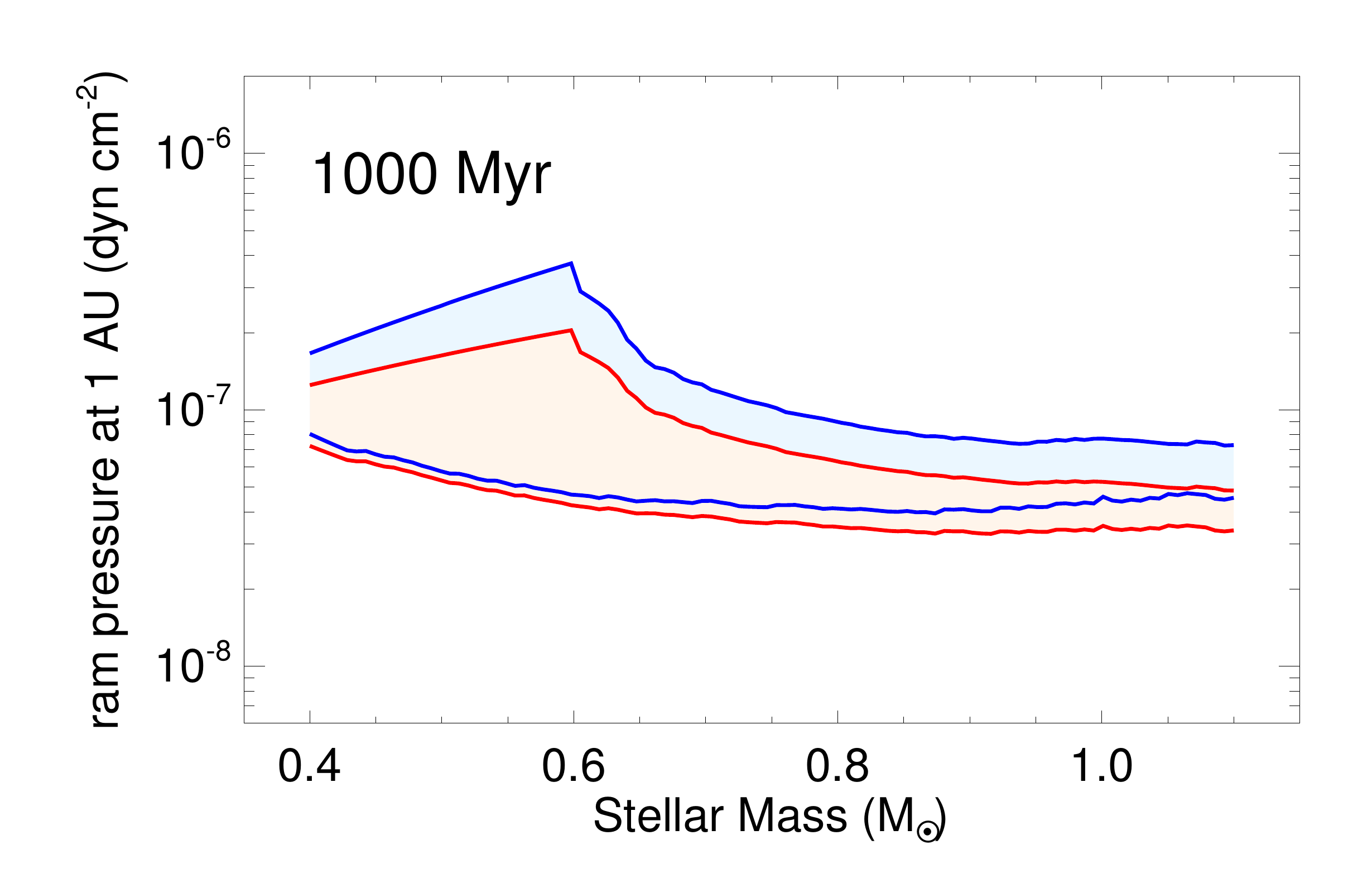}
\includegraphics[trim=5mm 5mm 5mm 5mm,width=0.44\textwidth]{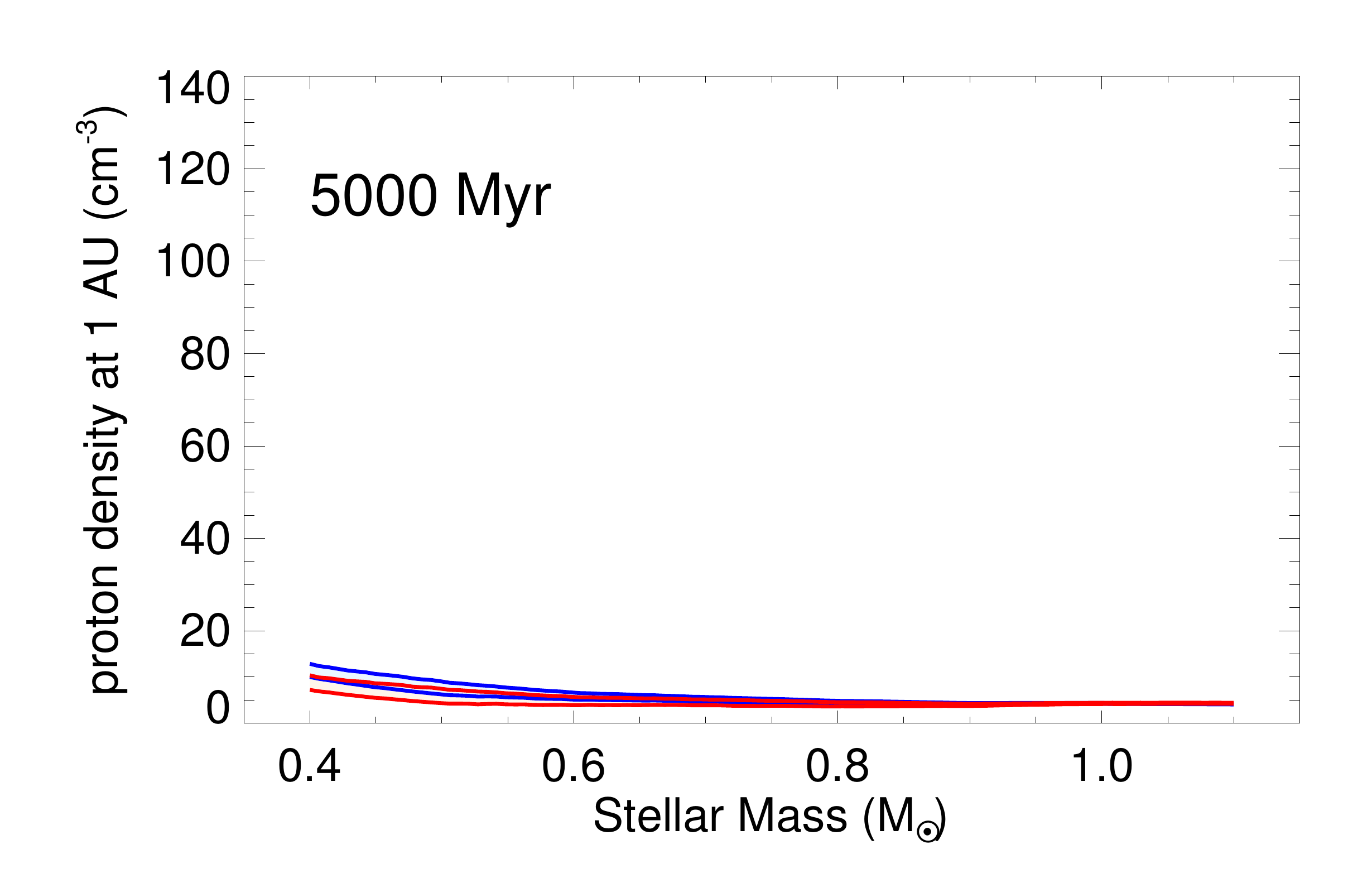}
\includegraphics[trim=5mm 5mm 5mm 5mm,width=0.44\textwidth]{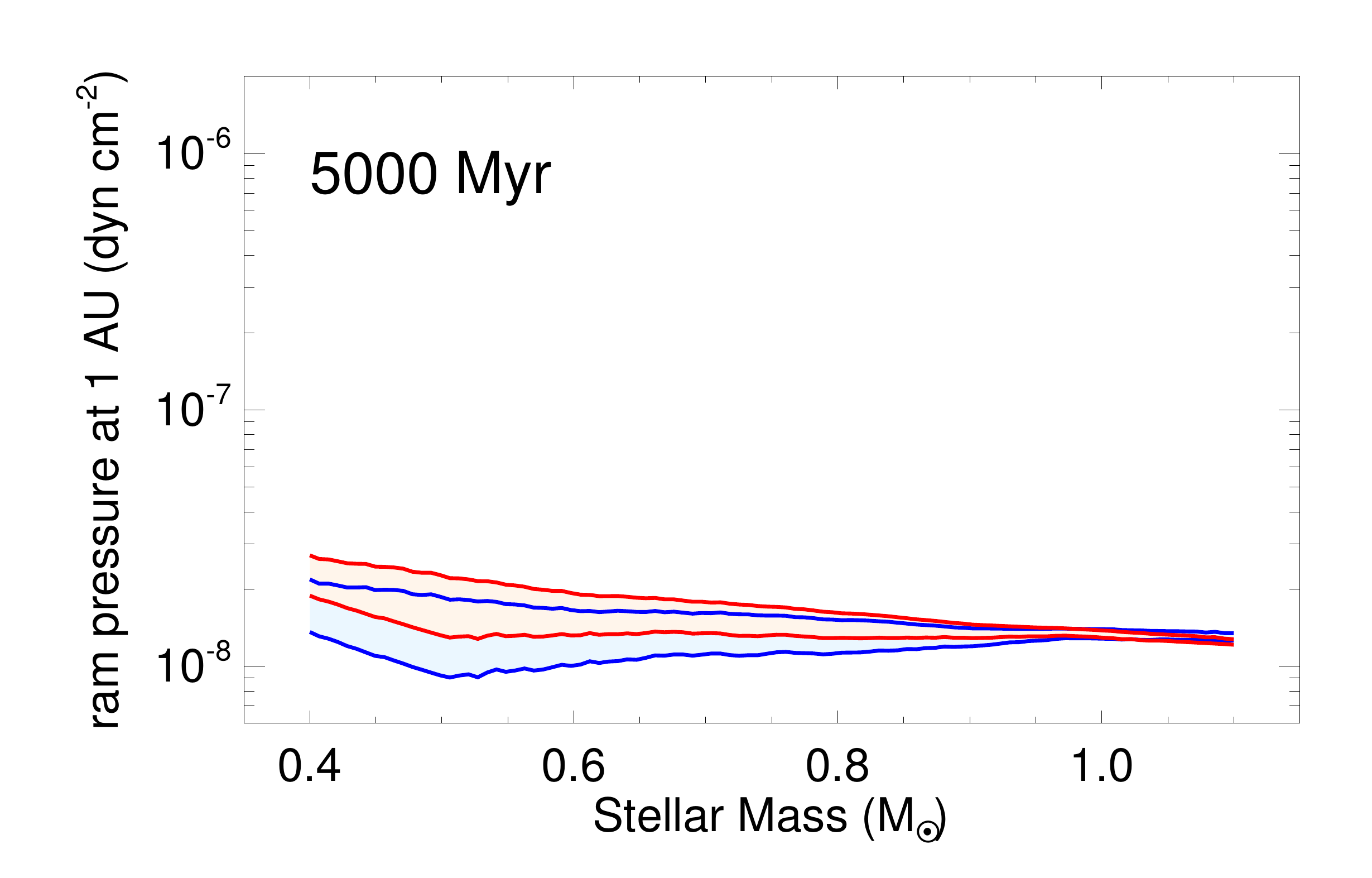}
\caption{
Figure showing proton densities (\emph{left column}) and ram pressures (\emph{right column}) as a function of stellar mass at three different stellar ages for the slow wind component.
In each panel, blue corresponds to Model~A and red corresponds to Model~B, and the lower and upper lines correspond to the 10th and 90th percentiles of the rotational distributions.
}
 \label{fig:winddenspercentiles}
\end{figure*}


We are now in the position to study the time evolution of stellar wind properties on the main-sequence for a range of stellar masses.
In Fig.~\ref{fig:superclusterevolution}, we show the time evolution of rotation periods for a distribution of $\sim$1500 stars from 100~Myr to 2000~Myr.  
In Fig.~\ref{fig:superclusterMdot}, we show the evolution of the distribution of mass loss rates for the stars in this sample.
The mass loss rates for different stellar masses at the 10th, 50th, and 90th percentiles of the rotational distribution at different ages are given in Table~\ref{tbl:percentiles2}.

The mass dependence of the mass loss rate depends strongly on the rotation rate.
In the unsaturated regime, low-mass stars have higher mass loss rates by a factor of a few than high-mass stars, at a given rotation rate.
Once all stars are in the saturated regime, the mass dependence of mass loss rate has been reversed, with the low-mass stars having the lowest mass loss rates. 
This is due to the low saturation threshold for low-mass stars.
Combining Eqn.~\ref{eqn:Mdotassumption} and Eqn.~\ref{eqn:saturation}, and assuming that $R_\star \propto M_\star^{0.8}$, gives 

\begin{equation}
\dot{M}_{\text{sat}} \approx 37 \dot{M}_\odot \left( \frac{M_\star}{M_\odot} \right)^{1.3},
\end{equation}

\noindent where $\dot{M}_{\text{sat}}$ is the saturation mass loss rate. 
This means that on the main-sequence, the maximum possible mass loss rate for low-mass stars is lower than the maximum possible mass loss rate for high-mass stars. 
This also means that the saturation mass loss rate for solar mass stars is \mbox{37~$\dot{M}_\odot$}, corresponding to \mbox{$5.2 \times 10^{-13}$~M$_\odot$~yr$^{-1}$} (in Table~\ref{tbl:percentiles2}, the saturation threshold for solar mass stars at 100~Myr is slightly lower than this value because in those calculations we take into account the slow increase of the stellar radius with age on the main-sequence).

At 100~Myr, most low-mass stars lie at the saturation mass loss rate and only a few have low mass loss rates. 
The distribution at higher masses also has a significant number of stars at the saturation threshold.
However, for stars with masses above 0.7~M$_\odot$, the distribution has a very strong cluster of stars with relatively low mass loss rates, corresponding to $\sim10^{-13}$~\mdot, which is not much higher than the mass loss rate of the current solar wind. 
For example, for 0.8~M$_\odot$ stars, the mass loss rates at the 10th and 50th percentiles of the rotational distribution are $5 \dot{\text{M}}_\odot$ and $10 \dot{\text{M}}_\odot$ respectively.
This is unsurprising given that these stars are only rotating a few times faster than the current Sun.
At 500~Myr and 1000~Myr, the low mass loss rate feature is more dominant, with a much smaller number of stars lying at the saturation value.
By 2000~Myr, no saturated stars remain at any mass, and all of the stars lie on the track of low mass loss rates. 
At this age, the mass loss rates of low-mass stars are generally slightly higher than the mass loss rates of high-mass stars.
This track is influenced by two competing effects: on the one hand, low-mass stars have higher mass loss rates at a given rotation rate than high-mass stars, but on the other hand, low-mass stars rotate slower than high-mass stars.
The result is only a weak dependence of the mass loss rate on stellar mass at a given age.

In Fig.~\ref{fig:windvelpercentiles}, we show the slow and fast wind speeds at 1~AU for stars lying on the 10th and 90th percentiles of the rotational distributions at 100~Myr, 1000~Myr, and 5000~Myr, based on Model~A. 
Along the 10th percentile track, there is almost no dependence on stellar mass.
However, along the 90th percentile track, the wind speeds are much higher for high-mass stars due to the higher saturation threshold.
The saturation wind speeds for 0.5~M$_\odot$ stars are $\sim$700~\kms\hspace{0mm} and $\sim$1200~\kms\hspace{0mm} for the slow and fast wind models respectively. 
At later ages, due to their slower rotation, the wind speeds are lower for low-mass stars. 
We do not show in Fig.~\ref{fig:windvelpercentiles} the wind speeds calculated from Model~B since they are approximately constant with age and independent of stellar mass.

The lower and upper dashed lines in each panel of Fig.~\ref{fig:windvelpercentiles} show the Michel velocities along the 10th and 90th percentiles of the rotational distributions.
At 100~Myr, along the 10th percentile track, the Michel velocity is much lower than the thermal wind speeds. 
On the other hand, along the 90th percentile track, the Michel velocity is larger than the thermal wind speeds of the slow component, and therefore are likely to determine the wind speeds, even in Model~A.
At later ages, the Michel velocity becomes much lower than all of the thermal wind speeds, and therefore magneto-rotational effects can be ignored.

In Fig.~\ref{fig:winddenspercentiles}, we show the time evolution of wind density and ram pressure at 1~AU for both Model~A and Model~B.
For simplicity, we only show the results for the slow wind model, though we find similar trends for the fast wind model. 
As in previous plots, blue corresponds to Model~A and red corresponds to Model~B, and the lower and upper lines in all panels correspond to the 10th and 90th percentiles of the rotational distributions respectively. 
The influence of magneto-rotational acceleration of the wind is shown as the dashed black line in the upper panels of Fig.~\ref{fig:winddenspercentiles}.
For both models, the wind densities and ram pressures at 100~Myr are higher than at 5000~Myr due to the higher mass fluxes.
At 100~Myr, the densities for Model~B are a factor of a few larger than for Model~A due to the lower wind speeds. 
On the other hand, the ram pressures calculated from Model~A are larger than the ram pressure from Model~B.
This is because the ram pressure, proportional to $\dot{M}_\star v$, has a stronger dependence on wind speed than the mass flux.

The evolution of the distributions of wind density and ram pressure is interesting for both Model~A and Model~B.
At 100~Myr, there is no spread in these properties for low-mass stars since they are mostly saturated.
This lack of spread is of course partly artificial since we assume one universal relation between mass loss rate and stellar mass and rotation, whereas in reality, there is likely to be some scatter about this relation.  
Going to higher masses, there begins to be a spread in the wind densities and ram pressures between the slowest and fastest rotators and this spread has a strong mass dependence. 
By 1~Gyr, due to the fast convergence of the rotation rates, this spread in wind properties has almost disappeared for solar mass stars.
On the other hand, a spread in the wind properties for low-mass stars has developed.
By 5~Gyr, since all of the stellar rotation rates have converged, there is almost no spread in the wind properties at any stellar mass.

Interestingly, at 5~Gyr, the wind properties do not have a strong mass dependence, and due to the slow evolution of rotation at later ages, also evolve slowly with age.
In studies that require only a crude estimate for the properties of an individual inactive star's wind, it would be reasonable as a first approximation to simply assume the measured solar wind properties, even if the star has a different mass to the Sun.
This is only true for slowly rotating stars, and it is unclear how well this extends to stars with masses below 0.4~M$_\odot$.
Of course detailed modelling of the winds of individual stars should be preferred when possible.


\section{Summary and Discussion} \label{sect:summary}

Stellar rotation and winds are fundamentally linked on the main-sequence. 
Stellar winds remove angular momentum from their host stars, causing their rotation rates to decrease as they age.
At the same time, the decrease in the rotation rates of stars likely causes large changes in the wind strengths.
In Paper~I of this series, we develop a stellar wind model based on scaling the solar wind to other stars.
In this paper, we develop an observationally driven rotational evolution model which we couple to our wind model.
We develop the rotational evolution model for two purposes: firstly, we use the model to constrain the wind mass loss rates of low-mass main-sequence stars as a function of stellar mass, radius, and rotation, and secondly, we use the results of the model to show how wind properties evolve on the main-sequence.

To constrain our rotational evolution models, we collect measured rotation periods for stars at different ages to determine the evolution of the distribution of rotation rates on the main-sequence.
Probably the major gap in our observational understanding of rotational evolution is the rotational distribution of low-mass stars ($M_\star < 0.8 M_\odot$) at intermediate ages ($\sim$1-2~Gyr).
In Fig.~\ref{fig:supercluster100}, we show the observationally constrained rotational distributions at 150~Myr and 550~Myr.
Statistics related to these distributions are given in Table~\ref{tbl:percentiles}.
At 150~Myr, there is a massive spread in the rotation periods of stars of all masses, though higher mass stars are more concentrated in the slowly rotating track.
By 550~Myr, the rotation rates of solar mass stars have mostly converged, and some convergence has taken place for lower mass stars.
The reason for this difference is likely to be the lower saturation threshold for mass loss rate and magnetic field strength for lower mass stars.

It is interesting to speculate about where the Sun might have been in the distribution of rotation rates at 100~Myr. 
At this age, approximately 70\% of solar mass stars are on the slowly rotating track.
The dominance of the slowly rotating track can be seen from the fact that the 10th and 50th percentiles are at 4~$\Omega_\odot$ and 7~$\Omega_\odot$ respectively.
Naively, we would therefore expect that the Sun at 100~Myr was a slow rotator, with a rotation rate of $\sim$4~days.
However, the rotational distribution contains a significant rapidly rotating tail, with the 90th percentile being at 50~$\Omega_\odot$.
There is a significant possibility that the Sun was a rapid rotator at 100~Myr.
This uncertainty is important because these scenarios correspond to radically different histories for the Sun's magnetic activity and therefore its levels of high energy radiation and the strengths of its winds.
It is common in the literature to use the scaling laws of \citet{2005ApJ...622..680R} to predict how the Sun's \mbox{X-ray}, EUV, and FUV radiation has evolved in time.
This corresponds approximately to the assumption that at 100~Myr, the Sun was rotating with a period of $\sim$2.7~days (the rotation period of EK~Dra), which is at the 75th percentile of the rotational distribution, and is therefore a relatively rapid rotator with a relatively high level of magnetic activity. 
If instead, the Sun had a rotation period of 5~days, its \mbox{X-ray}, EUV, and FUV fluxes would have been significantly lower, likely corresponding to the levels of emission for $\pi^1$~UMa in the sample of \citet{2005ApJ...622..680R}, which has an age of $\sim$300~Myr.
If the Sun had been an extremely rapid rotator at 100~Myr, it would have had a level of emission similar to that of EK~Dra, given that EK~Dra is close to the saturation threshold, but this level of emission would not have immediately decayed with age.
Instead, the Sun would have taken several hundred million years to spin down to the saturation threshold, and during this time, the emission would have remained at the saturation level.
Only after a few hundred million years would the emission have started to decay. 
These uncertainties are very significant for studies of the evolution of planetary atmospheres given the sensitivity of their upper atmospheres to the level of high-energy radiation from the central star (\citealt{2008JGRE..113.5008T}; \citealt{2010AsBio..10...45L}).

The basis of our rotational evolution model is a formula for wind torques derived by \citet{2012ApJ...754L..26M} from a grid of MHD wind models. 
The formula relates wind torque to the stellar properties, mass loss rates, and magnetic field strengths.
For the magnetic field strengths, we scale the solar magnetic field strength to other stars using the scaling law for the large scale field given by \citet{2014MNRAS.441.2361V}.
We then specify the mass loss rate by \mbox{$\dot{M}_\star \propto R_\star^2 \Omega_\star^{a} M_\star^{b}$}, where \mbox{$a \approx 1.33$} and \mbox{$b \approx -3.36$} are free parameters in our model that we fit to the observational constraints.
Another free parameter in our model is the mass dependence of the saturation threshold for $\dot{M}_\star$ and $B_\text{dip}$, which we assume is $M_\star^{2.3}$.
We choose this value since it is the closest to the weaker mass dependences derived for the saturation of \mbox{X-ray} emission. 
Interestingly, we find that the mass loss rate in the saturated regime depends on $\dot{M}_\star^{1.3}$, which is very different to the $\dot{M}_\star^{-3.36}$ dependence that we find in the unsaturated regime. 
These things can all be put together to derive simple scaling laws for the wind torque, as we show in Section~\ref{sect:rotevobestfit}.
In the unsaturated regime, the wind torque is $\tau_\text{w} \propto \Omega_\star^{2.89}$, and in the saturated regime, the wind torque is $\tau_\text{w} \propto \Omega_\star M_\star^{4.42}$ (where the constants of proportionality are different for these two scaling laws).  

Given that we have not considered the rotational evolution of fully convective stars (\mbox{$M_\star < 0.35$~M$_\odot$}), it is unclear how well our scaling laws apply in this mass regime.
\citet{2001ApJ...547L..49W} measured astrospheric Ly$\alpha$ absorption for the $\alpha$~Centauri system but was unable to achieve a measurement for the much lower mass Proxima~Centauri, and suggested that Proxima~Centauri has a mass loss rate that is at least an order of magnitude lower than the mass loss rate in the $\alpha$~Centauri system. 
Given the parameters of these stars, we would expect from our scaling laws that Proxima~Centauri has a slightly larger mass loss rate than both components of the $\alpha$~Centauri system combined, which is inconsistent with the results of \citet{2001ApJ...547L..49W}.
This could be evidence for a fundamental difference in the properties of the winds of fully and partially convective stars.
Although highly speculative, such an interpretation is consistent with the fact that the properties of global photospheric magnetic fields appear to differ significantly between these two types of stars, both on the main-sequence (\citealt{2008MNRAS.390..567M}) and on the pre-main-sequence (\citealt{2012ApJ...755...97G}), and with the significantly different rotational evolution that has been suggested for fully convective stars (\citealt{2011ApJ...727...56I}).

A major problem with stellar wind models is the lack of observational constraints on wind properties.
In the final section of Paper~I, we discussed promising methods for constraining wind properties, such as comparisons between astrospheric Ly$\alpha$ absorption and radio interferometric measurements of radio emission from the wind, or stellar observations during planetary transits. 
The rotational evolution of low-mass main-sequence stars can potentially provide the most important observational constraints on stellar wind properties. 
The formulae provided in this paper could be used as simple estimates for wind torques, which can then be used to constrain free parameters, such as the base density, in more complex MHD wind models, or to check the results when constraints on these parameters are already available. 
The scaling laws given above can be used to estimate the wind torques, but they are not likely to be realistic for the most rapidly rotating stars.
More accurate is to calculate the torque from Eqn.~\ref{eqn:modifyMatttorque}, using Eqn.~\ref{eqn:matttorque} to get $\tau'$.
Eqn.~\ref{eqn:Mdotassumption} and Eqn.~\ref{eqn:dipoleRossby} should be used as input into Eqn.~\ref{eqn:matttorque} for $\dot{M}_\star$ and $B_\text{dip}$ (where Eqn.~11 of \citealt{2011ApJ...743...48W} should be used to calculate the convective turnover times).
If the star is a rapid rotator, then Eqn.~\ref{eqn:saturation} should be used to calculate the saturation rotation rate, with $c=2.3$.

In Section~\ref{sect:windevo}, we combine the results of our rotational evolution model with our wind model developed in Paper~I to show how wind properties evolve on the main-sequence for stellar masses between 0.4~M$_\odot$ and 1.1~M$_\odot$. 
For the solar wind, we find a much weaker dependence of mass loss rate on age than predicted by \citet{2005ApJ...628L.143W}, with the highest mass loss rates at young ages that are only a factor of 20 above that of the current solar wind.
We also find no evidence for weak mass loss rates at young ages.
The strong winds at young ages, and the weak dependence of mass loss rate on age at intermediate and late ages, appear necessary for solar mass stars to spin down in a way that is consistent with observational constraints on rotational evolution.
We show that the early solar wind was likely to have been denser than the current solar wind, though the magnitude of this change depends in most cases on how wind temperatures scale to other stars, since temperature determines the wind speed.
This is not true for the most rapidly rotating stars, given that the wind speeds are determined by magneto-rotational acceleration.

The large spread in rotation rates for young stars leads to large spreads in the wind properties, though this spread is to some extent decreased by saturation.
Interestingly, due to the fact that most stars with masses $\sim$0.5~M$_\odot$ lie above the saturation threshold at 100~Myr, this spread is not present at these masses. 
As stars spin down, the spread in wind properties of solar mass stars disappears, but a small spread for lower mass stars develops.
However, by a few Gyr, this spread has completely disappeared due to the convergence of rotation rates.

Our motivation for this study is to provide important input into further studies that attempt to improve our understanding of the complex interactions between stars and the evolving atmospheres of potentially habitable planets. 
One interesting question in this area is the way in which the current Earth's atmosphere formed.
It is possible that the primordial Earth formed with a dense hydrogen atmosphere captured during the Sun's classical T Tauri phase (\citealt{1979E&PSL..43...22H}; \citealt{2014MNRAS.439.3225L}) which must have been lost early in the Earth's evolution.
It is easy to imagine how this could have happened if the Sun was born as a rapid rotator: in this scenario, the Sun would have remained highly active for a long time, driving significant thermal and non-thermal escape from the Earth.
However, not only was the Earth able to lose this original hydrogen atmosphere, but it was able to do so while keeping the thin nitrogen dominated atmosphere that it currently has. 
\citet{2010Icar..210....1L} estimated that at ages below $\sim$1~Gyr, the combination of a stronger solar wind compressing the Earth's magnetosphere and a higher EUV flux inflating the Earth's atmosphere leads to a significant portion of the atmosphere being exposed to the solar wind. 
They estimated that this would have caused so much non-thermal escape of atmospheric gas that the current Earth's atmosphere would have been removed in a very short amount of time.
A possible solution to the problem identified by \citet{2010Icar..210....1L} is that the Sun was instead a slow rotator in its early main-sequence life, and therefore did not have EUV emission and wind strengths significantly above current levels at any time after the ZAMS.

Although there are potential solutions to these problems that do not depend on the early Sun's level of activity, the above considerations show that the rotation history of the Sun can be a significant uncertainty in our understanding of the formation of the Earth's atmosphere. 
The Earth is the only planet that we know has formed in such a way as to be habitable. 
The origins and the evolution of stellar rotation, and how these connect to winds and high-energy radiation, are important factors that need to be considered when studying how often this is likely to have happened in other stellar systems.

\section{Acknowledgments} 

We thank the referee, Brian Wood, for his hard work on both our papers and for providing useful comments on the paper that have helped us to improve the original manuscript.
CPJ thanks Aline Vidotto for comments and help on the paper.
CPJ, MG, and TL acknowledge the support of the FWF NFN project S116601-N16 ``Pathways to Habitability: From Disks to Active Stars, Planets and Life'', and the related FWF NFN subproject S116604-N16 ``Radiation \& Wind Evolution from the T~Tauri Phase to ZAMS and Beyond''. 
This publication is supported by the Austrian Science Fund (FWF).


\bibliographystyle{aa}
\bibliography{mybib}

\end{document}